# Interaction Design for Socially Assistive Robots for People with Developmental Disabilities

by

**Xiaodong Wu**

B.Sc., Hope College, 2016

Thesis Submitted in Partial Fulfillment of the
Requirements for the Degree of
Master of Science

in the
School of Interactive Arts and Technology
Faculty of Communication, Art and Technology

© Xiaodong Wu 2019
SIMON FRASER UNIVERSITY
Summer 2019

# Approval

**Name:**                      **Xiaodong Wu**

**Degree:**                  **Master of Science**

**Title:**                        **Interaction Design for Socially Assistive Robots for People with Developmental Disabilities**

**Examining Committee:**     **Chair:**    Marek Hatala
                                                 Professor

                                     **Lyn Bartram**
                                     Senior Supervisor
                                     Professor

                                     **Carman Neustaedter**
                                     Supervisor
                                     Associate Professor

                                     **Wolfgang Stuerzlinger**
                                     External Examiner
                                     Professor
                                     School of Interactive Arts & Technology
                                     Simon Fraser University

**Date Defended:**           **Aug 14, 2019**

# Ethics Statement

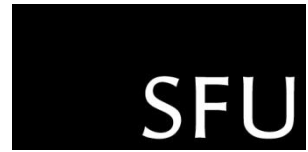

The author, whose name appears on the title page of this work, has obtained, for the research described in this work, either:

    a.    human research ethics approval from the Simon Fraser University Office of Research Ethics

or

    b.    advance approval of the animal care protocol from the University Animal Care Committee of Simon Fraser University

or has conducted the research

    c.    as a co-investigator, collaborator, or research assistant in a research project approved in advance.

A copy of the approval letter has been filed with the Theses Office of the University Library at the time of submission of this thesis or project.

The original application for approval and letter of approval are filed with the relevant offices. Inquiries may be directed to those authorities.

Simon Fraser University Library
Burnaby, British Columbia, Canada

Update Spring 2016

# Abstract

Social robots, also known as service or assistant robots, have been developed to improve the quality of human life in recent years. Socially assistive robots (SAR) are a special type of social robots that focus on providing support through social interaction. The design of socially capable and intelligent robots can vary, depending on the target user groups. In this work, I assess the effect of socially assistive robots' roles, functions, and communication approaches in the context of a social agent providing service or companionship to users with developmental disabilities. In this thesis, I describe an exploratory study of interaction design for a socially assistive robot that supports people suffering from developmental disabilities. While exploring the impacts of visual elements to robot's visual interface and different aspects of robot's social dimension, I developed a series of prototypes and tested them through three user studies that included three residents with various function levels at a local group home for people with developmental disabilities. All user studies had been recorded for the following qualitative data analysis. Results show that each design factor played a different role in delivering information and in increasing engagement, and there are more aspects of HRI to consider besides robot's graphical user interface and speech, such as proxemics and robot's physical appearance and dimensions. I also note that some fundamental design principles that would work for ordinary users did not apply to our target user group. I conclude that socially assistive robots could benefit our target users and acknowledge that these robots were not suitable for certain scenarios based on the feedback from our users.

# Acknowledgements

I would like to express my very great appreciation to my supervisor, Dr. Lyn Bartram, for providing insightful guidance and support for my study. I would also like to acknowledge Dr. Neustaedter for spending time reviewing my drafts and giving me invaluable feedback.

I must express my gratitude to my dear parents who sincerely care about my life here and offer spiritual support continuously. I wish to thank my girlfriend, Ping, for her company and encouragement.

I would also like to express my deep gratitude to Dr. Edgington and Dr. Abrahantes, my undergraduate research supervisors, for their patient guidance, enthusiastic encouragement and useful critiques of my past research work.

I am grateful to Ms. Keegan O'Toole, Ms. Bri Davidson, and other caregivers from the Camsell group home who gave me full support in the past two years. Without them, my thesis project would not go this far.

Finally, I must acknowledge *JDQ Inc.* and *Mitacs* who provided funding for this project through the Mitacs Accelerate program. I must give a big thanks to Mr. Jon Morris and Dr. Sina Radmard who not only supervised my internship and research but also provided guidance and support to my study.

# Table of Contents

# List of Tables

# List of Figures

# List of Acronyms

| | |
|---|---|
| **ASD** | Autism Spectrum Disorder |
| **DD** | Developmental Disabilities |
| **DDA** | Developmental Disabilities Association |
| **GUI** | Graphical User Interface |
| **HCI** | Human-Computer Interaction |
| **HHI** | Human-Human Interaction |
| **HRI** | Human-Robot Interaction |
| **SAR** | Socially Assistive Robotics |

## Chapter 1

# Introduction

## 1.1 Background

**Developmental disabilities** (DD) are a set of permanent and severe problems that are challenging many people nowadays. As the *2012 Canadian Survey on Disability (CSD)* has revealed, "160,500 (0.6% of Canadian adults) were identified as having a developmental disability" (Bizier, Fawcett, & Gilbert, 2015). This survey also indicated that 90% of adults with a developmental disability needed assistance with some kind of everyday activity, and 72.7% of them reported some degree of unmet need for at least one of these activities. Individuals with developmental could be considered as "less desirable patients" due to their difficulties with communication (Lewis et al., 2002), which leads to potential health care inequalities. Due to the complexity of developmental disabilities, these people urgently need skilled and knowledgeable health professionals and caregivers (Sullivan et al., 2011), but there is a huge gap for this need. That being said, not every one of these individuals has the opportunity to get daily care and systematic therapy.

In the past, there was some research work investigating the use of technologies such as speech generating devices in communication interventions (Rispoli et al., 2010) and robotic assistive therapy (Shamsuddin et al., 2014) for people with developmental disabilities. As HCI researchers, we are interested in exploring different technologies and designing meaningful interaction to help them. In recent years, many service robots, also known as assistive robots, are being developed and introduced to various user groups including the elderly. We see the potential of social robotics and determine to carry out our interaction research on the foundation of Human-Robot Interaction (HRI). We believe that socially assistive robots are a great medium for filling in the gap mentioned above, and that they can improve the life qualities of people with developmental disabilities.

To date, there is not much study of using assistive robots for people with developmental disabilities, which is a rather specific group of users. However, there is a large body of work investigating the application of sociable service robot for eldercare (Broekens et al., 2009; Baisch et al., 2017; Matsumoto et al., 2011). Although the elderly are not precisely our user group, these two groups do share many similarities and there is a notable overlap of users between these groups. Due to aging, the elderly tend to demonstrate decreasing functional levels, increasing dependency, and many of them get cognitive impairments of functions such as memory, judgement, and language (Baek et al., 2011; Udekwu et al., 2001; Wu et al., 2012). At the same time, people with DD often require more assistance to learn, understand or express information than others, because developmental disabilities can dreadfully affect their language and social skills (Developmental Services Ontario, 2016). Recent development in robotics has realized the assignment of sociable robots act as "considerate, flexible, and trustworthy caregivers" (K. J. Kim, Park, & Shyam Sundar, 2013). Socially assistive robots could be used for eldercare for two reasons: functionalities and abilities to provide affective assistance (companionship) (Broekens et al., 2009). Because of the similarities of the elderly and people with DD, we believe in the potential of using socially assistive robots to help and accompany individuals with DD, and to improve the quality of their lives. Our research team is trying to develop a new type of sociable robot whose characteristics are different from traditional robots – **socially assistive robots** (SAR). They are expected to have certain social intelligence to be capable of more than companionship with individuals and cognitive assistance, and thus to become a functioning part of the environment and community. As an HCI researcher in this team, I aim to design effective interaction for the robot to be capable of providing affective support, in addition to the functions as mentioned above.

This thesis presents a study of social interaction design of socially assistive robots specifically for people with DD. We explored the social dimension of HRI through three user studies. Our study started with an investigation on how caregivers communicate with their residents with developmental disabilities. Through this process, we analyzed the interaction patterns of both sides. The results obtained from this pretest study helped us with our research and design which we are presenting in this thesis. We gained an insight into our users' needs and cooperated with caregivers to find practical solutions to the challenges that both users and caregivers are facing every day. Therefore, our study could be significant for designing a supportive SAR that can improve the life qualities of people having daily challenges because of their developmental disabilities.

## 1.2 Users and the Environment

People with DD are a large user group in which there are substantial individual differences of cognitive levels, skills, characteristics, etc. In this study, we worked with a local group home of the Developmental Disabilities Association (DDA). There were around 5 residents living at the group home and 3 to 5 caregivers on site during the day time who provided residents with regular training, care, entertainment, and scheduling. Staffing was also provided at night to accommodate residents.

Residents had their own daily activities and they spent a lot of time together in the group home. At this assisted living facility, residents got customized living arrangements based on their needs. The most common types of assistance that residents needed were completing tasks like brushing teeth and getting dressed, and getting reminded of important routine like taking daily medication. In the morning, caregivers helped residents change and get dressed, take showers, and proceed to their social routine. Later, they might take residents out, do activities, involve residents in the community (e.g. fellowship, church, dancing). Caregivers also taught residents to play with technologies for entertainment, like Siri (a virtual personal assistant by Apple), tablets, and wireless headphones. Residents all got assigned tasks and chores to do in order to improve their independence and maintain their function levels. They also got regular training, like recognizing objects from flash cards. They have preferred (e.g. tea time), less-or-non-preferred (e.g. chores), and optional activities (going out for coffee). Caregivers typically used rewards as a primary method of motivating residents. Some of the rewards were more formal like snacks and entertainment time, whereas some were merely verbal motivators like "You did a great job today!"

At the group home, each of the residents had their own bedrooms where there were their favourite personal objects and medical equipment, if any. There is also a communal area (a kitchen and a dining room that is also a living room), and a quiet space for residents to relax – the Snoezelen room. A Snoezelen room is a calming, isolated multi-sensory environment where people get stimulated and relaxed. Residents can get training, mental developmental, and therapy in this room. Residents typically spent the majority of their indoor time in the living room, where they can watch TV and chat with each other or caregivers. They spent time in their private rooms when they want a rest or calmness. We will provide a detailed description of our study participants in Section 3.3.

## 1.3 Conceptual Framework

HRI is a multidisciplinary research area concerning the "analysis, design, modelling, implementation, and evaluation of robots for human us" (Fong, Thorpe, & Baur, 2001). Recent advance in the use of social interaction and affective expression in HRI (Bethel & Murphy, 2008) has benefited social robotics, a subdomain of HRI. Previously, social roboticists had determined the foundation of social robots being competent to provide service and companionship – that is, being functional and affective (Leite, Martinho, & Paiva, 2013). These two main features form a robust interface for the elderly to interact with technology which gives them not only care and companionship, but also enjoyment and dignity.

This overall research project is to study the social and affective impacts of social robots from two aspects: physical and non-verbal interactions. Since individuals with DD talk and behave differently from ordinary people and many of them have cognitive difficulties, SARs that are designed to communicate with them must have a robust interaction model to achieve a high communication quality. A graphical user interface (GUI) is an effective way to present information and thus is commonly used as a primary instrument of communication and interaction on socially assistive robots for the elderly with cognitive impairment (Pino, Granata, Legouverneur, Boulay, & Rigaud, 2012). Also, I investigated how the physical form and the change of a robot's spatial presence can influence the relationship between robots and people with DD. Therefore, this research consists of three primary research areas: robot behaviour, graphical user interface, and interaction design. (Figure 1.1).

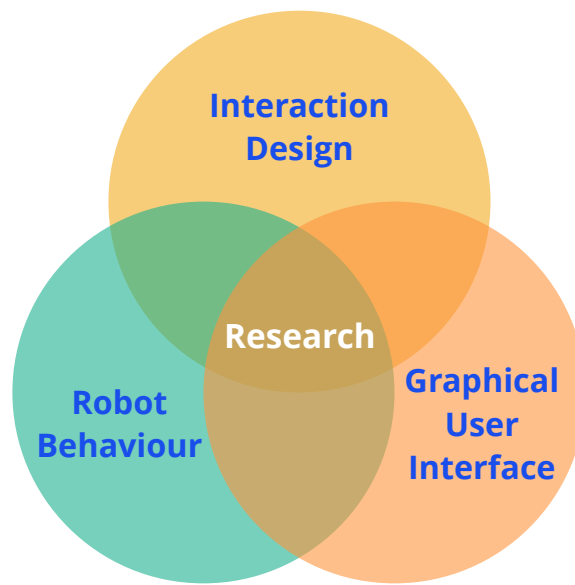

Figure 1.1: Three Central Research Areas

This project is essentially a study of human-robot interaction. Robotics is not the primary focus; it is, instead, the foundation of the research. To learn this connection between users (i.e. people with DD) and robots, I need to design social interaction and manipulate robot behaviour. *Robot Behaviour* cover technologies needed for this research, and elaborate on technical details. *Interaction Design* explores the connection between residents and robots. We need to know what the robots are designed for; thus, the purposes and functions of socially assistive robots have to be clarified here. *Graphical User Interface* solves communication problems. There is an overlap between interaction design and GUI. However, GUI should be considered as an individual category because of its significance – it is one of the main focuses of this project.

## 1.4 Thesis Problems

This thesis explores the social interaction design on robots assisting a specific user group: individuals with developmental disabilities. Although there have been many an investigation on the use of robots for service and companionship, empirical studies dedicated to our target user group are still extremely rare. This situation suggests that: *the HRI community is still in the rather early stage of exploring the application of robots serving people with progressive cognitive impairments*. Specifically, this work aimed to address the following problems:

- **Problem 1: We do not know how to design technical interactions for people with DD.** We are looking at a very specific use group who exhibit symptoms identical or similar to diseases like dementia and Alzheimer's and have distinct perception and cognition from the general population. This user group is unfamiliar to us, as we know little about their needs, behaviour patterns, function levels, daily lives, etc. Without a clear vision of user needs, we will be able to create neither artifacts nor interactive services beneficial to our users. Although we gained knowledge of developmental disabilities from our literature review, I found many of the resources provide little help to our thesis project due to different contexts (e.g. HRI for eldercare instead of DD) or lack of direct data (e.g. reports and surveys rather than interviews with caregivers or tests with people with DD). To sum, we do not know if we should design games, tasks, or entertainment media for residents. We also have little clue about designing graphical content to deliver information to residents with low cognition.

- **Problem 2: We do not know how to optimize the social interaction.** Currently, residents mainly socialize with other residents or the caregivers only. We are not sure how we can design and optimize social interaction based on residents' experience and habits. For social robotics, there are multiple aspects that we can choose to start with, including verbal communication, tangible interaction, visualization of information, etc. Apparently, we are unable to cover all of them in depth. Besides, our participant demographics may pose limitations to the factors based on which we can design user studies.

- **Problem 3: We do not know how to leverage robots' unique features to increase users' engagement.** Besides the humanoid form of SARs, robots' mobility is another critical feature that makes them distinct from devices like computers and tablets. We want to bring more trust and acceptance to the social interaction with residents and we are not sure if the use of motion would introduce a more positive experience. If so, how would we control the relevant parameters such as distance or speed?

## 1.5 Research Objectives

Developmental disabilities appear to have many similar symptoms to dementia as individuals with either of them tend to have declining mental functions in perception, thinking, and organization. We learned that there are multiple approaches that can potentially help with retaining a resident's brain functions (ASC, 2014):

- **Keep good social connections.** It is significant for residents to have meaningful and positive relationships which can help reduce their loneliness and emptiness. Residents usually stay in group homes or similar healthcare institutions; this makes it more challenging for them to stay connected with their family members. For this reason, always having people there whom they can talk or at least spend time with is critical.

- **Maintain a positive lifestyle.** Training resident's cognitive abilities plays a vital role during residents' convalescence. The lifestyle they are having will affect their emotions significantly. Thus, they should be guided to make their daily tasks more coordinated. This will also benefit their caregivers who often find it difficult to keep track of each resident's schedule.

- **Stimulate individuals' learning and thinking functions.** External memory aids like tablets or smartphones appear to contribute to enhancing residents' memories and reasoning skills. Using multimedia technologies as supplementary treatment is a common practice in many healthcare institutions such as the the Developmental Disabilities Association (DDA).

These recommendations on preserving mental functions show us the potential of having social robots engage with people with DD. With these tips in mind, we have the research aim to design efficient social interaction for assistive robots so that people with developmental disabilities can get assistance or meaningful interaction with the robots.

- **Research Objective 1: Design GUI prototypes to identify residents' challenges and needs.** I designed GUI prototypes based on caregivers' advice and feedback, and observed how residents behaved and reacted to our design. I selected several fundamental design principles and tested them in prototypes. I examined the effects and desired design of each aspect I selected. After studying the effect of each visual or interactive factor, I drafted preliminary propositions for designing robust human-robot interaction for users with visual or cognitive impairments.

- **Research Objective 2: Investigate how residents experience the technical interactions we designed.** HRI is far beyond a display on a mobile humanoid agent. I designed new GUI prototypes and compared the outcomes by deploying them on a computer and the robot. The essence of human-robot interaction, in the context of service or companionship, is to imitate a relation or interaction process that is similar to interpersonal Interaction. Hence, I compared the effects of different types of interaction. explored the social dimension of HRI, and identified lessons that we can learn from residents and caregivers' interaction.

- **Research Objective 3: Explore how motion and social distance affect trust and acceptance in HRI.** Trust and acceptance are of critical significance in social interaction. I observed how residents reacted when their spatial perception got stimulated and assessed their experience when the personal space was being "violated". I also compared their responses (e.g. gaze) when the robot keeps moving versus being stationary.

## 1.6 Organizational Overview

This thesis consists of seven chapters, including this first chapter presenting an overview of the entire project. I present the main research questions and objectives. I explore the solutions for the research question through three user studies in Chapter 5-7.

Chapter 2 describes several related studies in the realm of HRI, especially in social robotics. I have found projects based on frameworks similar to ours and identified design techniques that inspire me regarding the understanding of HRI. Those studies provide multiple perspectives to see and explore social robotics and validated theories and design principles that helped to construct our interface and evaluate the results. I also reviewed similar research work that adopted qualitative research methods and gained more methodological insights into the analysis of our results and the evaluation of our design.

Chapter 3 states the motivation of our research and summarizes our methodology of conducting a qualitative study.

Chapter 4 discusses our design iteration process. I identify the most critical scenarios that our prototypes should be designed for. Besides, I introduce the design principles that I followed or selected based on our preliminary interviews. I also explain how our studies were designed, which is critical for assessing the effectiveness of my design. I list multiple study variables, referred to as main factors in the context, which determined how our study was designed.

Chapter 5, 6 and 7 present three user studies, through which I discuss the methodologies that we used or created to achieve precise evaluations of my design at different stages of the iteration process. I demonstrate tools and methods that I used to conduct qualitative data analysis. I present my data analysis results, along with findings validating or dismissing the design principles that we followed. I also comprehensively summarized the interviews for all user studies, which is an indispensable part of our data. I discuss the results of qualitative data analysis and states the design implications that I obtained in the scope of my study.

Chapter 8 concludes the thesis, stating the strengths and limitations of the study and discussing the possible contributions that my work could bring to the HRI field. This final chapter also discusses possible work to do in the future, for people taking over this project or external researchers working on a similar study.

# Chapter 2

# Related Work

In this chapter, I review related literature and research work covering three primary research domains of our work: robotics, social interaction, and interface design (Figure 1.1). To start with, I investigate the demographics of our target users – people with developmental disabilities. Knowing who we are designing for is simply the question of top priority before we start the design process. Following the user population, socially assistive robots, the product which our interaction design is based on, are another essential research to explore. I list some of the popular social robots that are already available on the market. To have a deeper analysis, I then focus on the specific type of social robots providing assistance and service. Based on the previous research effort, I explore the dimensions of HRI, and attempt to find the most significant factors in our interaction design process: gaze, speech, gesture, and distance. Next, I examine dignity, trust, and acceptance, which are critical to long-term social presence and embodiment of socially assistive robots in our context of use. In the end, I conclude this chapter by reviewing the use of qualitative research methods in similar studies that focus on the elderly and people with mental impairments.

## 2.1 Developmental Disabilities

Developmental disabilities are a diverse group of chronic conditions as a result of mental or physical impairments. These conditions arise during the developmental period, may impact day-to-day functioning, especially in "physical, learning, language, or behaviour areas", and could persist throughout an individual's lifetime (Centers for Disease Control and Prevention, 2018). The specific causes of most developmental disabilities remain unclear, but the most common factors that were already identified include genetics and parental health. Developmental disabilities appear among all racial and ethnic groups and encompass intellectual disability, vision impairment hearing loss, learning disability, cerebral palsy, autism spectrum disorder (ASD), and Attention-Deficit / Hyperactivity Disorder (ADHD). In Canada, *developmental disabilities* and *intellectual disabilities* are synonymous terms. They may also be referred to as *learning disabilities* in the UK (Developmental Disabilities Primary Care Initiative, 2011).

It was estimated that in the United States there were 7 to 8 million people, or 3 percent of the population, having a developmental disability in 2005 (Ward, Nichols, & Freedman, 2010). In Canada, 1.1% of the national population aged 15 years and over (315,470 in total: 123,310 females and 192,160 males) have developmental disabilities as of 2017 (S. Morris et al., 2018). Unfortunately, it was reported that people with developmental disabilities are "an invisible population in Canada's mental health system"(Lunsky & Balogh, 2017). According to a study (Lewis et al., 2002) investigating the health situation of American adults with developmental disabilities living in communities, preventive services were noticeably inadequate and 36% of these people obtained psychotropic medication without any identifiable diagnosis.

Furthermore, there is also a tough barrier to attaining social inclusion for people with intellectual and developmental disabilities. Social inclusion is a substantial element of well-being for individuals with developmental disabilities, comprises interpersonal relationships and community participation (Simplican, Leader, Kosciulek, & Leahy, 2015). In addition, compared to the general population, individuals with DD "have poorer health and greater difficulty accessing primary care" and their "patterns of illness and complex interactions among comorbidities" are also different (Sullivan et al., 2011). Excellent primary care, however, could identify the specific health problems of people with DD and prevent their morbidity and suffering. Unfortunately, this kind of primary care is not accessible to everyone. Based on the facts mentioned above, we have got the motivation to learn about the DD community, and to help them by using existent technology.

## 2.2 Social Robotics

Social robots are a type of intelligent agents with certain degrees of autonomy. They are developed to imitate social behaviours to provide users with service, assistance, or companionship. Socially assistive robots are a more specific type of social robots whose primary aim is to provide assistance to users.

A comprehensive survey conducted by Leite et al. (2013) reviewed the current research on long-term interaction between users and social robots as of 2013. As Leite et al. indicated, longitudinal studies provide a pragmatic way of investigating long-term changes in user behaviour and experience. Their study found that social robots and virtual agents without strong social capabilities to engage users in the long-term made people lose their interests. This extensive literature review examined 24 publications and discussed robots for different purposes, including health care and therapy, education, public service, and home care. People tend to get attached to robots sharing our physical space, especially when they exhibit human-like patterns. Human-robot interaction for user care is still in the early stage, but there is evidence that people are willing to accept and interact with robots.

To investigate some of the principal research progress in social robotics, Breazeal et al. (2008) conducted a survey, arguing that the long-term objective of making social robots is challenging because they need to communicate naturally with people using both verbal and non-verbal signals. A wide range of social-cognitive skills will be a fundamental requirement for social robots so that people can intuitively understand them. This survey also studied socio-cognitive skills, suggesting that socially intelligent robots must be able to understand and interact with animate entities. That being said, social robots have to be able to recognize, understand, and even predict human behaviours in terms of hidden mental states. One of the most important thing for HRI is that the creation of social robots should be human-compatible and human-centric in the design process so that we can truly benefit from this technology.

Previous studies on social robotics have confirmed users' acceptance of therapeutic social robots in everyday scenarios, acknowledging the fact that users' satisfaction is determined by the user-technology fit (Baisch et al., 2017). Although most people probably have rarely got the chance to interact with a robot for a long time, the majority of people living nowadays have had access to smartphones, the Internet, or at least televisions. It has been almost a century since the television was invented and it was a remarkable milestone for multimedia. Because of televisions, we have learned how to process information in the form of text, images, audios and animations.

In the last decade, social robotics has become a popular research area to which roboticists and cognitive scientists pay close attention (Leite et al., 2013). Sociable virtual agents, in terms of physical appearance, can be divided into two groups: those with a display, or those humanoid ones without additional display. Many a robot with a display is similar to the television to some extent – these kinds of robots depend on external displays for better communication. Each of these two robot types has its strengths and weaknesses. For instance, robots relying on additional display can show much more information. In fact, many social robots, including those designed for eldercare, need external screens for communication.

## 2.3 Assistive and Service Robots for Eldercare

The application of social robots to elderly care was discussed in Broekens et al.'s article (2009). This article introduced a significant study as nowadays there is an increasing necessity for technologies that can improve the life quality of the elderly. Social robots are useful in eldercare because they can be a gateway to elderly interfacing with digital technology and also offer company to the elderly. Nevertheless, the effectiveness of this application has not been investigated much in the past. Broekens et al. compared several different social robots to investigate the effects of socially assistive robots on the health of the elderly. The results of this comprehensive research indicated that socially assistive robots have many functions and effects, including increasing health by decreasing level of stress, decreasing loneliness, increasing communication activity with others and improving the sense of happiness.

The ability to build social connections is not only a critical feature for social robots, but also one of the most fundamental dimensions of assessing robots' effectiveness. According to an evaluation study by Tsui et al. (2009), the most effective performance measures for service robots designed for the elderly consist of daily activities, response accuracy and time, mood, and quality of life. These measures overlap with those in the domain of Autism Spectrum Disorder (ASD), a multifactorial developmental brain disorder. Autism, as one of the most common types of developmental disabilities, poses severe functional challenges to individuals in terms of self-care, mobility, independence of living, and receptibility or expression of language (Institute on Community Integration, 2018). These challenges also stimulate individuals' needs for treatment, care, and service, most of which can be life-long.

Heerink et al. (2009) described their research on the influence of the robots' abilities to facilitate interaction on elderly users' of social robots. The experiments were conducted in eldercare institutions where robots with screens are employed. These robots can have conversations with users. The results of their experiment showed that participant felt more comfortable and expressive when they were confronted with the more socially communicative robots. This results also indicated that speech is an important part of social robots' integrity. The development of more appropriate acceptance models is needed in the future, especially conversation acceptance. Their experiments indicated that systems in which social abilities were incorporated were considered as more social. This is a meaningful point for social robots.

Although the need for social robot seems to emerge in recent years, this need might be more urgent than we expected. Wang et al. (2016) stressed the issues among the elderly population in Europe and discussed the solution using social robots. Eldercare is becoming a critical challenge due to the number of seniors living alone. Some social robots were designed to ultimately work in real conditions and help elderly users have more independent lives. The approach that Wang et al. took was to implement a speech user interface, extending the application contexts and making it possible to apply robotic service in many scenarios. They suggested that the most challenging aspect of personal robots is social communication, which is fundamental for elder-robot interaction. To achieve a low-complexity and user-centred speech interaction system, they employed a context-sensitive speech recognition approach to enable real-time flexible dialogue movement. This research had proved the significance of verbal communication in human-robot interaction (HRI).

## 2.4 Design of Socially Assistive Robots

When it comes to robot design, there are always moments when we have to decide whether or not to make an aspect more humanoid. Walters et al. (2008) suggested that users tend to be more intrigued by robots with characteristics and behaviours that differ from humans. However, this theory, in the field of social psychology, may not apply to the very user group which we are focusing on. Even for one single user group, there can be noticeable individual differences in terms of personalities and temperaments. Thus, decisions to attach personality traits to socially assistive robots have to be made cautiously with a subsequent assessment conducted.

An existing study (Zainon, Yee, Ling, & Yee, 2012) points out that users are more likely to struggle with understanding GUIs if the cognitive psychological principles and theories are not applied to the design properly or sufficiently, by conducting a series of empirical experiments based on theories such as Schema Theory and Gestalt Law. These experiments consist of multiple GUI design aspects including proximity, colour, placement of content, and type of icons (familiar vs. unfamiliar).

As part of our robot GUI design, we first need to determine the manifestation of the virtual agent, if any. Since our robot has a physical presence already, instead of only being a pure personified character in a computer screen, we wanted to augment the embodiment of the graphic imagery in a physical semi-humanoid robot. Wang et al. (2017) argue that 3D face stimuli expedite children's recognition of facial expressions, based on an experiment on 71 children aged between 3 and 6. This study had been inspiring to our project, partly because of the closeness of IQ between our user groups. However, our tests of dif-ferent portrayals of a given character did not show any evident disparity regarding users' acceptance, whether it was 2D or 3D.

Aside from the content that we design, i.e. GUI and other visual information on robot's display, the physical presence of the robot reminds us of a significant set of factors which interaction designers tend to omit: the physical parameters of the robot. These factors include the physical characteristics, such as height, size, and shape. They affect users' perception and understanding of the robot even before users start paying attention to the display or the audio prompt. For our users who are more sceptical about robots, ensuring our robot making a proper first impression is critical to our subsequent study. Irene et al.'s study (2013) explored the impact of robot's height in HRI (Figure 2.1), arguing that height differences influence people's perception and cognition due to the divergence of roles in human communication. For example, a taller person is likely to be perceived as more dominant, hence more "expressive, persuasive, and having more leadership qualities." Irene et al. proposed three hypotheses regarding the dominance of an operator using the telepresence robot for local participants. Through a controlled laboratory experiment with forty adult participants whose ages varied between 18 and 70, they found support for two of the hypotheses and discovered that participants had better experience and showed more dominance while the operator operated a shorter robot. Irene et al. also noticed local participants' increased gaze while speaking as the evidence of such dominance. In this case, the operator's role as the leader became the least persuasive while using a shorter robot. In the meantime, local participants' follower roles incline to get reinforced when an operator assigned leadership used a taller robot. Irene et al.'s work is significant to our research, as we are hoping to not make our social robot intrusive, aggressive or overly dominant, but have it maintain its authority in certain scenarios like giving commands for daily routines. As interaction designers, we have to exercise caution while endowing social robots with necessary qualities and "personalities", which are critical to the embodiment of robot-mediated connection.

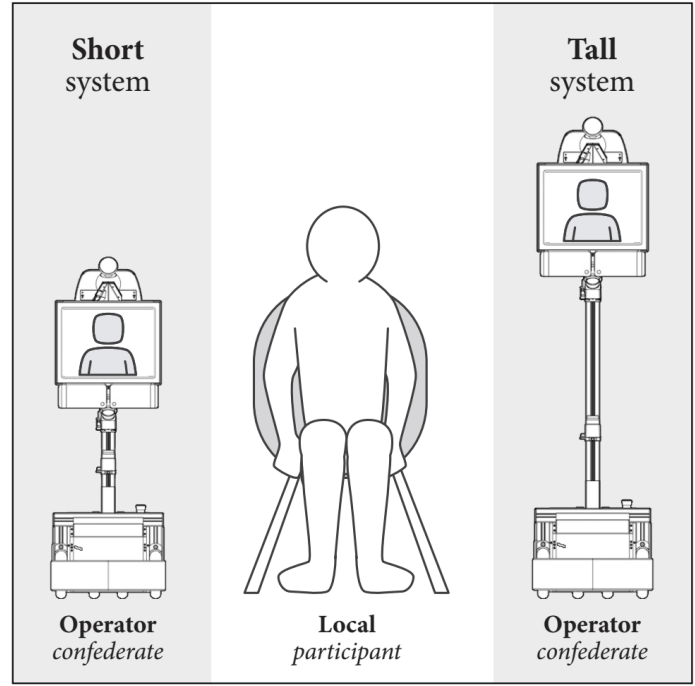

Figure 2.1: Example of height differences between telepresence robots and a participant (*Image copied from Rae et al. (2013) with author's consent*)

As the relationship between users' engagement and the social attributes of assistive robots is the primary topic of this study, we need to create a robust performance benchmark to assess the impact of each attribute. In the area of HRI, there are various benchmarks to evaluate the performance of a robot system. Some of them mainly focus on such technical aspects as scalability and imitation (2007). To us concentrating on the sociability of assistive robots, we determined to construct an efficient and practical benchmark from a less technical perspective. We aimed to employ this evaluation model to detect and measure *Aether*'s impact on both individuals with developmental disabilities and their caregivers. The following fundamental usability criteria had been taken into our evaluation of the overall effectiveness of our design: efficiency, effectiveness, engagement, and user satisfaction (Quesenbery, 2001; Georgsson & Staggers, 2016). These elements are essential for assessing the usability of a product or service used by users in a specified situation, according to the latest ISO standards for ergonomics of human-system interaction (International Organization for Standardization, 2018). The interesting challenge was the tension between the expectations and the allowances of the effect the robot has on the caregiver practices.

In order to have the robot start and maintain a healthy and lively connection with target users, the first factor that we should consider is the environment. Our users, i.e. the residents and their caregivers, live or work at a group home, which is essentially an open space. Although each resident has their own bedroom, they do share most of the space and facilities together. The public area creates a positive atmosphere that is beneficial for residents' functional levels. The whole point of providing on-site health care to people in need is to improve, or at least maintain the functional status which impacts one's subjective perception of life quality (Udekwu et al., 2001).

Similar to our project, a study conducted by Glas et al. (2013) identified three HRI scenarios which they designed and implemented studies based on: multi-robot coordination, context-aware service, and personalized service. The last two scenarios are strongly connected to the scope of our study. Glas et al. devised a robot system framework (Figure 2.2) that endows their social robot with five essential skills. This framework includes five key modules responsible for those essential functions or skills. Glad et al. first identified the requirements of deploying the robot in desired environments, and constructed modules dedicated for each category of scenarios such as navigation and human-robot communication. The robot they designed is able to perceive, identify, and react to individuals, and to analyze, recognize and respond to users' behaviours or inquiries. Besides those essential functions ready for HRI, this robot also owns capabilities of localization and path planning, which are critical for a service robot being able to move around in the real-life environment.

Glas et al. (2013) implemented a pivotal service mechanism, allowing the robot to allocate services to related modules. This makes the robot highly efficient concerning the role as a service agent. The robot is able to not only provide on-demand services but also proactively anticipate users' needs and schedules. It should be noted that autonomy marks the maturity and imitation of a virtual agent. In other words, autonomy is an indispensable prerequisite of achieving realistic characters. Besides, being capable of offering personalized services is a shortcut to having the robot befriend humans.

## 2.5 Adaptive Factors for Robot Appearance for Different Situations

Users must be able to understand (comprehension) and emotionally engage with (affect) the robot. As in human-human communication, the requirements of the situation will dynamically impact the ways in which the robot should communicate with and support the residents. if a robot does not adjust its communication style to the particular situation at hand, this can lead to confusion and misunderstanding in the short term, and contribute to anxiety about the robot and a lack of trust and confidence in it in the longer term, ultimately leading a feeling of disengagement with the robot. This is likely to be highly variable with respect to individual residents in the home whose individual cognitive and affective responses will rely on different levels of ability and anxiety in the particular contexts in which they are interacting with the robot. We will explore how visual appearance needs to be tailored to each user and context, using specific scenarios and use cases identified by the DDA staff.

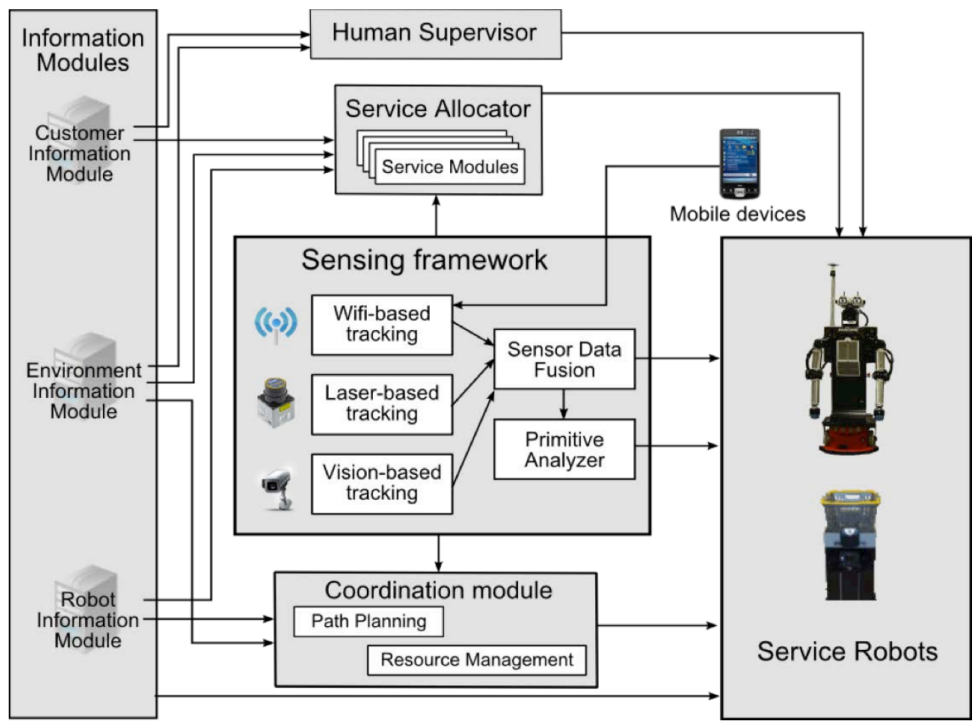

Figure 2.2: Overall Architecture of Glas et al.'s proposed Network Robot System framework (*Image copied from Glas et al. (2013) with author's consent*)

Lee and Yoo (2017) explored the interaction design of companion robots assisting the elderly with major tasks. They examined emotion interacting system of robot (Figure 2.3). They found that facial figures are perceived as a stimulus to users. They conducted a user test with 8 elderly individuals whose average age was 69, using Wizard of Oz method to remotely control the robot during the study. Pre-recorded voice and facial figures were prepared on the robot. The results suggested that robot-initiated interaction generates a better emotional atmosphere, and that information delivered through emotional robot face could advance higher memory retention rate.

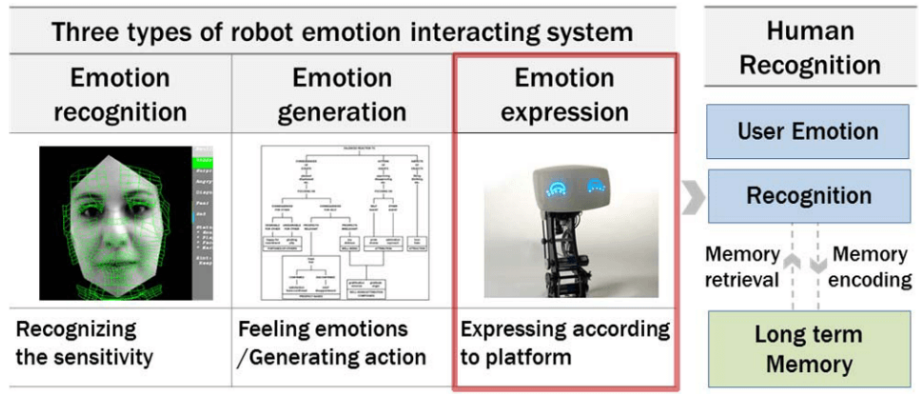

Figure 2.3: Emotion interacting systems of robots (*Image copied from Lee and Yoo (2017) with author's consent*)

## 2.6 Types and Dimensions of Interaction

In 1997, Takeda et al. (1997) summarized three most common types of interactions between humans and robots while they were investigating how robots communicate and collaborate with people: intimate interaction, cooperative interaction, and interaction through instruments. As this study was so dated that a bulky standalone computer (referred to as an instrument) needs to be used for computation and recognition in addition to the robot, the last type is no longer applicable to the research nowadays. The computer embedded on the robot is computationally sufficiently powerful to process most of the tasks we needed; we also have cloud computers for additional functions. Nevertheless, this research team pointed out that the unique trait that robots have which make them outweigh virtual interfaces: the ability to have direct, intimate and cooperative interaction through gesture generation and motion.

Gockley et al. (2005) designed *Valerie the Roboceptionist*, a permanent installation in the entranceway to a building at Carnegie Mellon University. The robot has some basic features like giving directions and checking weather information. It was built from a mobile base with a moving monitor mounted on top which has a graphical human-like interface. The researchers argued that the graphical face is very expressive and more reliable than a mechanical face. Also, future maintenance is easy on graphical face on a screen because changes can be made easily to it. The authors acknowledge that long-term human-robot social interaction can be improved and enhanced based on human-human social interaction. For speech on social robots, they should avoid being a "motor-mouth" and try to provide more natural dialogues with users. Valerie is a good attempt to develop a social robot that can interact with people over a long period of time. To enhance interaction, designers of social robots should be aware that these robots should be used solely to convey information. A good and ideal situation is that social robots can encourage users to engage in dialogue repeatedly.

Zhao (2006) explored the emerging impacts of humanoid social robots and human-humanoid interactions. The interactions between robots and humans were categorized into several different types, including utilitarian and affective robots. The author believed that in the immediate future, the connection between intelligent machines and humans will bring profound impacts on our lives, and that there will be a time for a revolution of humanoid robots. These robots are becoming more and more important in our lives because they are autonomous, interactive and human-like. The author compared this human-robot interaction with computer-mediated communication (CMC). Social robots have extended the domain of human expression, communication and discourse into this computerized world. Zhao argued that the rise of a synthetic social world calls for a new conceptualization of society because humans and robots have a higher level of communication and interaction. The social relationship between humans and humanoids will be a new research area as social robots develop.

## 2.7 The Effect of Gaze Interactions

To create the best HRI experience, researchers usually investigate and compare several dimensions of HRI, which is a necessity for designing robots that are capable of taking a beneficial role in people's daily lives (Breazeal et al., 2008). Studying the dimensions of HRI also helps designers determine the requirements on robots' social skills (Dautenhahn, 2007).

For the social dimension of HRI, much research (Ruhland et al., 2015; Kirchner et al., 2011; Zaraki et al., 2014; Admoni & Scassellati, 2014; Li & Chignell, 2011) tried to explore the application and affect of gaze and gesture, two important modalities in HRI. Among these studies, a humanoid robot was designed by Prange et al. (2015) to construct multi-modal human-robot interaction based on human gaze and robot gestures. They implemented an eye tracker on a robot to constantly monitor the gaze of residents with dementia. An object needs to be placed on a customized table with marked coordinates for the robot to track. However, this research failed to produce satisfying results. The integrated camera of their robot supports low imaging quality, resulting in poor recognition accuracy. Besides, the eye tracker technique they used could not produce stable gaze estimation.

This experiment was substantially limited to verbal and gestural interactions and the test scenarios are confined to lab environments where certain types of equipment are required.

Gaze can be used as not only state-display cues on a robot to indicate the robot's conversational and operating states (Breazeal et al., 2008), but also determining evidence to judge the extent of the user's engagement. People in a social condition would gaze straight at the conversation partner, whereas in a less social condition they would turn away from the partner (Heerink et al., 2009). When humans communicate in person, they are "face to face": gaze imputes attention, concern and engagement (Ruhland et al., 2015; Admoni, 2016). However, it may also be disconcerting and disruptive and create anxiety. We will explore how simple "gaze", in the form of turning the robot screen towards the user, affects the communication. This will be particularly important in group situations and scenarios where non-facial visual representations are used.

Simple facial expressions or even lighting patterns may afford non-verbal communication that can support and enhance verbal conversations. While there are sophisticated systems in research for fine-tuning facial expressions for avatars (Shayganfar et al., 2012), human perception is optimized to seek the canonical (simplest) form and the well-known set of simple iconic Chernoff faces serve as the basis for a wide application of mediated face-based expression (Donath, 2001), notably Emoji and other icons. We will explore the most useful set of facial expressions and how simply they need to be rendered. There may also be situations when the robot should not "have a face", as in when fading into the background, or when showing other relevant information (as in examples of exercise or engaging pictures.) To what extent do we need to provide additional perceptible feedback on the robot that might mimic, or augment the purpose of, facial and expression communication?

## 2.8  Speech - Why is verbal communication indispensable?

Because of our focus on behaviour patterns and graphical user interface at this stage, we need to find an efficient way for human-robot communication. Sugiura et al.'s work (Sugiura et al., 2015) presents a new method for speech synthesis for robots based on Hidden Markov models (HMMs). It focuses on natural and friendly synthesized speech for robots because speech communication has been a very popular research area for human-robot interaction conferences and competitions. The authors compare traditional Text-to-Speech (TTS) systems to non-monologue speech synthesis systems and found that TTS systems tend to be very unnatural and unfriendly, partly because they are designed for text reading instead of communications. This result is inspiring for my research because TTS is the easiest and most commonly used method for speech generation in various devices, including robots. Also, Sugiura et al. point out that monotonous intonation tends to prevent novice users from knowing that the robot is actually asking a question. Their research team built the new non-monologue speech synthesis system and the results have shown that the performance of their method almost approached the theoretical upper limit. They have also proved the effectiveness of applying non-monologue speech synthesis to robots. Furthermore, their system is based on cloud-based technology, making it easy to collect a speech synthesis corpus for robots and to maintain the system in the future.

Our project is not to design a speech interface. Instead, it is an exploration of how we can evaluate people's perception of language in the verbal form. Cha et al. (Cha, Dragan, Forlizzi, & Srinivasa, 2014) stated that there seems to be a gap between robots' true capability and perceived capabilities. Their research focuses on the speech pattern. After conducting an online study to explore the effects of speech on perceived capability, they found that physical failures of a social robot negatively influence participants' evaluation of only the robots' physical capability, and speech can positively affect participant's ratings of the robots' physical capability. Thus, the significance of speech is clear for social robots. Knowing the properties of robots can help designers expect users' need and behaviours. This paper is an initial study to explore how robot behaviours influence users' perception of robots' capabilities; the results have shown the impact of speech in this process as it has many levels and functions.

To date, speech recognition accuracy is still a technical challenge to computer scientists developing language processing engines. Even if our robot employs a state-of-the-art speech recognition technology with a 5% word error rate (Novet, 2015), it is still difficult to have the robot understand people with developmental disabilities. The current speech recognition technologies are generally built for users without accent and in regular tones, in an ideal and quiet environment. For people with developmental disabilities, it is hard to predict their tempers. Sugiura et al. pointed out that natural communication between humans and machines is an extremely demanding task due to challenges of speech synthesis and the size and quality of corpus. As verbal interaction is an indispensable part of our robot system, it is a must to ensure that the speech from the robot is smooth, friendly, and efficient. There are two main approaches to address this problem: using pre-recorded audios and synthesizing the speech using Text-to-Speech (TTS).

A careful investigation is needed into the speech patterns of people with DD and their caretakers. Later, we can decide whether we should replicate or recreate the communica-tion experience, or take a different approach. The development of domestic user-interface robot was discussed by Kröse et al. (2003) who developed a robot capable of maintaining a natural interaction through speech and emotional responses. This natural interaction with the user was made possible using a mechanical head conveying emotion information. This robot is a service agent that can carry simple conversations with users and provide information in a natural way. The head of this robot can express six basic emotional facial expressions. Though facial expression will not be my primary research topic, it is a significant part of my study because users tend to look at the robots while having speech conversations. This research also introduces their software development framework which is, according to the authors, a "state-of-the-art software tool" to implement distributed robot applications. This paper manages to show a connection between two "service to humans" paradigms. Further research emerging between these two fields will be conducted in the future.

To help elderly individuals with Alzheimer's disease and dementia, Rudzicz et al. developed a mobile robot to support users' daily living (2015). This work explored the application of speech interface in the human-robot interaction, in which their user persona is similar to ours. In their robot system, they used a commercial Text-to-Speech (TTS) service and its default settings. The verbal dialogues were applied to task scenarios like "go to the kitchen" and "boil water." During the user study, Rudzicz et al. observed that elderly users tended to ignore the robot's verbal behaviours which oftentimes brought confusion to the interaction, and this disregard pattern constituted 40% of the behaviours while participants interacting with the assistive robot. Rudzicz et al.'s preliminary examination of speech interaction in social robotics proved the feasibility of integrating verbal behaviours into an assistive robot. However, a limitation of this work was the lack of prosodic tuning– they kept the default parameters of the TTS service without modification, resulting in effective verbal communication. Also, just like many studies of verbal interaction, Rudzicz et al.'s design was also constrained due to the accuracy of speech recognition. Overall, their research pointed out the potential challenges of employing speech in HRI and concluded that the design and imitation of HRI tasks need to be based on real-life tasks that users are already familiar with.

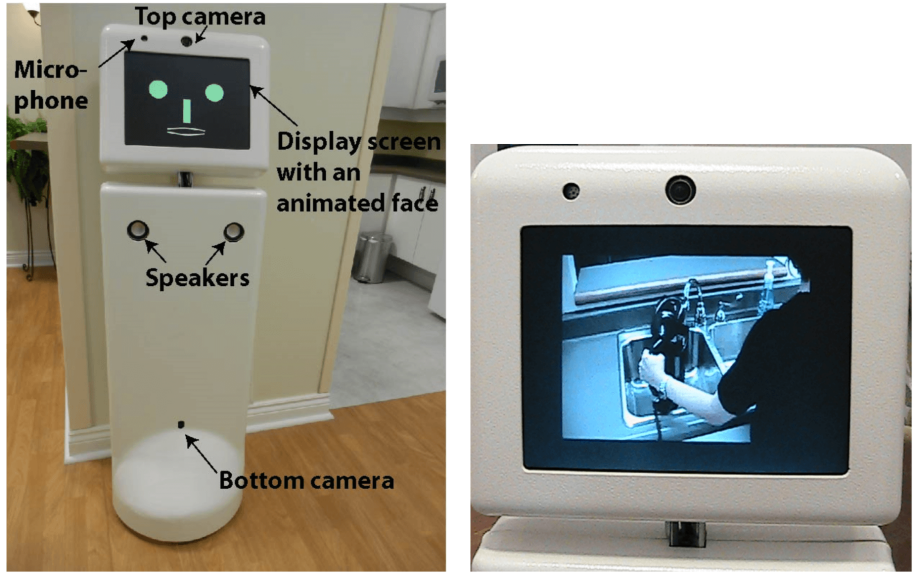

Figure 2.4: The prototype robotic caregiver (left), showing a video prompt in the display screen (right) (*Image copied from Rudzicz et al. (2015) with author's consent*)

## 2.9 Proxemics

In psychology, *proxemics* is a research area related to epistemology, referring to the study of perception and use of space (1968). The concept and application of proxemics exist generally in our daily lives: furniture in the room, interpersonal communication, checkout queues at the supermarket, etc. Hall (1990) categorized four proximity zones (Figure 2.5) based on the distances and their effects while exploring the dynamism of space. This finding was concluded based on interviews and observations and has been one of the most fundamental theories of proxemics (Bainbridge et al., 2008; Brandon, 2012; Tapus & Matarić, 2008; Yamaoka et al., 2010). Hall acknowledged that the study had certain limitations due to participants' personalities and environmental factors such as noise, and thus the numerical representation of each zone was "only a first approximation." Nevertheless, this work is still inspirational for our study design and it gave us a good starting point to explore the effects of social distance in HRI.

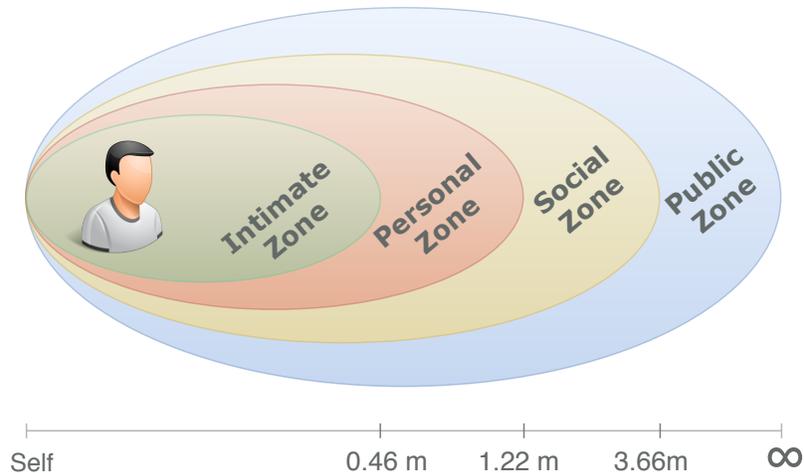

Figure 2.5: Four Proximity Zones

Although social robotics was not even an infant realm of HRI during Hall's time when he focused on the spatial relationships of humans, many HRI researchers have benefited from Hall's study. An example is Yamaoka et al. (2010) who devised a model for information-presenting robots' spatial adjustments based on observations of human-human interaction scenarios. Yamaoka et al. simulated four settings of placing a robot, with different proximity and orientations to participants. They designed a user study with 22 undergraduate students, using a questionnaire as the evaluation method to investigate users' impressions of each setting. They highlighted the significance of the robot's and participants' fields of view (FOV).

The effects of social distance on human-robot interaction were investigated by Kim and Mutlu (2014) through two user studies. They explored proxemics from three aspects: *physical distance*, *power distance* (robot as supervisor or subordinate), and *task distance* (cooperative or competitive human-robot relationship). Individuals' experiences and perceptions were analyzed to evaluate the effects of social distance. Kim and Mutlu found that user experience (UX) got enhanced only when the nearby robot acted as supervisor, opposite to the distant robot which needed to act as subordinate. Also, Kim and Mutlu designed tasks for participants to complete with a robot and noticed that users' performance declined when the robot got close to users.

Although Kim and Mutlu's study gave us inspirations for designing our user study, we found their methods not applicable to our study overall for the following reasons: First, they examined users' perceptions based on the results of task performance, and users' experience using questionnaires. In our case, we found that the evaluation of task performance is not the most appropriate approach to assess the perceptions of individuals with developmental disabilities. Task performance, however, can be used as a preliminary and low-cost attempt on the diagnosis of dementia. Second, like many of other researchers working on social robotics studies, Kim and Mutlu used a questionnaire on participants to analyze their satisfaction, comfort, pleasure, and likeability. Unfortunately, the traditional survey approach does not work on our participants due to their verbal and cognitive challenges.

Equation 2.1 shows Lasota et al.'s (2014) implementation to control robot's speed based on the distance between the robot and the user. In this equation, $\alpha$ stands for the percentage of decline in the robot's velocity, $d$ represents the current distance between the robot and the user. $d_{stop}$ refers to the distance at which the robot should cease to move, and $d_{slow}$ is the distance at which the robot's acceleration becomes negative. $\beta$ and $\gamma$ are coefficients that calibrate the velocity function. They implemented a real-time safety system that is able to convert regular industrial robots to a human-safe platform. This implementation did not require any hardware modifications on a robot, but does require exact user localization. Overall, they met the initial objective to achieve safer HRI. A limitation of this work is that the implemented safety system did not work as well when a human moves, due to the constraint of trajectory prediction.

$$\alpha(d) = \begin{cases} 1 & d < d_{stop} \\ 1 - \beta(d - d_{stop})^\gamma & d_{stop} \leq d \leq d_{slow} \\ 0 & d > d_{slow} \end{cases} \quad (2.1)$$

## 2.10 Dignity, Trust and Acceptance

Sharkey et al. (2014) studied the connection between robot care for the elderly and dignity. There is a capability approach (CA) that is introduced as a different account to evaluate the significance of having human dignity in a human-robot interaction scenario. The objective of this research is to explore approaches which enable robots to take care of the elderly. They found that some forms of robot intervention can increase older people's access to central capabilities. The CA was reflected upon as a framework for their assessment. The CA allows a valuable and balanced perspective on the connection between the dignity of the elderly and robots. It can be used as a good starting point for study in the future because they improve our understanding of using robotic technology to help maintain humans' dignity as they age.

The key of our study is to build a robust connection between users and the robot. Most research problems we are trying to solve revolve around this fundamental pivot. One of the phases of building connection in HRI is to create trust. Hancock et al. (2011) conducted a meta-analysis of factors impacting trust in HRI (Figure 2.6). They aimed to assess and quantify the affects of the robot, human, and environmental factors on the perception of trust. Their examination collected 29 empirical studies, of which 11 did experimental analysis and 10 did correlational analysis. Their analysis revealed that the most significant factors in the development of trust in HRI are robot attributes and performance. They also concluded that environmental factors correlated somewhat to trust in HRI , whereas human-related factors had little effects.

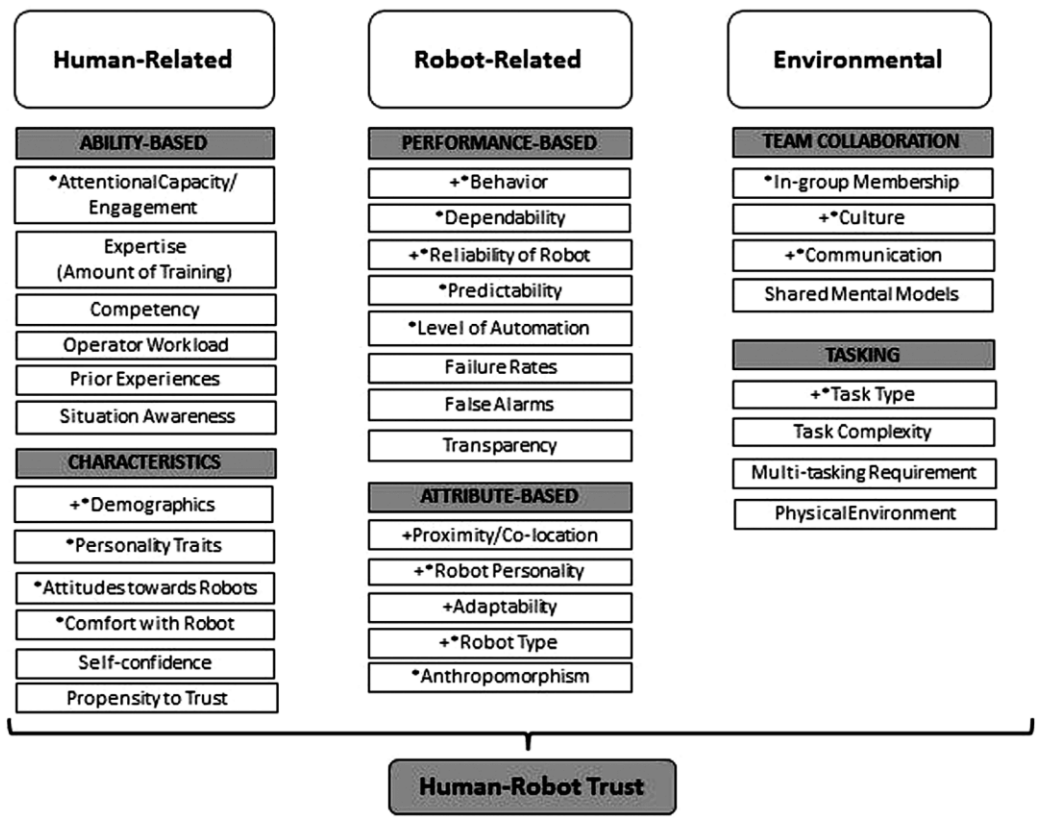

Figure 2.6: Factors of trust development in human-robot interaction. These factors, included in the correlational analysis (*) and the experimental analysis (+), were identified a priori via literature review and subject matter expert guidance. *(Image copied from Hancock et al. (2011) with author's consent)*

## 2.11 Qualitative Analysis of Users' Experience

Qualitative methods are one of the most common type of research methods, apart from quantitative and mixed methods. Quantitative research uses numerical or other quantifiable data from experiments, surveys, measurements, etc to obtain objective results mostly leading to conclusions that can be generalized. By contrast, qualitative research usually employs interviews, usability tests, and field studies to comprehend the background and experiences of users (Mortensen, 2018). Qualitative methods are usually "more exploratory" and aim to gain new insight into the individual's or user group's "experiences" (Mortensen, 2018). A notable property of qualitative methods is that to some extent researchers consider themselves as a co-creator of research results. Thus, results and conclusions obtained from studies are not wholly reproducible and objective. Compared to quantitative approaches, qualitative methods provide researchers with a great opportunity to delve into a topic or domain that is unfamiliar to them. When not having enough resources or time, they can be more explorative using qualitative methods.

A qualitative study by Manzi et al. (2017) examined users' experience with their design of a cloud robotic system supporting the elderly. Their robot collects environmental data and cloud resources to support users' independent living. The researchers defined the desired services with a focus group of 19 elderly participants and ran a real-life test with two users for five days. Based on users' qualitative feedback, they validated the robustness of this cloud robotic system. Their analysis was derived from users' impression and experience and reflected on the following five categories: aesthetics, anxiety, reliability, ease of use, and utility. Manzi et al.'s study took an inductive approach for data analysis and consists of three phases: create categories by coding raw data, connect the categories with corresponding research objectives, and develop a theory. These three phases form the foundation of the grounded theory approach, which is a common method for carrying out qualitative research focusing on theory development. Manzi et al.'s work has been inspirational for the design of our studies, as we also used inductive, thematic analysis and axial coding for our data.

# Chapter 3

# Research Motivation and Approach

In this chapter, I state the principal aims of my research project and present the design iteration process, including the following stages: understanding participant demographics, defining design requirements, selecting methodologies for user studies, starting and iterating exploratory designs. The subsequent chapter will further explain our interaction design considerations and process.

All of our GUI designs had been implemented on *Aether* (Figure 3.1), a socially assistive robot developed by JDQ Systems Inc. as part of its 3Spheres Robotics Project (JDQ Systems Inc, 2015). *Aether* was developed by a research team in charge of engineering, product design, and software development with the support of the Developmental Disabilities Association. This thesis project mainly focuses on social interaction design in HRI through exploring the robot behaviour and the robot visual communication with respect to affect and cognition. As visual communication is a critical component of this design project, I investigated how to adopt user's real-life experience to my GUI design and fulfil the potential and enhance the embodiment of the robot's graphical display in order to to enhance user's engagement.

As a side note: we would like to clarify the wording for reference to the participants of our project. We worked closely with three residents of a DDA group home in Richmond, BC, Canada, and several caregivers and staff there as well. In this thesis, we refer to three residents with DD as "residents," "users," and the caregivers as "caregivers," "staff". Due to the fact that residents need caregivers' company or assistance during our user studies, and we also rely on caregivers' feedback as an important source of data, we consider both residents and caregivers as "participants" in our user studies. Nevertheless, we see residents who directly interact with the robot as the "primary participants" and their caregivers as "secondary participants."

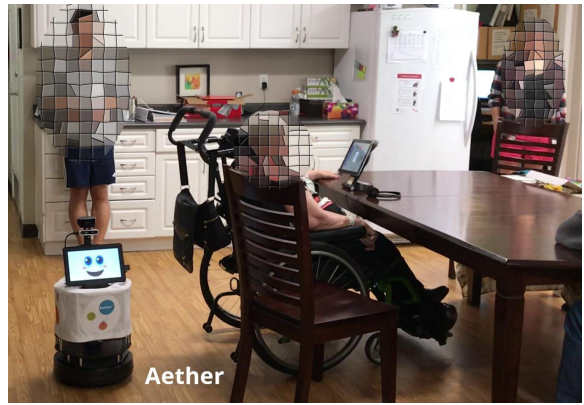

Figure 3.1: *Aether* the SAR at the group home

## 3.1 Functions Expected to Happen

As design practitioners developing a product, firstly we need to understand the specific goals of the development process. A pragmatic representation of these goals is the functions of the product. in other words, knowing the functions we need to fulfil would partly answer the question: *What are we designing?* We had a general idea of the functions being needed for socially assistive robots in the context of group homes. Although these ideas, extracted from the background research, are relatively abstract overall, they do provide a good start to further investigation and prototyping. From the focus groups with caregivers, we found the highest demand of the following functions for socially assistive robots:

- **Scheduling** Due to dementia, individuals suffering from developmental disabilities tend to have confusion and anxiety, the latter of which can be easily triggered by any minor imperceptible change of the environment. These individuals also have the need of knowing what events or activities are on their schedules.

- **Prompting** Individual with developmental disabilities also need help for task analysis. Simple tasks such as washing dishes can be challenging to them; some of them require extremely specific instructions given as simple steps. Robots are a perfect tool for logical and patterned information, given they can get clear inquiry commands as input.

- **Companionship** Sometimes, a robot does not have to do anything prove its value. One of the functions of group homes is to provide a cozy environment for residents to reduce their anxiety. For this reason, there are decorations and gadgets set up in the Snoezelen room to relax residents. Providing that a socially assistive robot can gain trust from residents, it would be another bonus for delivering stimuli at the group home.

Admittedly, these three functions were based on our exploratory background research. We had to get more primary data to determine the functions to be designed and tested.

## 3.2 Design and Development of Aether

As illustrated in Figure 3.2, a monitor was installed on Kobuki (Kobuki, 2015), a mobile robot base, along with other sensory and computing components. A small, yet powerful computing unit was installed on the base, running ROS on Ubuntu 17.04. *Aether* could thus be controlled remotely via SSH (Secure Shell) protocol. The entire system was about 20 inches tall, with a replaceable monitor mount to ensure that we could change the display module based on our design needs in the future. Initially we used the display (12.3" (2736 px x 1824 px)) of a Surface Pro 6 computer as the monitor; later, we installed a dedicated 10.1" (1024 px × 600 px) LCD screen on Aether.

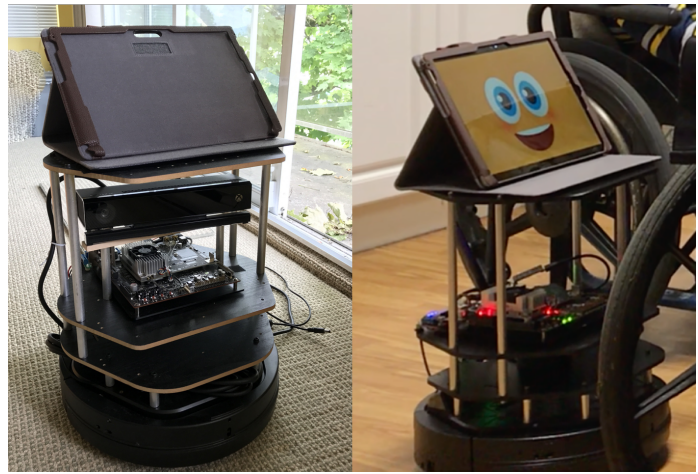

Figure 3.2: The setup of *Aether* without (left) and with (right) the display

As this thesis project was a progressive study, major changes had been made irregularly for the long-term objective. For instance, there was a 6-month gap between User Study 2 and User Study 3. The research team working on this social robotics project kept updating the hardware and software framework, making *Aether* more powerful and capable. Before we were running the third user study, the following changes had been made:

- **New Hardware** Thanks to the Intel® RealSense D435 Depth Camera, *Aether* can now perform localization and facial recognition faster and more accurately. Its 1920 x 1080 RGB Resolution and global shutter sensor enhanced the reliability of facial recognition whether it is indoor or outdoor, day or night.

- **Cloud-boosted Framework** We used a MiFi router that enables *Aether* to keep its connection to the cloud all the time. All terminals and servers we used to operate this robot system now connect to the same hotspot, eliminating the firewall barriers in any group home.

- ***Amazon*-backed processing engine** We integrated the Alexa Voice Service (AVS) to our platform, bringing Alexa's capabilities including custom skills to *Aether*. We could then give verbal commands directly to Alexa, like "Computer*, ask *Aether* to move one meter forward."

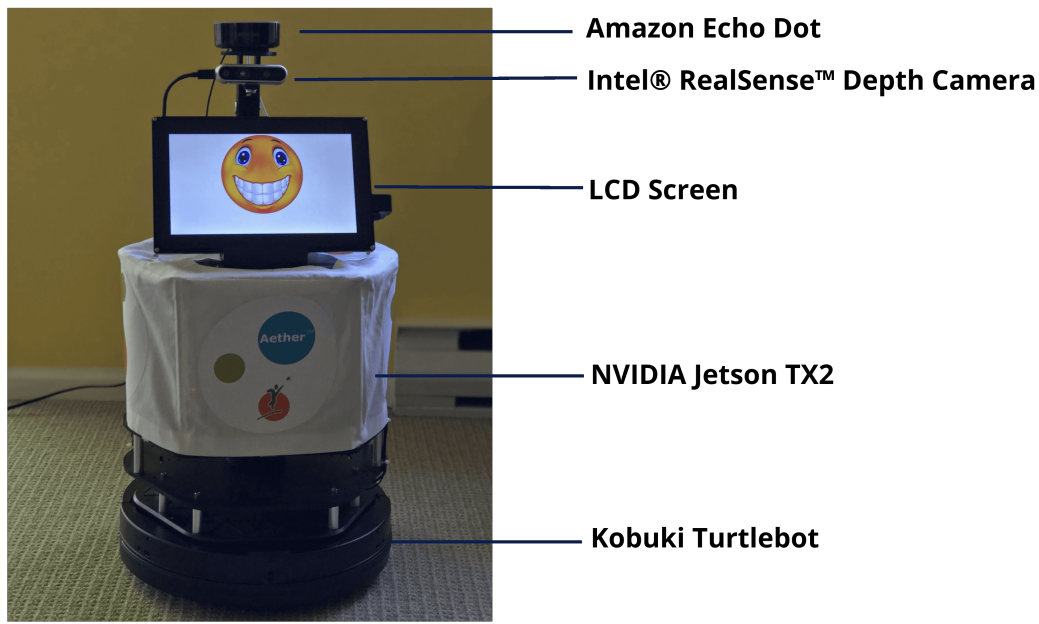

Figure 3.3: New Robot Setup of Aether

*"Computer" is the signal word we used to replace the default "Alexa"

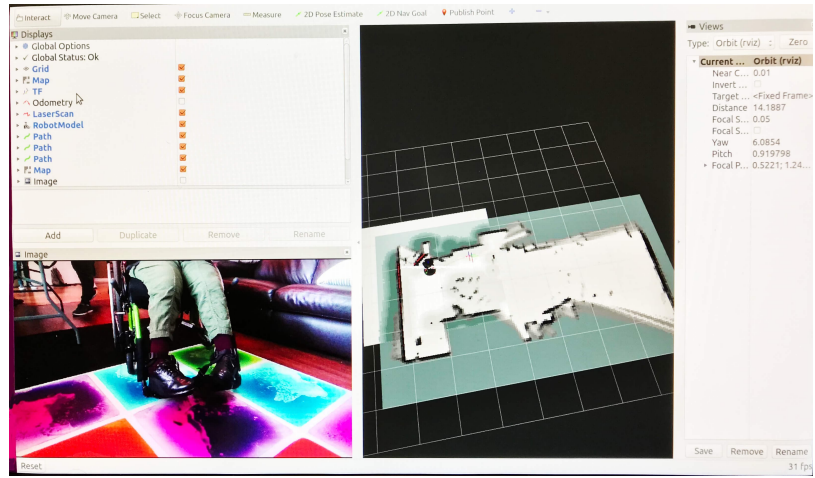

Figure 3.4: Control panel interface of *Aether* from the server, showing *Aether*'s views, localization progress, and other technical details

## 3.3 User Demographics

Knowing who we are designing for is the first question to ask with the product or service to design in mind. Our target users can be divided into two different groups: residents at the group home and their caregivers. In user-centred design (UCD), we need to have a clear comprehension of our users before discussing user experience. In the meantime, we need to have users as part of the evaluation of the overall user experience, which requires long-term follow-up and close observation during pilot studies.

Three residents at a local group home managed by DDA (Developmental Disabilities Association) participated in our study, along with a few caregivers. The number of caregivers observing or participating in each user study varied based on the schedule of the group home and the need based on the design of our user study. All resident participants were female adults with various degrees of developmental disabilities, which meant that they also had different levels of cognitive abilities, and verbal and physical skills. Rather than generalizing our assumptions and conclusions, we mainly aimed to find the significant factors differentiating between different types of interaction due to the small sample size.

Participant 1 (R1*), aged 50, had rather limited verbal and physical skills. Resident 1 also had some degree of delusion with symptoms such as giggling abruptly by herself. She had receptive language difficulties resulting in her restricted comprehension of simple words and phrases like "TV."

Participant 2 (S1), aged 54, had good verbal skills, but she was not very responsive all the time. There were occasions when she could answer some simple questions with full sentences, but only with caregivers whom she was familiar with. Resident 2 had several typical symptoms of dementia, including problems with focusing and paying attention, immature judgement and decision making, and limited reasoning skills. She could stare at an object for a long period of time, but it was difficult for researchers to understand if she was indeed focusing on the object or just in a trance.

Participant 3 (C1), aged 74, was able to verbally communicate with both companions and strangers. She could be a bit over-talkative occasionally and tended to give answers that can please questioners. Although she had to sit in her wheelchair all the time, she was still able to provide information using signs, gestures and facial expressions. Resident 3 did not have good executive functions, or planning and sequencing skills. She also had challenges in short-term memory, which was often a sign of dementia. Due to the declined thinking and reasoning skills, oftentimes she had confusion and troubles in solving problems and following storylines.

We needed caregivers in our study because the residents were not a user population that we could inquire in the normal way. Caregivers who observed our studies while accompanying the residents were the other level of participants in our study. These caregivers, who were also observers and advisors, all had training at DDA and their work experience varied. Almost all caregivers were at their late 20s or 30s, and 90% of them were female. The managers at the DDA group home were also caregivers, with more experience with residents and more systematic knowledge of developmental disabilities.

Designers sometime make assumptions and take it for granted that their users will understand the purposes of the design and get expected user experience. However, we are faced with a challenging user group who cannot articulately convey their thoughts and experiences to us. Hence, we need to take precautions when speculating about our users' perception and experience. This is not easy, as it may not be feasible to inquire about users' feelings and comments directly, due to their cognitive impairments. Therefore, we could conduct multiple short user studies while iterating and improving our design. Besides, we need to search and identify viable methods of data evaluation. From there, we could possibly achieve generalization from our work, which could benefit future researchers in this community.

---

*The abbreviation indicates each participant using the first letter of their names. Each unique abbreviation is used in following chapters and *Appendix*.

## 3.4 Qualitative vs. Quantitative Research

We had chosen qualitative research methods as our primary approach to collecting and analyzing data, for the following three reasons:

First, the small sample size of our user group had made it impractical to draw accurate, generalized conclusions. Due to the nature of this study, we are unlikely to have a much larger user group in the future. Qualitative methods can lead us into more in-depth observation and interpretation of individuals' experiences and perspectives. For instance, we can do open-ended interviews which are based on an individual basis. Second, our study is based on exploratory observation instead of instrumental measurement; we did not apply any sensor-based technology to users, such as eye-tracking heat map generator. Thus, there are not many numerical or statistical results we could collect and analyze. Lastly, our results for the user studies had multi-dimensionally crossing relations that would lead to more meaningful clues using qualitative coding rather than descriptive analysis.

Therefore, qualitative research methods can better fit our research.

## 3.5 Research Methodology and Methods

Due to the challenging nature of our resident participants, standard usability tests do not work in this environment. Instead, we employed several different qualitative methods, including participant observation, and interviews and focus groups with caregivers.

One of our research goals requires us to understand the communication patterns of people with different degrees of developmental disabilities. To achieve these goals, we needed to collect data through interviews or literature reviews before starting the design process. Figure 3.5 shows the time frame of this project. In order to get as much background information as possible, we decided to have several preliminary interviews with people who are users or related to users. Due to the fact that most residents cannot talk or phrase thoughts in explicit language, we could only interview caregivers. The information we obtained from the caregivers gave us enough clues about the need of our user

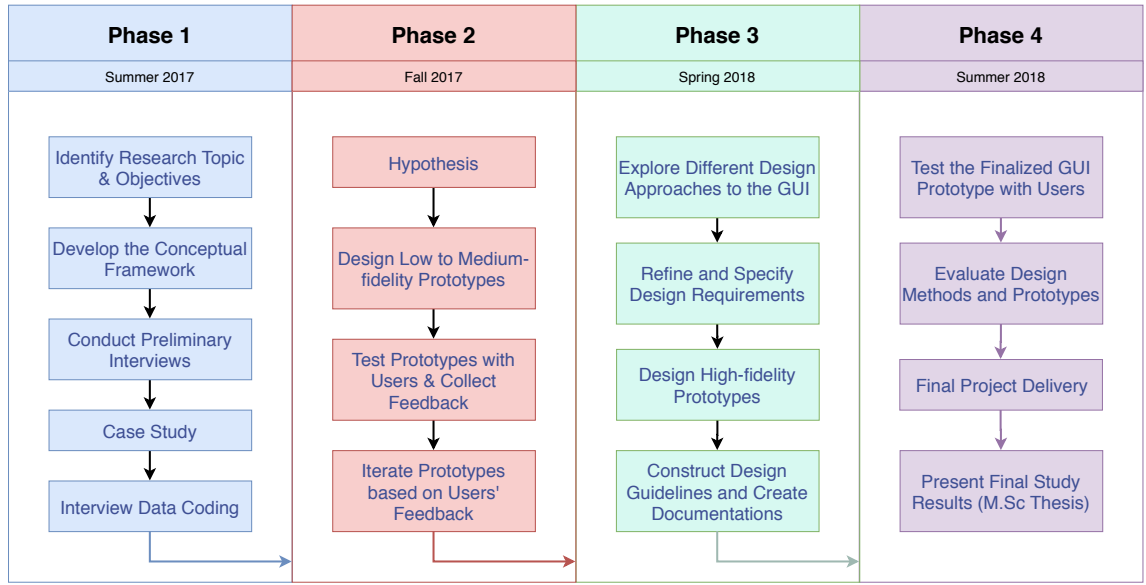

Figure 3.5: Overall time frame of this thesis project

group– residents and their caregivers. We then drafted several designs of the GUI and either tested them or inquire about them in more interviews with caregivers. We did not only test the visual design, but also aspects including sound, proximity, and overall user experience (UX). We conducted three user studies in one year, exploring the foregoing facets of interaction design.

### 3.5.1 Focus Groups and Interviews with Caregivers

Focus groups are used as one of the research methods in this study. Previous research in similar situations suggests that focus groups are a reasonable method to use in terms of collecting data and getting overviews of the interviewed group. For instance, Wu et al. (2012) conducted three focus group sessions in the process of designing an assistive robot for the elderly. The proposed idea for these focus groups was to provide engineers with some suggestions when they design the robot. Their results showed that these focus groups helped designers consider the social context of the robots and the elderly, because these groups allow participants to share and discuss their ideas of assistive robots. Therefore, the focus group method used in our research project can be informational and helpful as well.

We used focus groups as a way of reviewing with caregivers to get a better understanding of the resident population. The preliminary interview was held in a residential branch of DDA (Developmental Disabilities Association), where we had several residents with developmental disabilities and their caregivers. A focus group "needs to be large enough to generate rich discussion but not so large that some participants are left out." (Eliot & Associates, 2016)

Our open-ended interview included 5 interviewees in total, including three caregivers, one office manager, and an external supervisor and counsellor of this project. The interview was semi-structured. A general outline of interview questions was used, but some other questions were generated spontaneously during the interview and they were recorded as well. The interview lasted about 50 minutes. During the interview, I took notes and highlight the vital points that I got from the interview.

We also conducted focus groups with two caregivers three times to gain more insights into the communication approach for residents. We learned about the visual language tools that caregivers used.

For the data we collected from focus groups, video and audio recordings brought us a high degree of reliability and validity. They were helpful for identifying who is speaking and for improving the accuracy of the research by replaying sessions during analysis. The audio files were transcribed or coded as textual data in addition to the notes recorded during the interview. We then printed out the transcript and conducted line-by-line open coding with a pen. This approach enabled us to get a preliminary descriptive framework of the data that we collected.

### 3.5.2  Observational Studies and Cognitive Testing with Residents

It will be a challenge if we want to collect feedback from residents, because of not only the verbal skills but also their cognitive abilities. For some of them, they will just choose whatever comes the first as "the best one" when we present a few selections. Therefore, we moved our focus to residents with higher cognitive and verbal abilities, and also caregivers. All experiments and interviews were conducted on an individual basis.

Due to the difficulty of getting feedback from residents, we relied on standardized data and objective observations instead. Residents' verbal capabilities were determined based on their verbal dexterity indicating whether they are able to explicitly express their feelings and thoughts. Residents who were verbally capable tended to have a much lower cognitive impairment. For them, we assessed their memorization of certain tasks displayed on the GUI. This approach was inspired by two established tests: Nonverbal Medical Symptom Validity Test (NV-MSVT) and Story Recall Test (Green, 2008). NV-MSVT, without any textual information on the screen, contains only images and takes 5 minutes to run. Although these two tests were originally developed to appraise severe memory impairment for people who might have dementia and Alzheimer's disease, they do have many common characteristics to our study including the focus on users' perception and the use of stimuli.

# Chapter 4

# Interaction Design: Challenges and Methods

Interaction design, as the core component of our study, determines the feasibility of employing socially assistive robots to provide service or companionship to individuals with developmental disabilities. In this chapter, I describe the procedure of our interaction design process and list a few early-stage prototypes from which we found further design implications for subsequent user studies. After we conducted focus, we learned of the visual language tools that caregivers used. We then defined and selected the most common and scenarios that we could implement to our interaction design prototypes. We reflected on several topics mentioned in Chapter 2 - *Related Work* and got more practical perspectives of design considerations such as visual hierarchy and speech through interviews with caregivers and small pretests with residents. The following two chapters will elaborate the design process of each of the three user studies dedicated to exploring specific aspects of HRI.

    We do not know how to integrate the characteristics and needs of people with DD with our interaction design on a socially assistive robot. Applying the knowledge of our user group to the design process is an indispensable but complex step. We are not certain whether commonly accepted design principles apply to our target user groups. However, due to mental impairments, our target users can react differently than common users. Thus, we need to assess the fundamental design principles, especially from the visual aspect. These design principles (Marcus, 1995) include organization (navigability and consistency), economization ("doing the most with the least"), communication (meaning can be transmitted and perceived), symbolism, etc. Certainly, these above-mentioned principles are only examples of GUI design solely from a perspective of visual communication.

As we were designing for a user group that was unfamiliar to us, we rely on caregivers' advice and feedback when we started or modified our prototype or study methods. Caregivers in the group home all had formal training and residents spent most of their time with them every day. Hence, caregivers, especially experienced managers and caregivers, knew what would work and what would not. They could help us not only identify the flaws in our study design, but also evaluate residents' experience – this is critical to our study.

## 4.1 Current Visual Language Tools

It is important to know users' current experience in the process of interaction design. To make designs or provide service for residents, we first should know how they perceive information at the group home. In other words, we need to investigate how caregivers deliver information to them as well. During the interviews, the caregiver demonstrated to us the tools that they used to deliver information. Currently, at the group home, two primary methods of communication used for the residents are the information board and an iPad app with simple symbols and icons, to communicate simple instructions and feedback. These visual language tools are being used by caregivers because of language deficiencies.

Figure 4.1 shows how caregivers show information to residents at the group home. Each section of the information board consists of several pictures, usually in a specific sequence. Those pictures are also being used to train residents regularly in the form of flashcards. Although these photos also come with textual labels, residents are only expected to be able to comprehend the graphical information. The texts are for caregivers' use only, allowing them to index, locate, sort, and print conveniently.

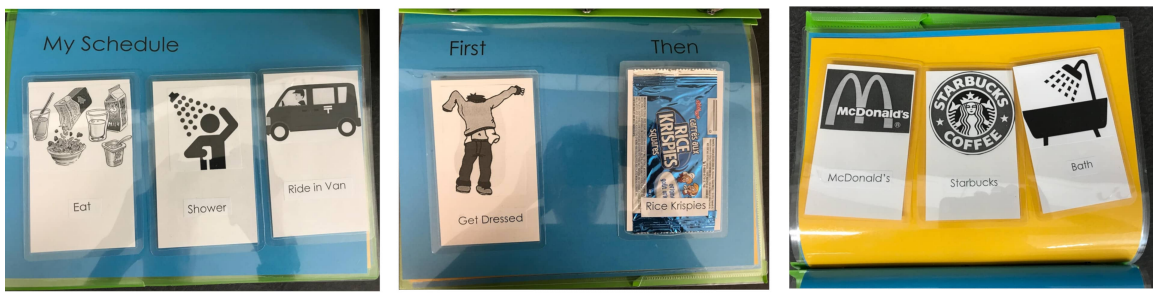

Figure 4.1: Information Board at the Group Home

Besides the information board, Figure 4.2 and Figure 4.3 show how residents view their schedules and activities on an iPad. These iPad apps were initially developed for children instead of individuals with cognitive impairments. However, they are still being used at DDA because children and residents at group home have similar cognitive abilities. Although not all residents use these iPad apps, and the activity information there is more for the purpose of training residents of cognition and memory, we as interaction design researchers can still get inspiration from the design of these apps.

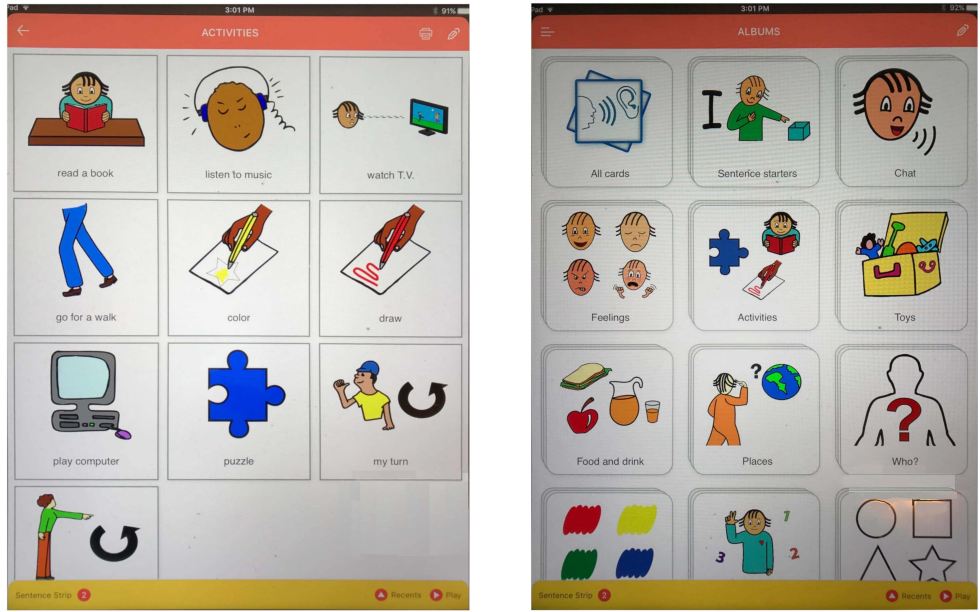

Figure 4.2: Activity and Album View on iPad

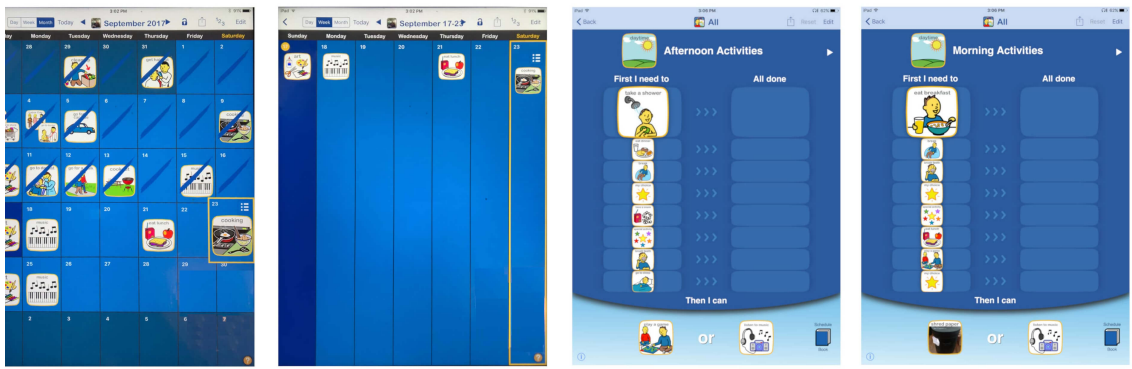

(a) Calendar 1    (b) Calendar 2    (c) Task View 1    (d) Task View 2

Figure 4.3: Calendar and task views on iPad

The use of iPad at group home has already shown that multimedia devices can be of good use for people with mental impairments. Thus, for us to expand this communication and leverage the robot, we analyze the design of these sets of software or devices and find concepts and methods that we can adopt to our design.

## 4.2 Defining and Selecting Scenarios

For robots that are competent to function and engage users, there are four indispensable prerequisites: self-awareness, self-reliance, communication, and adaptability (Fong et al., 2001). A robot needs to be able to recognize different scenarios and have fallbacks in the event of reaching certain bottlenecks. This aspect is critical as essentially robots are designed to serve humans. Each phase of the entire operation process needs to be well-thought-out as an assurance of safety and stability. The robot does not necessarily have to talk to users but has to convey information in verbal, haptic, graphical, or other sensory forms. It also needs to meet the needs of most users in a user group, rather than being limited to a specific one or two. For socially assistive robots, this adaptive feature is especially significant as the users have substantial differences in skills and levels of perception.

We started the design process from several preliminary interviews with the caregivers and identified some of the most common scenarios to consider in our further user studies: giving prompts, scheduling, entertainment, etc. Given the large number of scenarios, we started from two simple, but the most common scenarios: scheduling and reminding users of upcoming events and activities, and giving them prompts. Rewards is also an important and common scenario at the group home and it fits the role of socially assistive robots. We then used these three most essential scenarios in subsequent user studies.

**1. Reminders & Scheduling**

Scheduling and giving reminders and alerts to residents is an essential part of caregivers' daily jobs. Most of the reminders have regular and periodical times; still, it is an unneglectable burden for caregivers especially when a group home gets understaffed. Common events that need to be reminded to the residents include medication, doctors' appointments, and other scheduled activities. Medical-related events are the top priorities over other activities. However, they happen to be the most challenging type of reminders as on average there are 8 different medication administration times every day for all the residents and the schedule depends on the individuals. For schedules events, currently caregivers use an iPad app which is similar to the regular Google Calendar app, but with larger font size and more pictures indicating different types of activities. Making responses to residents' inquiries is another troublesome task for caregivers as they need to give timely and accurate answers to residents' questions, most of which are schedule-based. Each resident has their unique timetables. The alerts and notifications can be categorized into two main types: immediate and urgent events (e.g. taking medication), versus scheduled and secondary activities (e.g. tea time).

**2. Prompts**

Another critical part of caretakers' jobs is to provide residents with prompts, including both verbal and non-verbal ones like gestures. These prompts are indispensable, ensuring residents' safety and healthy development. Currently, caregivers also rely on pictorial cues for prompts. Residents at DDA group homes get training before and during their stay. At the beginning of the training, residents get detailed "step-by-step" instructions for tasks like laundry and chores from their caregivers. Later, caregivers reduce the frequency of giving prompts or skipping one of the steps. This is to enhance residents' capabilities of *chaining*. The concept of *chaining*, referring to the ability to build meaningful and accurate connections between things, symbolizes one's independence and features in intellectual development as an indispensable part.

> *The big thing we try to focus on is we don't want it to be just verbal - just 'do this, do this, do this.' But having that sort of breakdown, we can just show them a picture of shoes, and they would get their shoes, because it's a lot easier to feed those prompts eventually, than it is to have this constant 'Do this, do this' and they become sort of, I guess, reliant on those verbal prompts when they're constantly getting them.* - P1, Caregiver, Female

Caregivers usually need to inform residents of their upcoming tasks. For a number of tasks, caregivers need to break simple tasks into very specific steps. For example, when it comes to a resident's turn to do laundry, the caregiver says prompts aloud and explicitly, like "first, open the door." Not knowing how to do routine tasks is not the primary issue; the real problem is that residents do not want to do them. In this case, the caregiver tends to give rewards or penalties (usually in the form of not giving rewards) in order to stimulate residents to complete their jobs. There are a variety of motivators which work differently for each individual. For example, a resident at a DDA home really likes bananas, whereas another one is more interested in music from the '60s and '70s. As part of this reward scheme, it is also necessary to keep track of residents' progress and let them know how far they are to get the rewards. There are also times when a resident need to be prompted for behaviours: a resident might be craving people's attention and recognition, so he/she talks too much. It would the caregiver's job to let this resident know when to talk. Another example is that some residents always leave the table awfully messy, and in this case, their caregivers need to remind them to eat slower and keep the table clean. Thus, a fundamental function that our assistive robot should have is to prompt residents for activities and behaviours.

Verbal prompting is one of the most common and effective types of prompts. However, it may not work for everyone. A case in point is that a resident does not respond to the caregiver's verbal requests and hints and the caregiver has to point to things that need this resident's attention using his/her hands. Studying the effect of different forms of communication and prompting is critical for developing an efficient and intuitive robot which can guide and assist residents. Another common way of prompting is pictorial prompt, which is more versatile and customizable. We can use or create pictures, animations, and videos to deliver messages to residents. Factors that we need to consider and evaluate include what kind of graphical forms and styles we should have, which may vary depending on residents and types of incidents. Physical prompting is also often used at group homes, as real objects are more approachable and comprehensible than any other kind of illustration.

**3. Rewards**

Apart from prompts and reminders, rewards are also given to residents on regular or irregular bases based on the type of activities. For example, there are daily rewards for cleaning up, and additional rewards if residents follow instructions. Rewards potentially play a critical role in increasing and maintaining residents' engagement in HRI.

> *The idea is that you will create a program to motivate – it doesn't really matter what the reward is; the point is that they get it every single time they do it. The next time, they will get it every other time after they've been consistently doing it for getting that reward,* - P1, Caregiver, Female

## 4.3  Design Considerations and Challenges

After learning how caregivers present information at the group home, we have defined the factors that matter the most to residents' learning and interaction patterns with the robot.

The following factors have been taken into our considerations while developing the GUI and interactions.

### 4.3.1 Design Workflow

Figure 4.4 shows our design workflow. For participatory design, we need to learn about users' experience and participants' feedback. Through the iterative process, we modified or recreated designs based on the results of user studies.

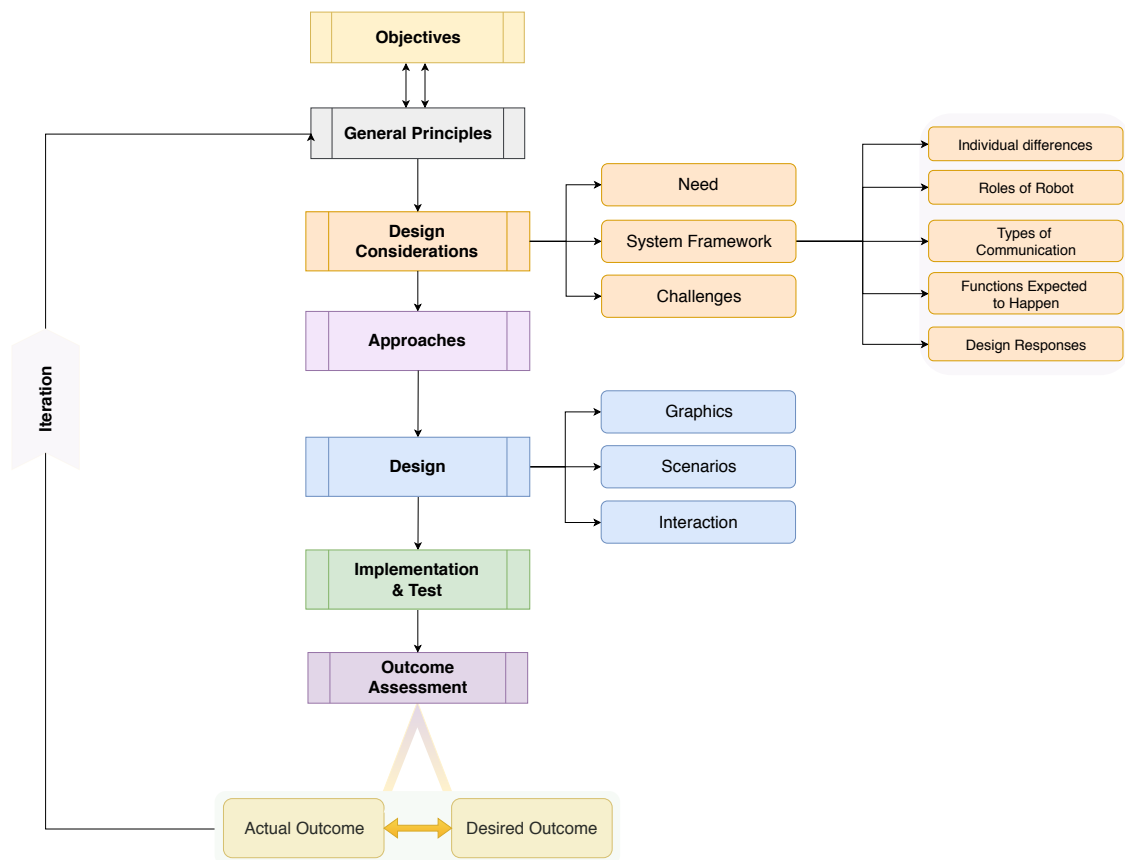

Figure 4.4: Design workflow

### 4.3.2 Character Types

In order to design GUI prototypes for users, we need to run a pilot study to narrow down the scope of basic elements, based on users' direct feedback and secondary results from their caregivers. The pilot study is a part of our iterative prototype design process, aiming to refine the most fundamental factors affecting users' emotions and level of interest by using techniques of behavioural analysis. Although our prototype is mainly designed for residents, we also consider caregivers as an unignorable user group as they will need to interact with the socially assistive robot as well. We frame the study within certain conjectures like users prefer one type of character to the others. More specifically, we are eager to understand and find out:

1. Do users prefer one type of character to the others? If so, is this preference noticeable? What are some of the strong evidence supporting our conclusions of their preferences?

2. Having all characters show similar neutral-like emotions, is there any distinct change of user's response when we switch the characters?

The details about this pilot study, including the procedure and results, are presented in Section 5.1. This pilot study required us to have a focus on participants' views particularly their direct feedback and evaluation of engagement based on our deductive observations. We needed to find evidence to support our interpretations of users' behaviours.

### 4.3.3 Personification of Robot

Many researchers have made endeavours to categorize robot personality and there was evidence that people spend more time gazing at robots which behave actively, suggesting that they tend to be more interested in extrovert robots (Mubin et al., 2014). This coheres with the psychological discovery (Hendrick & Brown, 1971) that the introvert has more heterogeneous relationships with people: the extrovert is likely to be only attracted to gregarious persons because of their similarities, whereas there is no negative indicator when the introvert interact with the extrovert. However, it is also worth pointing out that previous studies on personality psychology have come to two outwardly paradoxical theories: similarity attraction and complementary attraction (Brandon, 2012).

Whether it is similarity attraction and complementary attraction, we will have to explore and revalidate this by ourselves for the following reasons. Firstly, our users might behave differently from their true internal personalities due to cognitive impairments. We cannot conclude a linear relationship between their personalities and the engagement with the robot. Secondly, we have created a catalogue of existing residents' profiles, including their habits, personal preferences, always-up-to-date schedules and etc based on previous observations and interviews. Utilizing these data will improve the quality of human-robot interaction as the robot would be able to pinpoint users' needs and desires. Lastly, we have not determined whether this robot can serve only one or multiple users at the same time. As a virtual agent, the robot is being considered to be capable of interacting with all types of user. Therefore we need to set up a default personality "value" for the robot in the event that a new user's profile has not been indexed in the database.

### 4.3.4 Robot Behaviour

Designing robot behaviour is a critical part of our HRI study. Thus, we need to when and how the robot approaches users. Also, it is vital to integrate emotion expressions to the robot's personification, because of the significantly positive correlation between the robot's capability of expressing emotions and the quality of human-robot interactive communication (Leyzberg et al., 2011). We considered 12 common emotions that the robot can be equipped with to enhance its personification and ranked them according to the frequency of each expression in everyday scenarios.

**Visual Elements of GUI and Affective Communicative Intent**

"Robot's affective communicative intent" (Figure 4.5) refers to the delivery of message showing the current status of a robot in a humanoid way. It lays the foundation for effective interaction and high engagement. Symbols and derived animations were used to reflect different communicative intent. The most common emotional communication include: "Smiling," "Thumbs-up," "Cheerful," "Sleeping," and "Neutral." It should be noted that a same emotion could be repeatedly used for different scenarios. For example, the "cheerful" emotion can be triggered when someone's birthday comes, or simply when the caregiver is going to give the resident a ride outside. There are three phrases to test communicative intent:

- **Evaluate the frequency of each emotional communication based on the most common scenarios identified previously**

  Previously, we had identified the most common and essential scenarios that residents underwent every day. These scenarios included medication reminders, activity reminders, daily tasks, etc. There are many other types of emotional communication besides these common ones listed above. Not all need to be implemented due to users' cognitive challenges. They can only memorize and learn to identify a few states, so diminishing users' learning curve is significant.

- **Test the effectiveness of the graphics**

    Keeping a good consistency in graphical styles is the prerequisite of achieving an accurate evaluation of different communicative intent. When a user fails to understand a symbol, it is likely that the very image/video being used is not clear enough, then we will need to find alternative resources to retest that emotional communication.

- **Assess the necessity of using animation for each emotional communication**

    There is generally an assumption that that animation is engaging and fun, but some of the emotional communication can be more effectively represented using static images. We also need to take the robot's motion into consideration while analyzing each emotion.

**Social Interaction Actions and Affective Communicative Intent**

A socially assistive robot should be more than an intelligent digital agent. We explored the possibility of endowing social robots with humanoid characteristics. Depending on the stage of the interaction (approaching, beginning, interacting, and ending), the socially assistive robot exhibit a feeling or emotion that could be perceived by users. In this case, users will consider the robot more like a social agent, instead of a movable computer. We defined the interaction procedure (i.e. four stages), identified the robot's affective state(s) for each stage, and designed actions for the robot, as shown below.

1. **Approach the User**

    (Greeting and Happiness)

    The robot gets called by the user and then approaches him/her. It keeps a minimum distance of 70cm to reduce users' anxiety due to close proximity. This proximity parameter can be set to different values by caregivers based on users' personalities and behaviour patterns. This first step is critical to engage the user's interest and attention. The robot approaches with a neutral-like smiley expression and a verbal greeting. After the robot comes close enough to recognize the user's face, it says a personalized greeting, for example, "Hello John, how are you doing today?" or "How can I help you today?" This also applies to situations where the robot needs to come to the user actively to give reminders.

2. **Begin Interaction**
    - Give notifications

| Robot State | Example | Use | Frequency |
|---|---|---|---|
| Neutral | 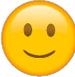 | A neutral emotion is being used as a default expression for any other cases not included in the following states. It simply implies a waiting status. | 5 |
| Smiling | 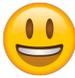 | A smiley face tells a user that the robot is reading the user's input (e.g. voice or touch), and indicates that a conversation or other interactions are in process. | 4 |
| Thumbs-up | 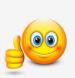 | A thumbs-up symbol gives user encouragement, compliment, or consent. | 3 |
| Double Thumbs-up | 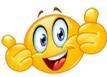 | A double-thumbs-up symbol is a simple variant of the thumbs-up symbol. It is only being used for particular occasions. | 1 |
| Thumbs-down | 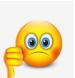 | This icon will be rarely used. It only applies to denials and dissatisfaction, which can trigger users' disgust and hostility. | 2 |
| Cheerful | 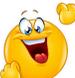 | It indicates things that are pleasant, exciting, or appealing. This state may be triggered on various conditions as each user might have distinct stimuli. | 4 |
| Sleeping | 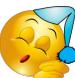 | The sleeping state is a symbolized and humanoid way to show the current unavailability, for example, when the robot is charging. | 1 |
| Hands Pointing Out | 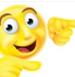 | This state is critical as it shows a secondary icon on the screen indicating another task, event, or activity. It bridges the gap between a humanoid imagery and icons. | 4 |
| Surprised | 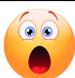 | The "surprised" image will be rarely used, suggesting a surprise feedback. | 1 |
| Confused | 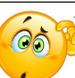 | When the robot is looking up information or even fails to understand the users' commands, a confused emotion will show up. | 1 |
| Shushing | 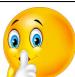 | The shushing motion is specifically designed for users who need prompts when they should be quiet. | 2 |
| Saying Bye | 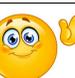 | After a user finishes interacting with the robot, it is time to say bye. | 3 |

Figure 4.5: Proposed catalogue of affective communicative intent

(Relaxation, Politeness or Enthusiasm)

As the robot is not called by the user but it needs to give timely prompts, it is vital for the robot to reduce users' fear and psychological resistance. Thus, a more evident smiley emotion needs to show up on the display, with gentle verbal prompts.

- Inquire about users' intention and need

    (Anticipation and Patience)

    When the user shows the intention to call the robot over, the robot needs to indicate the fact that it is ready for interaction and communication. A smiley face with eyes gently blinking suggests that the robot is waiting for users' instructions and inquiries.

3. **Interaction**

    - Deliver messages

        (Calmness or Enthusiasm)

        The robot shows an evident smiley face for simple answers. When additional symbols and icons need to be used to refer to activities, events and objects, a delighted face with two hands point to the icon(s) will be used in addition to the icons.

    - Provide rewards

        (Excitement and Motivation)

        The robot shows a cheerful face with hands clapping.

    - Give prompts

        (Desire and Anticipation)

        More specific emotions with gestures might be used depending on the type of prompt. Negative affective states should be avoided to reduce users' fear and anxiety.

4. **Terminate Interaction**

    - The robot moves away

        (Satisfaction)

        The robot shows the gentle smiley face and says bye to the user, and then moves away.

- The robot calls the caregiver

    (Satisfaction)

    In the event that the robot is unable to assist the resident or he/she needs the caregiver for other reasons, it will switch back to the smiley face with eyes gently blinking, suggesting that the caregiver will be here shortly and that please wait patiently. After the caregiver comes to the resident, the robot will say bye briefly and move away.

### 4.3.5 Visual Hierarchy

Visual hierarchy is the sequence in which users perceive information from a user interface (Davies & Tutton, 2016). It starts from real objects and gets stepped up to more abstract line drawings and text. This list (Figure 4.6) shows the order of different types of visual representation based on the difficulty level of understanding the information. As indicated in the figure, real objects are the simplest form of visual communication for people to understand, whereas abstract writing being the most complex form. This hierarchy, however, does not work for every individual. Some people may find photos easier to understand than miniatures.

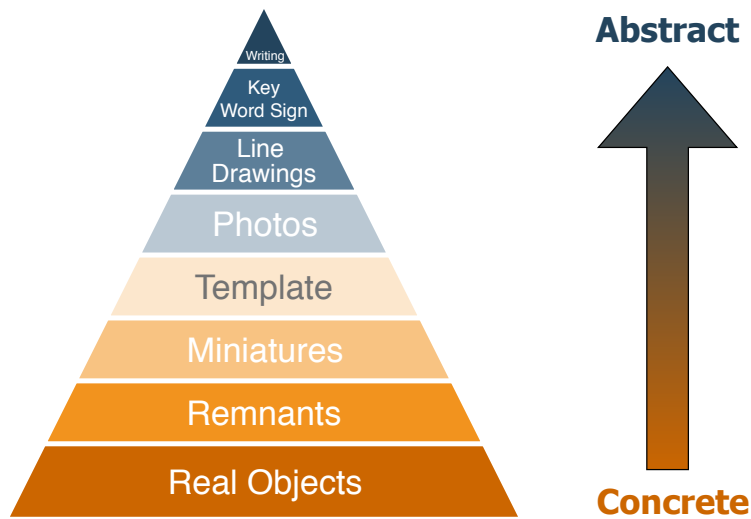

Figure 4.6: Visual Hierarchy

### 4.3.6 Speech in HRI

We used IBM's *Watson Text to Speech*, a TTS service equipped with deep learning technologies to synthesize life-like voice. We selected one of the U.S. female voices that sound natural and friendly. To compensate for our users' limited verbal skills, we further reduced the speech speed to 70%.

Besides the technical details of synthesizing speech, we also need to think about other aspects of verbal communications. First, the words we use have to be simple enough. Second, the verbal instructions have to be extremely precise. From our preliminary focus group interview, we learned that the wording and directives need to be concise and precise:

> *It (the robot) needs to be having a time limit and be honest with that. So if you say that "you're brushing your teeth for 5 seconds," it counts out load 5 seconds till it is finished. It's encouraging trust that what people say is what they mean. It also lets them see that... it's something unpleasant that we're going to do, and then it will be done.* - P1, Caregiver, Female

We learned that the challenges of verbal interaction for assistive or companion robots mainly come from two factors: accuracy and the interaction between human and robot. Unlike other mature interface technologies, speech interfaces incorporate some unique features which require designers to carefully investigate and think out the special relationship between users and the robot terminal.

Nevertheless, while we know that speech is extremely important to our study, it did not form the basis of our further studies, but we cannot separate speech from other pieces.

## 4.4 Summary

In this chapter, I listed the defined scenarios for designing interaction prototypes, outlined the design workflow, and classified a series of design considerations. The concepts and principles I obtained in the course of identifying the foregoing are critical to the interaction design of socially assistive robots. In the subsequent two chapters, I will illustrate how I apply these findings or ideas to my design, through three user studies.

## Chapter 5

# Designing GUI for Socially Assistive Robots: User Study 1

This chapter presents a user study we conducted to explore visual aspects of HRI through GUI. Aside from visual communication, we also investigated other facets such as verbal and physical interactions. I further investigate the social dimensions of HRI in depth in User Study 3 (Chapter 7).

Our research consisted of several studies carried out on site with the staff and residents in the group home under the supervision of the care staff. These studies incorporated several qualitative and observational methods common in interaction design (Fong, Nourbakhsh, & Dautenhahn, 2003), including staff interviews, resident observations, and prototypes to quickly simulate different types of robot interactions with the residents. We designed and executed several studies that explore visual and gaze-based communication with residents with the objective of determining design requirements in the 3 areas described in the chapter of Introduction (Figure 1.1).

## 5.1 User Study 1: Defining the Relevant Visual Factors

Our user study was motivated by a study from (Glas et al., 2013) who designed and implemented a framework providing social robot services in five ways. They first identified scenarios for developing a social robot and then determined key considerations and design requirements. Then they developed a robot system allowing the robot to complete tasks in real environments.

We had a conjecture that anthropomorphic pictures and animation could enhance users' engagement in HRI; we were also curious about how residents would respond.

Through this user study, we aimed to define the most significant visual factors for interaction design. The results should directly inform and guide the design of the robot's visual interface. We expected the following outcomes:

- A set of validated guidelines on what kinds of visual representation are affectively and informatively useful and appropriate for resident-robot communication given defined scenarios;

- Validated identification of the effect of gaze on resident acceptance through User Acceptance Testing (UAT), and engagement and validated conditions where gaze interaction is and is not useful/effective/necessary;

- An initial identification of factors in robot visual interface that should be customized to the resident and the scenario.

Prior to this user study, two caregivers recommended that we simply present all character options to residents, and also warned us about the limited amount of responses that we expected from residents. A senior caregiver also advised we create a GUI demo based on residents' real-life experience in the group home. The content of the demo had to be or have something that residents were very familiar with. Besides, the verbal prompts from the robot were required to be concise and simple for residents to understand.

## 5.2 Participants

The resident participants remained the same as previous pilot studies (Section 3.3 - *User Demographics*). Three caregivers observed this study. Two caregivers participated in the in-depth one-on-one interview after the study.

## 5.3 Procedure

The first stage was to run a small pilot study to test static images and animated videos (2D vs. 3D) with residents and caregivers. Residents' facial expressions and lengths of focus span were recorded while they were looking at the display. We asked residents to point out the very character which they would prefer over others and inquired of caregivers with their perspectives and feedback and document their responses. The second stage was to select one of the characters to continue the prototyping process. As shown in Figure 5.2, different layouts were presented with either an animated avatar or an event icon, or both.

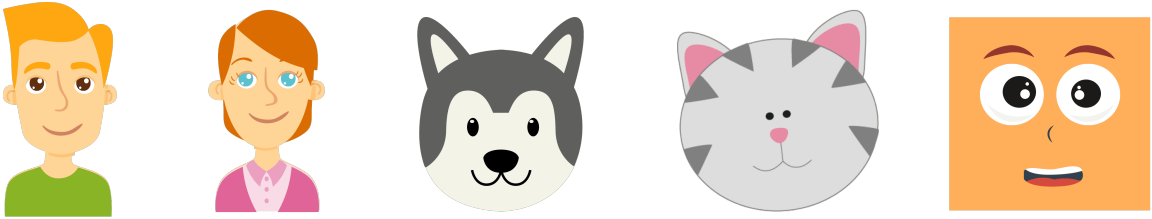

Figure 5.1: Five animated characters to be tested

Five animated characters (Figure 5.1) were shown to users one by one in the following order: male, female, dog, cat, and emoji-like face. Users had 10 seconds to look at each of these static images with a one-second pause between each image, and then they were asked to point out their favourite among all. A series of short silent video clips were played for residents in the same order as the static images. These animations brought vivid patterns to characters, making the graphics more appealing. Sound was not included as we wanted to eliminate the effects of other factors at this stage. All our studies could be stopped anytime if a user showed any sign of resistance or discomfort.

The same process was applied to caregivers while they were assisting residents during the demonstration. After this test, we asked caregivers a series of detailed questions as supplemental data:

1. Could you help us interpret this resident's preference and reactions?
2. Do you have a personal preference among these five characters?
3. Which character do you think might be the most suitable for the similar user group in general?

We also created several layouts (Figure 5.2) based on the type of elements. These layouts were not shown to the residents, but we asked one caregiver for critiques and suggestions after the user study.

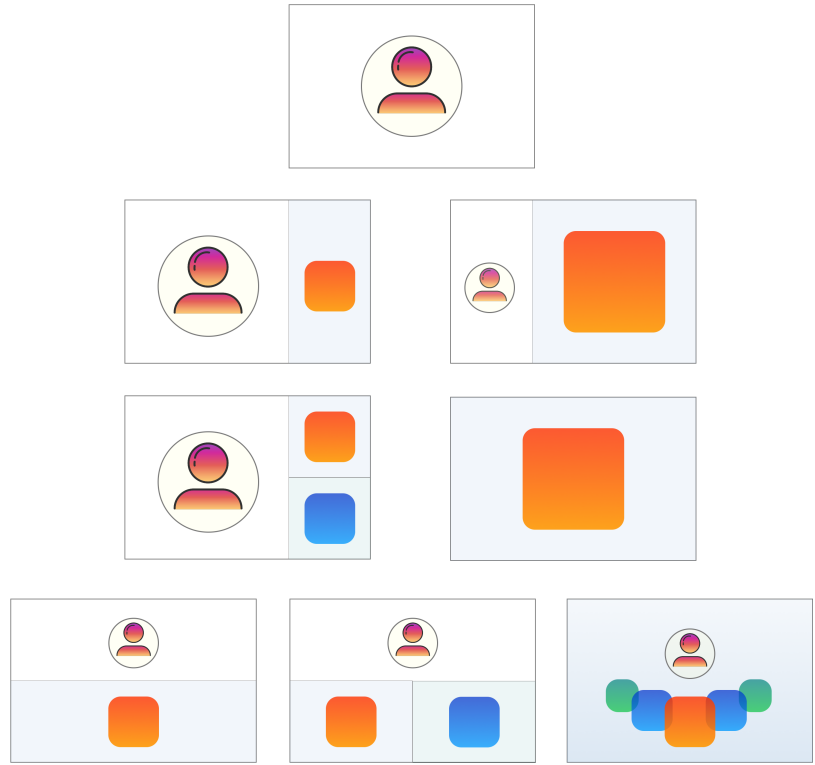

Figure 5.2: Different layouts of GUI prototypes

**1. Residents** The experiment was designed as follows: the robot's display showed 3 simple tasks and a reward after completing these tasks. This setup was close to the real-life scenario at the group home where caregivers routinely remind residents of their upcoming events and duties and motivate them using rewards like donuts. The difference was that they did not need to physically complete the tasks, and would not get the reward. (Residents were not aware of this prior to the study.) Each task was displayed for 15 seconds, accompanied by a female voice stating the task verbally. The verbal prompts were as concise as "Do laundry." or "Clean the table." All icons and symbols used had been taught to them by caregivers previously as part of their development program. Residents had unlimited time to recount this whole procedure, with all 4 icons showed in front of them. They needed to not only identify the reward, but also the order of three tasks.

**2. Caregivers** Two caregivers were accompanying residents while we were running tests. After the tests, we had a one-on-one interview with each caregiver, asking their feedback and interpretations of residents' behaviours. The following questions were asked in our open-ended interviews:

- What do you think of the overall design of the robot's GUI?
- Did you notice any moment when the resident is especially interested, focused, or disinterested?
- Is there anything that we can improve?

## 5.4 Method of Evaluation - User Acceptance Testing

User Acceptance Testing (UAT) validates a product's stability in users' normal interaction and reveals the issues that should be fixed by the next phase. UAT is a significant segment of software development, and thus we should cautiously choose the appropriate approach to deploying UAT. The most common approaches include basic UAT, beta tests, friendly user tests and parallel use (Cimperman, 2006). Due to the nature and time frame of our study, we chose observational *in-situ* test as the main method of validating our prototype designs. It allows real end users to use the product while researchers pay close attention to the interaction process and record any issue found.

## 5.5 Data Collection and Analysis

The entire study with residents was recorded using the webcam on the monitor. I focused on qualitatively analyzing where and how long they were looking, and the change and movement of their focus especially when there was animation or transition happening on the display. The screen-captured video and webcam video were compiled into one video file using Final Cut Pro, and then we imported the video to NVivo, a qualitative data analysis (QDA) software package (QSR International, 2018), and highlighted the significant moments (frames) based on participants' change of emotion, gesture, or gestural and verbal response. The majority of the codes are descriptive for participant observations. During the study, I also took notes on observations about residents' responses. The notes helped me code the video footage during data analysis. The interviews with two caregivers were audio-recorded using a smartphone. After the user test, I continued to transcribe the interviews with caregivers. The textual data was later imported to NVivo. I used analytical open coding for the interview data and on-site notes of observation and found the following results.

## 5.6 Results and Discussion

None of the three participants gave a clear indication of what their favourite character was. From the captured videos, we did not observe any obvious disfavour or liking for any of the five characters. However, for two of the participants, the Emoji character seemed to be more understandable compared to the rest. Although it is fairly easy for ordinary people to tell the type of each character, it was not the same case for our users. For participant C1, she could not tell the third character was a dog, and participant S1 hazarded a guess at the fourth character and figured it was a cat with the prompt of her caregiver.

As for the male and female characters versus the Emoji character, our background research shows that most social robots with display did not use any humanoid, gender-specific characters, the only exception being *Valerie the Roboceptionist* (Gockley et al., 2005). For the Emoji-like character, we found it was more attainable to implement different emotions based on it, and it can be easily converted to a 3D character in CrazyTalk$^{TM}$. Besides, the Emoji-like character is gender-neutral, allowing different personalities and voices to be implemented on the robot in the future. One caregiver recommended the Emoji character for future designs based on her knowledge of the residents' acceptance, and the other caregiver did not have any personal preference. For 2D versus 3D, our test of using 2D and 3D characters did not show any divergence of users' preference or acceptance. We observed that residents did not seem to really care about the personification of the character because they did not show any favour or disgust for the 2D and 3D character. Besides, they had issues identifying animals. In conclusion, we did not find anything that discouraged us from using the simplest form of visual representation from this study.

In a previous study from Wang et al. (2017) who conducted an experiment with 71 children from 3 to 6 years old, the researchers argued that 3D faces preserve more information of the face geometry and lead to improved performance of identity identification compared to 2D faces. Given these conclusions are correct, more information could be a burden for our users to process, compared to children in this experiment. Also, we were mainly concerned with the performance of emotion recognition based on our selected character type as the identity of a character can be easily be learned through daily training of users and users' frequent interaction with the robot. Therefore, we minimized the amount of information in the character and chose a 2D, static Emoji-like character for *Aether* to show the anthropomorphic features.

The data we collected from the interview was very informational and was used as a guide for our following design. The counsellors and employees working at the DDA group home provided us with some insightful information and perspectives in terms of applying assistive robots to help residents with developmental disabilities.

For the layout options (listed in Figure 5.2), we had defined the best ones based on the interview with caregivers.

> *If you think about it, we read left to right; we teach everything (that) is left to right. For all of these guys, that's what they've learned. Everything we've shown has been reinforced that left and right; so the most important information they're going to look at is going to be on the left. You want whatever they're going to derive the information from to be the largest.* - P1, Caregiver, Female

In a GUI consisting of more than one element, the primary element should be placed on the left, in a size that is significantly larger than the rest. This is a design principle that works for both the ordinary population and people with DD.

> *If this face is not giving them any information, having it over there that's the identity of our robot. This will give them all the information, that's just sort of an extra piece. The primary reason that I wouldn't want it on this side is especially for T1 and R1, if they're looking at it, they're probably going to get distracted and they're never going to look past it.* - P1, Caregiver, Female

In the interviews, caregivers pointed out that residents with higher cognitive capabilities were more likely to be attracted by animated graphic elements, and they could still understand the visual information. Caregivers were concerned that people with limited cognitive abilities, by contrast, might find the movements or transitions of elements disturbing and confusing. For them, reading a static image had already been a challenge. For this reason, we need to categorize users into two groups again based on their function levels, as we did for symbols.

The layout of the GUI needs to be simplified further so that it would not cause any confusion or distraction for residents. Even a basic layout with two simple elements, the character (face) and the activity/event symbol, can create extra information for residents to perceive and process. Therefore, ideally, there should be only one element displayed at the same time. For residents with higher cognitive function levels, having two elements at the same time can also be acceptable as the transition between the character and other symbols can be smoother.

A limitation for this study was the medium we used to select a desirable character: an LCD screen. Although it undoubtedly makes sense to run a test on a display that is eventually going to be used in the real product, for our users the screen posed a visual challenge. "People with dementia or developmental disabilities are extremely sensitive to light." (P1, Caregiver, Female) A lit-up screen, in spite of its quality, has flickering that gives residents extensive stimuli. Thus, the test could produce more ideal results if we had used high-contrast black or grey-scale pictures printed on a piece of white paper.

> *Basically, we want a dark (colour) with a light (colour). So mixing blues and greens together - just don't do it. If you're going to stick with a colour, try to go as straight as you can. So I want to get fancy with like let's mix Azure or something in there, like go RED, go GREEN – go something that is very visible to them. So we're having one colour sort of dominating the image. These are all on the white backgrounds. I wouldn't do yellow and white - it's really bad. If you are doing a darker background, stay away from any white images because they'll be seen black on white which is really easy for us, but seeing white text on black is extremely difficult for these guys. So try not to reverse those colours.*
> - P1, Caregiver, Female

We had got an initial concept of how colours affect residents' perception and experience. We made design decisions after this user study, including reducing the number of colours and shades on a screen. The selection of colour needs to be cautious as some of them can potentially trigger negative experience. There are some safe choices for colours, as mentioned by the caregiver above. We could start from there and try to explore other possibilities through subsequent user studies.

# Chapter 6

# Exploring the Social Dimension of HRI: User Study 2

In this user study, I investigated the dimensions, especially the social dimension, of HRI. For socially assistive robots, the social dimension of HRI is not only indispensable but also one of the most important dimensions. A clear awareness of human behaviour and traits across social, cognitive, affective, physical dimensions is a prerequisite for SARs being capable and competent companions for people (Breazeal et al., 2008). Because this thesis project mainly explores the social interaction design of SARs, we want to focus on investigating the social dimension of HRI. Inevitably, we will also encounter other dimensions as we study the affect and behaviour of participants. Although these other dimensions are not our primary focus, they are still critical to our study because of the links between them and the social dimension. Thus, we record relevant findings and seek the connections with the social dimension.

To better understand the social dimension of HRI, especially for our user group, we need to know how users currently socialize. Dautenhahn (2007) argues that it still remains unclear whether robots are able to fulfil the "social-emotional" dimension in human-human interaction (HHI). This is also a question that we are investigating in this user study: what are the differences of the social dimension between HRI and HHI? We are eager to know this gap between HHI and HRI as it is the key to a more acceptable, engaging, and interactive level of HRI.

Between HHI and HRI, there is HCI as the transition. In the past, Li and Chignell (2011) found that interpersonal behaviour (i.e. HHI) impacts on the way that people interact with computers (i.e. HCI); and this finding had been applied to HCI design. Li and Chignell extended this discovery by applying the same concept to HRI design – this inspired our user study. We believe it is worthwhile to examine HCI in addition to our exploration of HHI and HRI. Hence we compare HHI, HCI and HRI and aim to find out how we can base HRI design on HHI and HCI.

Following User Study 1, we continued our exploration of GUI design of SARs; we carried out User Study 2 and used the GUI and speech the primary types of interaction to solve the research questions defined above. We designed a short GUI demo showing residents their scheduling events and a reward to observe the interaction process and outcome. This study was divided into three stages (Figure 6.1) based on the types of interaction: human-human interaction (HHI), human-computer interaction (HCI), and human-robot interaction (HRI). We observed the communication between caregivers and residents, then conducted two Wizard of OZ (WOZ) tests by controlling a computer and a robot remotely to play the GUI demo. We observed and recorded these two interaction procedures to assess participants' experience and compared them to the first observation (HHI).

In this chapter, I first describe the procedure and findings of each stage and then discuss the questionnaire feedback. The core participant group of the first three parts was residents, and that of the questionnaire part was caregivers. In Section 6.5, I explain how we analyzed the data collected from each user test and further investigate the coding results. As user study 2 consists of several complementary and linked tests, I first present the coding results of each part through coding the video footages separately; then I show the overall coding results for all parts combined in Section 6.6.1. The overall results were obtained after I imported all data to NVivo for thematic analysis. At the end of this chapter, I compare HHI, HCI, and HRI, discuss design implications and summarize findings on visual factors.

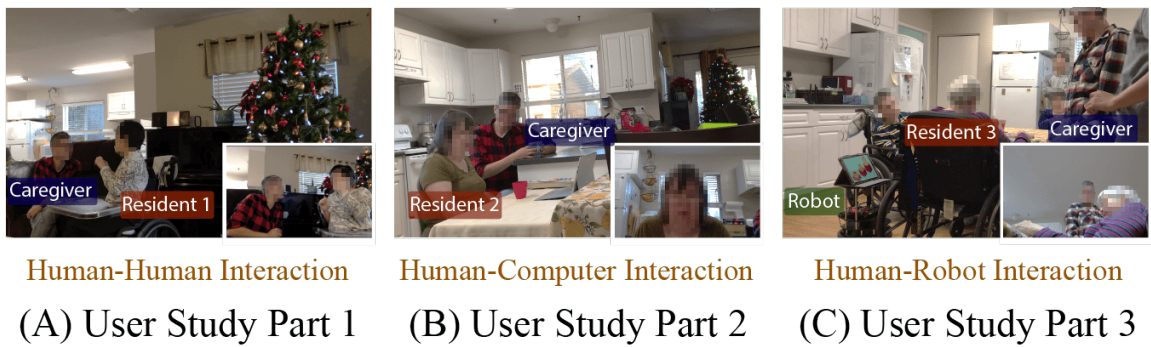

Figure 6.1: Three stages of the third user study, corresponding to three types of interaction

The resident participants in this user study are the same as the previous study. One caregiver who was the manager of the group home participated in this study.

## 6.1 Part 1 - Observation of Residents and Caregiver's Communication (HHI)

### 6.1.1 Procedure

The first part of the study was completely observational. A caregiver was asked to talk to each of the residents for about 3 minutes. There was not any specific topic or question that the caregiver had to mention. In contrast, a caregiver could bring up any topic that she/he might feel appropriate to attract residents' interest. She started the conversation with a greeting, "Hi, XXX, how are you today?" Then the caregiver tried to bring up topics that might be interesting for residents. For example, "Do you want to go out later?" and "Should we order more tea? What kind do you like?" The majority of the dialogues were caregiver's questions and residents' responses. Residents were not obligated to make any response and were free to ask questions or talk about anything they preferred. The caregiver pointed to objects and the surrounding area to attract and guide residents' focus. She did not use any other widgets (e.g. flashcards or toys) in this phase. The caregiver's approach to handle the conversation, along with the study environment, was kept the same as the participants' daily experience. Therefore, there could be distractions or unexpected events occurring in the environment, and the researcher would not intervene in the event that anything happens. The researcher hid behind the participant to observe and record the study.

Since the caregiver did not use any visual prompt in this user study, and the following two parts (i.e. HCI and HRI) involved a GUI demo, it is not appropriate to compare them at the same time. Therefore, we only explore interaction behaviour and patterns from a general level. In the following part (Section 6.2), we still compared different types of interaction from their overlap in this study: speech.

### 6.1.2 Data Collection and Analysis

The entire user study was recorded using two cameras. One of the camera captured the facial expressions and gestures of residents, and the other one recorded the participants in the room setting. I conducted low-level coding through watching the recorded study videos three times. At this stage, I only marked the noteworthy moments and observations, especially regarding participants' attentional states. I explored when and at what circumstance participants would pay attention to the caregiver and even give a reply. I also observed their emotions during the communication and got the caregiver's help to identify and inter-pret their emotions and responses after the study by rewatching the recorded study video. I applied the same basic coding approach for the other two parts (HCI and HRI), and then conducted a formal open, axial coding for further thematic analysis.

### 6.1.3 Results

When the caregiver sat next to residents and maintained a conversation, she got a significant amount of verbal responses. As long as residents could perceive and understand caregiver's questions, they usually gave a response. When having the communication, residents generally paid their attention to the caregiver, especially for S1 and C1. R1 was easily distracted by other things in the room like always – this was expected. Even for R1 who tended to be distracted easily, she responded to caregiver's questions, followed the derivatives, and repeated the words of her interest many times. S1 held her gaze at the caregiver for the majority of the conversation and answered most of the questions. She used a lot of gestures to show caregiver her responses. C1 was very talkative and used much body language to explain to the caregiver. C1 was fully engaged in the communication and tried to maintain gaze even though C1 and the caregiver were sitting side by side. Contrary to S1 and R1 who only responded to the caregiver, C1 also actively asked the caregiver questions. Residents tended to give some thought to what the caregiver said, and even repeated the word of their interest.

### 6.1.4 Discussion

Part 1 shows an ideal scenario in which the robot can have meaningful interaction with residents. In other words, HHI suggests the desired outcomes that the robot can bring. We observed that even for a caregiver, it is not feasible to get a response from each resident due to their personalities and cognitive capabilities. However, we did learn how to design and improve the social interaction between *Aether* and residents from observing this HHI process. For example, the caregiver knew each resident's interest and lifestyle well so that she could ask questions that could trigger residents' attention, curiosity, and reflection. The caregiver used a lot of questions to hold the conversations. We can have *Aether* imitate this approach to social interaction.

## 6.2 Part 2 - GUI Demo on a Display (HCI)

While we want to base our HRI design on HHI, we need to know how HCI and HRI differ. Residents were accustomed to sitting in front of an iPad or computer screen at the group home. We are hoping to understand what makes the social interaction with robots unique. Solving this question will also help us meet the next research question in subsequent User Study 3.

We then conducted a series of Wizard of Oz (Portet et al., 2013) experiments. The WOZ technique is a prevailing solution to iterative prototyping (McTear, Callejas, & Griol, 2016). It is commonly used to test the prototype to discover the drawbacks of the current design by observing users' reactions and acceptance.

For the GUI demo, the senior caregiver (who was also the manager) suggested that we keep the "task-reward" scenario and that we use the symbols from Board Maker. She told us that most group homes for people with DD use board maker, a special education program developed by *Mayer-Johnson*. Board maker contains thousands of symbols supported in 44 languages. Caregivers use it for scheduling and communication and print some of the most common symbols to train residents daily, reinforcing residents' memory and cognitive functions.

### 6.2.1 Procedure

A 12-in Microsoft Surface 5 computer (11.50 x 7.9 inches) was placed on a table in front of the participant with the keyboard detached. A three-minute-long GUI demo started to play automatically shortly after the researcher initiated the demonstration. The GUI prototype contained three tasks (getting dressed, having lunch, and doing chores) and a reward (tea time with the caregiver) (Figure 6.2 (a)) as an imitation of the "To-do Board" being used at DDA group homes. The demo also had verbal narrative to describe each of the tasks in a few words (e.g. "First, get dressed.") and that they would get a reward after completing the tasks in succession. The demo started with a neutral face (without emotion) picture zooming in slowly. After 5 seconds, this picture disappeared and a smiley face picture began to show up. Each transition took 10 seconds in total. Later, the face picture, centred on the screen, became 50% smaller and moved to a separate area on the right side. This area was 1/3 of the total screen size and was designed as a dedicated region for the anthropomorphic representation of the robot. The left side of the screen showed all tasks vertically aligned on the same page first, and after 5 seconds, it played each individual task and the reward in order with a 10-second duration. The tasks and reward information was extracted from residents' real lives in the group home, and they were not required to physically carry out the tasks in this user study. Instead, residents were simply expected to pay attention to the demo. A caregiver sat next to the participant but remained silent the entire time to help reduce the participant's anxiety. The caregiver may give a gestural prompt to help the resident focus on the display if the resident gets distracted.

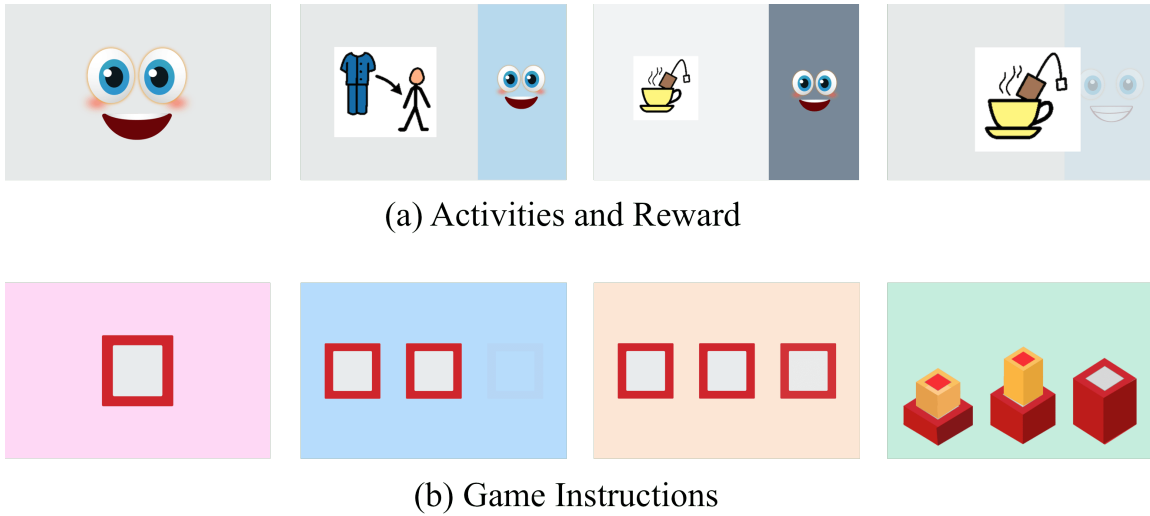

(a) Activities and Reward

(b) Game Instructions

Figure 6.2: Screenshots from User Study 2

### 6.2.2 Results

In this Wizard of Woz test without a robot, R1 only looked at the screen very occasionally for a short period of time, at the beginning of this test. She was looking at other things in the room most of the time, and her focus moved to the screen for a second when the caregiver pointed to the screen silently. R1 was not interested in listening to the verbal information from the robot. S1 was staring at the screen most of the time and sometimes got distracted by other people passing by her. There was no verbal or gestural response from her. C1 was engaged in the demo, and nodded many times to acknowledge the understanding of information from the demo. She talked to herself to organize the information. C1 had the most response among all three resident participants.

### 6.2.3 Discussion

Although the primary focus of this study was the visual aspects of robot design, we also noticed the significance of verbal communication, which led to positive feedback most of the time. Users were notably more responsive when a word or phrase corresponding to their interest was mentioned. This finding suggests that verbal and graphical stimuli can be combined together to maximize engagement.

Following Part 1 (HHI), we found that as we were moving to a technological infrastructure, the social interaction was fundamentally affected. The changes of virtual and physical embodiments had evident impact on residents' perception and acknowledgement of the virtual agent that attempts to convey information and elicit users' reaction. We began to reflect on this change and found an essential factor for social interaction of HRI: trust. In this small community, communication and trust are of critical importance. Introducing technology to this community has altered the conventional way that residents experience and engage with the social community. This is partly the reason that residents had more gaze and less disengagement compared to Part 1 (HHI). We further explored how technology impacts on residents' affective communication in Part 3 (HRI).

## 6.3 Part 3 - GUI Demo on the Robot (HRI)

### 6.3.1 Procedure

The computer was mounted on the robot, playing the same GUI demo as before. The researcher hid behind the participant while controlling the movement of the robot. One primary goal of this part was to explore the effect of the robot's limited physical abilities by observing the user's emotional changes, especially when the robot was approaching and leaving the user.

To further assess the impacts of design elements on HRI, we developed an additional GUI prototype to provide board game instructions specifically to C1, the resident who had the highest verbal skills and cognitive abilities amongst all at this group home. We controlled the robot to approach the resident and stay about 10 inches from her. A *Melissa & Doug* shape sequence sorting set was jumbled and prepared on the table in front of the user. This simple educational toy was initially designed for children of ages 3- to 7-years old to improve their reasoning and math skills. As this GUI demo showed step-by-step instructions (Figure 6.2 (b)) to the game, it would be possible to evaluate the effectiveness of HRI based on objective metrics such as completion results. We can also investigate which part of the design has flaws according to the "roadblocks" that the user encounters.

Two cameras, including a handheld camera and a webcam, were used to capture the entire study process alongside participants' facial emotions and gestural responses.

### 6.3.2 Results

R1 was extremely excited to see the robot at the beginning, based on her gaze, grin and body language. However, this attraction only lasted 15 seconds and she got distracted. After the robot moved closer, she got attracted again, and actively inspected the robot for 5 seconds, and then turned to something else. There was no verbal response from her. S1's response was basically the same as that in Part 2. She was visually tracking the robot when it moved. There was no response, either. She seemed to be anxious and exercised caution when *Aether* was moving closer towards her. C1 spent 95% of the time focusing on the robot, and thought it was amusing when *Aether* moved close to her and bumped into her wheelchair.

For the additional game test designed exclusively for her, she was very occupied with the game and could not have much time to look at the robot. She answered to the robot's questions and then looked at the display briefly. After the game, she looked at the robot and pointed to the display during the replay. She was talking a lot as if she was discussing the game with the observers around her. Although C1 failed to complete the game, she appeared to really enjoy communicating with *Aether*.

### 6.3.3 Discussion

Clearly, the robot attracted more attention that the computer and the caregiver. Although this was not the first time that residents interacted with *Aether*, they were still very excited when it was introduced to the community. A fundamental difference between a computer and a robot is the physical embodiment, which plays a critical role in social interaction. Because of the robot's ability to move and its physical form, residents paid more attention and had more interest.

The robot is not simply a different type of technology – it is a new form. This form refreshes residents' perception of the social agent from HCI, and increases their acceptance and trust. HRI acts as a bridge between HHI and HCI, and reduces the drawbacks of them. For example, in Part 3 (HRI) residents had substantially more gaze and interest compared to Part 1 and Part 2, but in terms of getting responses, Part 1 (HHI) still has the definite advantage.

The results of getting verbal responses are polarized between HHI, HCI, and HRI. Even for R1 who tended to be distracted easily, she responded to most of the caregiver's questions, followed the derivatives, and repeated the words of her interest many times. C1, often being quiet and introvert, actively asked questions – these two reactions did not happen in Part 2 (HCI) or Part 3 (HRI) of this study. Besides, residents tended to give some thought to what the caregiver said, and even repeated the word of their interest. This repetition did not happen in the other two parts either.

After coding the data from all 3 parts together, we have a deeper comparison of HHI, HCI, and HRI in Section 6.7.1.

## 6.4 Questionnaire

As it was difficult to create a single benchmark to numerically assess the design factors that we are investigating, we asked two caregivers to fill out a questionnaire consisting of eight concise questions after the user study. We also conducted an open-ended interview with these two caregivers to get more detailed and specific perspectives and suggestions for the user study. Both caregivers were female full-time employees with years of experience of caring for residents with developmental disabilities, and they observed the entire user study.

The eight questions were as follows: On a scale of 1 - 10 with 10 being the most effective, positive, or of the highest degree, how would you –

- Q1: describe the overall experience/process of delivering instructions?
- Q2: describe the verbal instructions?
- Q3: describe the graphic illustrations (shapes, colours, etc)?
- Q4: describe the animations and transitions between scenes?
- Q5: evaluate users' interest and engagement in this demo?
- Q6: evaluate the social attributes of this robot (during this demonstration)?
- Q7: assess the friendliness of this robot?
- Q8: assess the reality of this robot? (Does the appearance look real? Do these shapes and graphics correspond the reality?)

### 6.4.1 Results

Both of the caregivers gave similar feedback (see Fig. 6.3) on most facets of the interaction design. It should be noted that there was a substantial divergence of opinion about *Verbal Instructions* and *Overall Experience* between these two caregivers. In the subsequent interview, one of them pointed out that the overall design neglected residents' incapabilities to reference. Specifically, an object or activity need to be available for residents right when the robot mentions it. Otherwise, it would pose confusion for them due to their limited perception and understanding. Overall, two caregivers' impressions of other aspects align with each other.

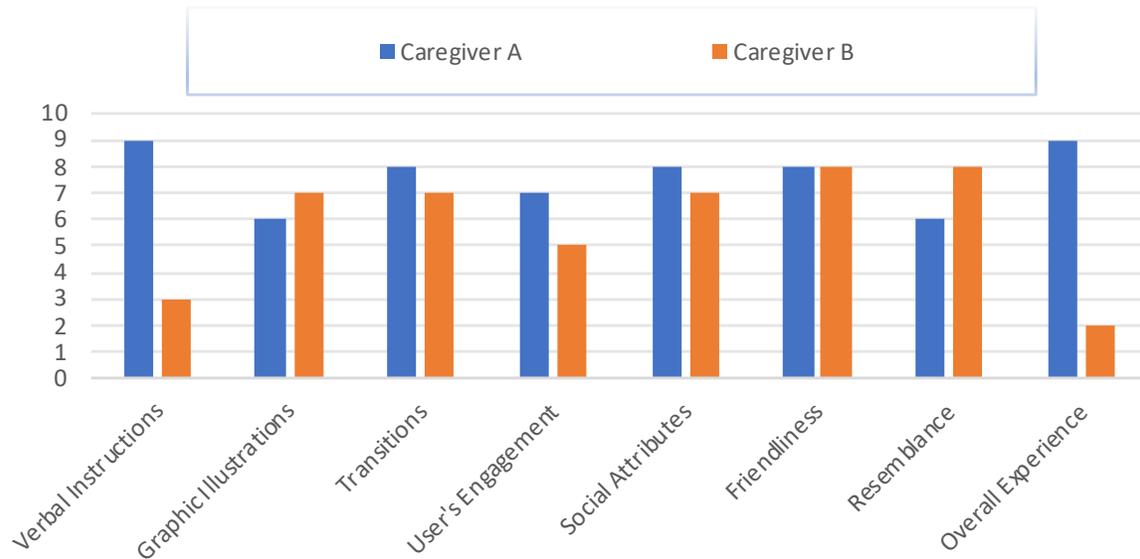

Figure 6.3: Interview questionnaire feedback from caregivers

From the interviews with two caregivers, we collected some valuable information for further analysis. Their communication method was inspiring for our design. We will elaborate the findings in the *Results* section.

## 6.5 Data Analysis

All video and audio recordings had been compiled into a single video file, showing each stage of the study from two different perspectives on one screen. This compiled data file was then imported into NVivo (Figure 6.4) to process and code textual and multimedia data. Each key frame of the video had been noted with a concise description. Then, along with residents' response being classified, each of these highlighted frames was categorized based on the form of communication, the type of information, and the kind of interaction. We transcribed the after-study interviews with caregivers and included them as the second source of data. We used open, axial coding to examine the data reflectively, identify relevant themes and topics, and then refine and relate them. This inductive and deductive method of data analysis helped us build patterns and find support from the data. After cod-ing all the data we collected, we created a comprehensive node matrix to cross-tabulate the observations of this study at each stage.

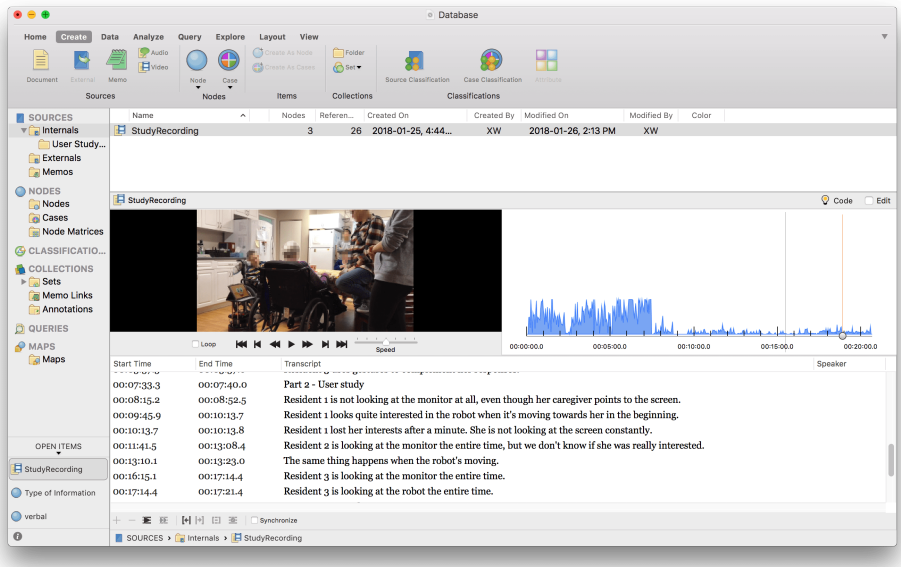

Figure 6.4: Coding recorded videos of user study 2: the *transcript* section refers to each key frame we coded

By highlighting texts and using colour stripes (Figure 6.5), we coded and reviewed our data in NVivo. This straightforward representation of coding made it easy to analyze and compare the nodes.

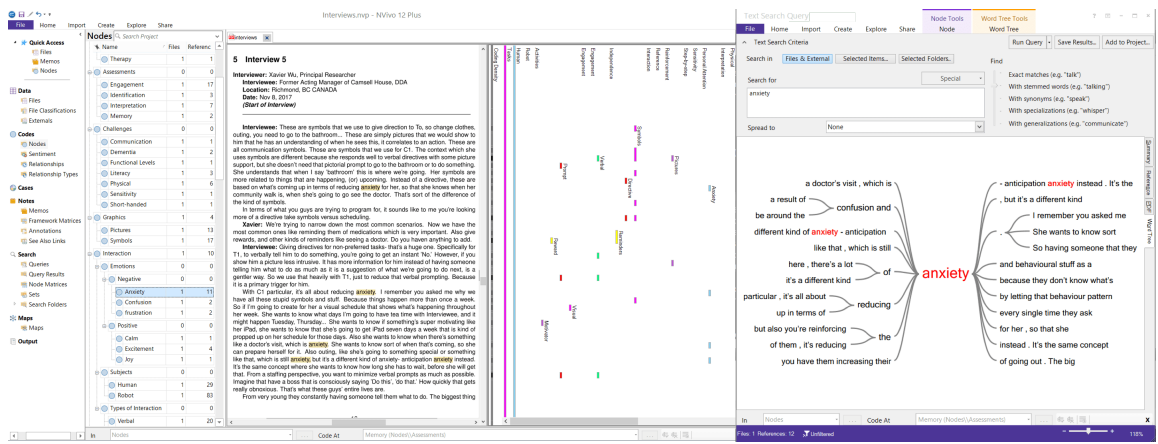

Figure 6.5: NVivo interface for highlighting and coding in coloured stripes and for creating a word tree

### 6.5.1 Evaluation / Dimensions of HRI

The following themes and topics, derived from previous interviews with caregivers, are taken into our consideration when evaluating and understanding the data from Part 3.

**Type of Information**

- Introduction
- Question
- Instruction
- Reminder

There are four essential types of information provided by the robot, in either a graphical or verbal way, or both at the same time. Transmitting information, as one of the fundamental functions of service robots, needs to be conducted efficiently and strategically. The type of information presented from the robot, along with the order and quota of each category, is a delicate task that designers have to contemplate.

The most common type of information is *introduction* which usually first appears as an opening - i.e. a self-introduction such as "Hello, my name is *Aether*, your personal assistance robot." It is noteworthy that there tends to be a misconception that *introduction* stands for the act of having people learn about something for the first time. In fact, the classification *introduction* here is a broader and more general definition referring to any kind of statement, description, or explanation. It is to emphasize the process that robots present or deliver information. For instance, giving feedback to information inquiries such as personal schedule or weather forecast falls into this category.

Questions can be used to provide users with options and receive inputs, thus robots can pose questions to users to enrich the content of communication. Depending on users' verbal capability and level of cognitive impairment, the quantity and type of questions can vary. There should always be a fallback for the expected responses in the event that a user cannot give any valid feedback or the feedback cannot be interpreted by the robot.

As one of the goals of this project is to support caregivers' routine duties which are oftentimes "mechanically" repeating instructions to residents, we have to assess the validity of different approaches to conveying instructional information. Apart from this, giving *instructions* is an ideal scenario for us designers to test the effects of our design elements, because it includes all essential components of bilateral communication: messages, con-text, and feedback. Usually, researches can observe users' reactions and feedback immediately while testing prototypes.

Similar to instructions, reminders are another routine procedure for caregivers, which are simple yet significant. On average, the caregivers at the DDA group home need to remind each resident to take medications 8 times a day, which can be a difficult task when the group home gets understaffed.

**Engagement**

- Interest
- Gaze
- Reflection
- Verbal Response
- Gestural Response
- Repeat

The level of engagement can be judged by the following criteria: interest, gaze, reflection, response (verbal, gestural), and repeat. Typically we can determine the overall interest of users based on observation and caregivers' judgement. For instance, when a user looks only at other objects or persons rather than focusing on the robot, we can determine that this user's interest is none at that moment.

Usually, gaze is a manifestation of a user's engagement or thinking. Whether he/she is gazing at the screen or the robot body, it is usually a good sign of a user's engagement. However, although this indication is easy to be observed, its validity needs to re-examined to ensure the positive relationship between gaze and engagement. For some users with severe cognitive impairment, they might keep looking at an object in front of them for a long time. Hence, we need to compare users' behaviours and their personal record, along with caregivers' analysis, in order to achieve accurate results.

Verbal and gestural responses mostly happen as reactions to the robot's questions. Sometimes for residents with noticeably low verbal abilities, repeating after the robot when it mentions some keyword that triggers their interest would be evidence that the robot appeals to them. A case in point is that one of our users repeated the word "TV" multiple times after the robot briefly mentions it in a question.

**Disengagement**

- Disregard
- Distraction

We wanted to investigate not only the factors resulting in users' engagement, but also those causing their disengagement, dislike, hostility, and disgust. In contrast to the manifestations of engagement, the symptoms of disengagement are much more evident and sparse. Generally, a user is not engaged in the interaction with the robot if he/she is ignoring the robot or gets distracted by trivial things.

**Type of Communication**

- Verbal
- Graphical
- Physical

At this stage, the robot is only equipped with a few main components: the robot base, a display with speakers, and other sensors. This restrains the types of communication: verbal, graphical, and physical. The robot is capable of playing verbal prompts and other audios, and can move around in the room. The physical movement is one of the key factors differentiating robots from other devices like computers and iPads. We wanted to find out if users were attracted by the robot's movement to some extent.

## 6.6 Results

### 6.6.1 Overall Coding Results

This section presents the coding results of all parts of User Study 2. Table 6.1 summarizes the results of multiple matrix coding queries. This table has been categorized based on four dimensions: communication approach, type of information, form of interaction, and type of response. We wanted to have a deeper understanding of the data, and thus we coded the data from all parts together without differentiating HHI, HCI, or HRI, except for the "form of interaction" dimension shown in Table 6.1.

Table 6.1: Matrix coding results of the user study

| Interaction Attribute | Response | Engagement | | | | | | Disengagement | |
|---|---|---|---|---|---|---|---|---|---|
| | | Gaze | Gestural Response | Verbal Response | Thought | Repeat | Interest | Disregard | Distraction |
| Communication | Graphical | 7 | 3 | 1 | 1 | 0 | 5 | 1 | 2 |
| | Physical | 2 | 0 | 0 | 0 | 0 | 2 | 1 | 1 |
| | Verbal | 5 | 5 | 11 | 4 | 3 | 3 | 3 | 4 |
| Information | Instruction | 4 | 2 | 1 | 0 | 0 | 3 | 0 | 2 |
| | Introduction | 2 | 2 | 1 | 0 | 3 | 0 | 1 | 1 |
| | Question | 0 | 2 | 10 | 5 | 0 | 0 | 2 | 2 |
| Interaction | Human-Human Interaction | 1 | 2 | 10 | 3 | 3 | 1 | 2 | 3 |
| | Human-Computer Interaction | 2 | 1 | 0 | 0 | 0 | 1 | 0 | 1 |
| | Human-Robot Interaction | 4 | 2 | 1 | 2 | 0 | 3 | 1 | 1 |

Table 6.2 lists all nodes for all parts of this user study we created and coded, along with their quantities and levels. We categorized all nodes into four different levels or themes. This table provides a detailed summary of the core concepts our study, and also suggests their connections. We had eight main themes: *Approaches*, *Assessments*, *Challenges*, *User's Need*, *Objectives of the Project*, *Interaction*, *Principles to Follow/Test*, and *Defined Scenarios*. Those themes established the flow of our entire study together.

Figure 6.6 shows the visualization result of an example word. "visual", in the format of the word tree. The word tree layout enables us to see the context of a word or phrase when we are focusing on a single node or theme. This kind of visualization highly summarizes the occurrence of a word or phrase, and more importantly – the connection of it to other nodes. For example, I started to notice "visual hierarchy" and "visual field" when I examined "visual", and I found that it is users' "visual impairment" that results in their ability to only perceive lower-level elements in the "visual hierarchy", and thus we need to make sure whatever we want to show them is in their "visual field." By using word tree visualization, we also managed to analyze other primary nodes as well.

Table 6.2: Number of codes from interviews into identified themes

| Name | References | | Name | References |
|---|---|---|---|---|
| **Approaches** | 0 | | **Interaction** | 10 |
| Directive | 17 | | Emotions | 0 |
| Motivator | 35 | | ‣ *Negative* | 0 |
| Penalty | 2 | | • Anxiety | 11 |
| Prompt | 26 | | • Confusion | 2 |
| Reward | 38 | | • frustration | 2 |
| Therapy | 1 | | ‣ *Positive* | 0 |
| **Assessments** | 0 | | • Calm | 1 |
| Engagement | 17 | | • Excitement | 4 |
| Identification | 3 | | • Joy | 1 |
| Interpretation | 7 | | Subjects | 0 |
| Memory | 2 | | ‣ *Human* | 29 |
| **Challenges** | 0 | | ‣ *Robot* | 83 |
| Communication | 1 | | Types of Interaction | 0 |
| Dementia | 2 | | ‣ *Verbal* | 20 |
| Functional Levels | 1 | | ‣ *Visual* | 21 |
| Literacy | 3 | | **Principles** | 0 |
| Physical | 6 | | Chaining | 1 |
| Sensitivity | 1 | | Hierarchy | 6 |
| Short-handed | 1 | | High Sensory | 6 |
| **Need** | 0 | | Processing Time | 4 |
| Activities | 45 | | Reference | 11 |
| Company | 2 | | Reinforcement | 10 |
| Personal Attention | 8 | | Step-by-step | 7 |
| **Objectives** | 7 | | **Scenarios** | 1 |
| Engagement | 17 | | Critical Scenarios | 0 |
| Independence | 10 | | ‣ *Reminders* | 16 |
| Task Completion | 0 | | ‣ *Scheduling* | 1 |
| | | | ‣ *Tasks* | 29 |

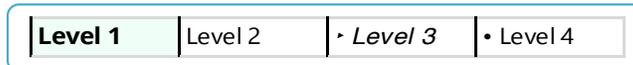

| **Level 1** | Level 2 | ‣ *Level 3* | • Level 4 |

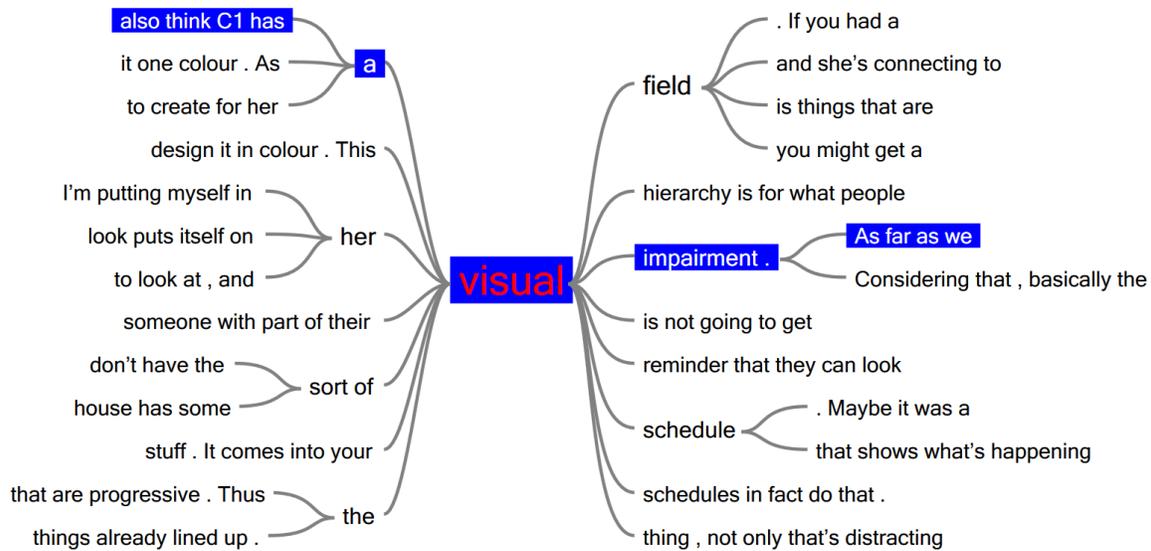

Figure 6.6: Word tree visualization for the word "visual"

In addition to word trees, I also conducted a cluster analysis in NVivo (Figure 6.7). After listing all the nodes, I found it difficult to start grouping and categorizing them, because of the number of the nodes and the multiple dimensions of their structure. Cluster analysis provides an efficient way to group nodes or sources, and to explore their relationships and differences based on nodes' similarities and patterns. For example, we can see that "confusion" and "dementia" are connected tight from Figure 6.7, and so are "reinforcement" and "reminders". When I go back to the data source and find the precise connection, I can find that people with "dementia" usually have low cognition and thus frequent "confusion", and "reminders" from the robot need "reinforcement" (i.e. repeat more frequently) when they get delivered to the resident. However, cluster analysis has its weakness: there is not a context. Thus, I combined all these methods of analysis together to get a comprehensive understanding of our codes.

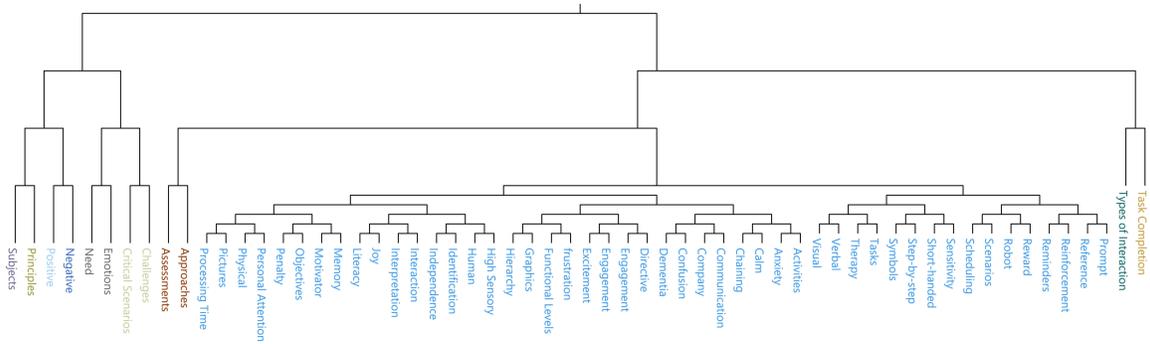

Figure 6.7: Items clustered by coding similarity

### 6.6.2 Challenges for Residents

The existence of the robot had also posed a challenge of dual reference to the residents. Dual reference refers to the capability of processing two information inputs simultaneously. Although ordinary users would usually not have issues with handling multiple information streams at the same time, it was the opposite way round to our users. This dual-reference challenge indicates the significance of context in HRI. During the user study, we found it very difficult to have residents understand and interpret the visual information when the activity or object mentioned was not present next to them. We learned from the caregiver that communication based on an existing context would help residents build connections between the perceived information and reality.

> *...that dual-reference thing is going to constantly work against you. So if you're doing a scheduler and literally the only thing it's giving you is, say, "In 15 minutes, you have to get ready to go for a walk." That's all they (residents) are focusing on at that time. They're going to get information from a robot and that situation. Anytime you're going to try to use the robot while they're trying to engage in an activity, you're going to face the challenge of dual reference.* - P1, Caregiver, Female

The finding on the dual-reference challenge also underlines the difference in user needs between different user groups. Ostensibly trivial phenomena like this could make a huge difference in HRI. Therefore, the robot's functions should be more intelligently integrated into the usage scenario in the future. In fact, *Aether* was designed to able to store the location information of items and this would make it possible to augment the virtual human-robot communication with objects from reality.

> *The pictures you've chosen are fairly concrete. If we're looking at, for example, Aether showing pictures of things like a washing machine? There's no context for that. So that would be something that we'd have to teach. For example, today I showed her a picture of a forest. She was unable to tell me what it was. We are looking at concrete things. It's a bunch of sticks in the ground right now. I said,"They have trunks." She was telling me that must be an elephant then. So we're looking at her ability to visually interpret, plus that sort of like that rational thought of what could this possibly be.* - P1, Caregiver, Female

Nevertheless, there is a "shortcut" to solve this issue. We tried to improve and alter our graphic design to make users understand the GUI, but soon we realized that we could also improve UX through better use of their existing knowledge. Instead of creating new graphics, we could use pictures and drawings that residents are already familiar with. Therefore, for the most common and basic symbols, residents will struggle much less while trying to comprehend the visual information. We will try to use these symbols in *Aether*'s GUI for future user studies.

As mentioned above, it is critical to allow residents to link what they see and what they know. This connection not only reduces negative emotions (frustration, anxiety, confusion), but also enhances the relationship between users and the robot. Using the symbols that residents are already familiar with is not the only option. Our design should base on residents' real-life experience so that they can quickly find a context. For example, we used a "sleeping face" to indicate the robot is not available or offline. This makes sense to the general population but is not the best design choice for residents.

> *This is a gesture that we use with her when she's interrupting which is sort of an ongoing challenge with her. So she understands when she sees a hand come up, that she has to stop and wait. And this is sort of a universal 'Don't touch' symbol that we use for anything from 'danger' to 'it's not yours' that this symbol itself they know that it's a NO – we can't use it. So any of these would be probably a better option in terms of showing that the robot is unavailable... But instead of having this sleep mode, you could have the face with a hand up, or the face with a break symbol next to it. So they understand 'OK we still have the robot but right now this is like the status.' That's not saying that they can't learn this, but in terms of having something that right now they understand, this would be the direction I would point to.* - P1, Caregiver, Female

We also investigated what else constrained users' engagement in HRI. Besides the visual design aspects, the physical representation of the robot also played a key role in HRI. I observed that R1 was always looking around and got distracted by her surroundings all the time, no matter whether the robot was present or not. The reason is that R1 can't reference, so she does not have the ability to concentrate on something, look directly at it and engage in it.

> *Even me trying to get R1's attention, you see in the video that I'm shifting my body around, I'm putting myself in her visual field. If you had a robot that could put itself on her visual field, you might get a reaction from that way. But with it as low it is as it is on the floor- it's not something that she's actively going to look at, and her visual field is things that are moving up here. I am not even sure how well she's able to see in terms of down there. We cannot really assess her vision. - P1, Caregiver, Female*

To be more specific on the robot's physical limitation, it was just too short to be seen by some residents. Individuals with DD can't split reference. For the general population, people can multi-task and move their focus among different things. This is basically impossible for people with DD.

> *For C1 the robot talks to her from the ground, and the activity beyond the floor... Her scope of things that exist are things that are now in what she can see and touch. Her feel of the things that exist at that moment is that puzzle. Robot? (It) doesn't exist because it's not in her field of vision. You also think C1 has a visual impairment. As far as we know, (Interviewee points to an object), this, there she can see. For her to be able to look at this, get the information in reference to the robot – she can't. - P1, Caregiver, Female*

> *I honestly think the robot is just too short. It's simply out of their field of vision. When we do reference training, as I said, with R1, you need to come into their field of vision, and for C1 that robot needs to be three feet taller so that's a big barrier that you already have there. You can see how quickly she saw there; it was hard for her reference both of them, then she's like "I'm just gonna ignore it." - P2, Caregiver, Female*

Thus, in the future, *Aether* needs to be built taller. This had been an engineering challenge due to the size of the robot base. Having it overheight would destabilize the body.

A possible solution is to only increase the height of the display, but this requires more research from our engineering team.

### 6.6.3 Findings on Visual Factors Affecting Users' Perceptions

Through this user study, we had the basic visual factors apply not only to web and app design, but also to the development of our GUI. Three of these factors are ubiquitous in our life: symbol/icon, layout, and colour. We can see the effects of them everywhere: currencies, websites, product packages, etc. The use of animating elements has been proven to be able to reduce end-user disruption and cognitive perturbation, and enhance understandability and trust in interaction (Dessart et al., 2011).

- Symbol

    One significant goal of our design is to reduce users' anxiety. This is also an essential prerequisite for keeping users engage in the interaction with the robot. Graphic symbol as one of the key factors to the visual design substantially affects users' perception of information that the robot is delivering. As textual information will not be used in our GUI design due to the verbal challenges of our user group, symbols and icons indicating tasks and objects are undoubtedly the primary approach to visually transmitting information.

    Nevertheless, symbols may not always be a positive factor. Some users are fully capable of understanding verbal directives, and thus reading graphics in addition to verbal prompts is unnecessary. They might get more confused or anxious because of this extra means of instruction. Also, although most of the users can understand the symbols with which they have been trained previously, they may dislike these icons. The direct outcome of using graphics to them will be that they would not even look at the graphics. Therefore, it is imperative to classify users into two basic groups based on their perceptive capabilities and interest in graphics before applying symbols.

- Layout

    We held a conjecture that the dominant area on the display should be reserved for symbols and icons instead of the robot's virtual face, and this guess was confirmed by a caregiver. The residents tend to focus on things that are simple and large, and like most of the people, they read from top to bottom, left to right. Hence, Layout D (Figure 6.8), as an example, is the most appropriate layout for a scenario requiring two zones or sections.

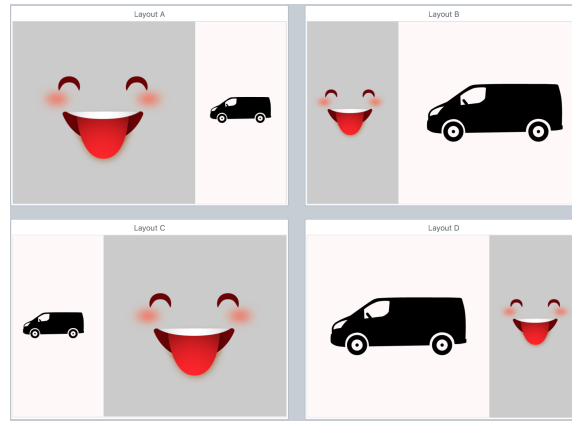

Figure 6.8: Four Basic Layouts for a Simple "Symbol + Face" Scenario

- Colour

It has been affirmed that most of the residents have an awfully hard time reading white symbols on a black background. Previous research from psychiatrists and visual analytical scholars have also found that different colour palettes produce effects on users' perceptions and trigger emotions differently (Song & Yamada, 2017).

> *When your screen transitions, the back colour changed. Keep it white or keep it one colour. As a visual thing, not only that's distracting but also it affects their ability to focus on it. Because they're already struggling to focus on something in the first place, once you start moving those things, they are really subtle to us. People with developmental disabilities tend to have a lot of sensory issues* - P1, Caregiver, Female

> *Your actual animations were good. Just the background colour– keep it one colour. The one screen where you go to the red cube on the orange background, you need to make that orange background a very very pale orange or white. Just because that's not high contrast.* - P1, Caregiver, Female

- Visual Hierarchy

This study confirmed our design conjecture about visual hierarchy (Figure 4.6). Based on interviews from caregivers, we learned that although residents comprehend visual depictions in the same order as most people, residents can be noticeably more sensitive to specific types such as real objects which are easy to reference. Regarding the UX design for residents, large, contrasty, and clear photos are necessary, but they do not guarantee residents' understanding as not all of them have good vision. The finding of this visual hierarchy inspired us while we are selecting symbols and designing visual illustrations.

> *This visual hierarchy is for what people are able to see and understand. This is what we use, especially for assessing really young kids with developmental delays and stuff. Even pictures on a page where we're seeing it as a coffee cup and it's really obviously a coffee cup for us, it's a completely abstract concept to them; it just looks like lines on the page.* - P1, Caregiver, Female

### 6.6.4 Interaction Design

When the social robot tries to construct a context for users, in addition to visual aspects, tasks presented by the robot need to be divided into baby steps, which means that each step should only refer to a single action. In the context of GUI design, there should be only primary action on each screen (Porter, 2017). The real challenge of applying this principle into our design is the short-time memory of our target users, who will get lost immediately if there is not a strong coherence and continuity of graphic information. Hence, we need to leave enough time for users to process the information between each little steps.

The caregiver pointed out a design flaw in our user study: the interval time between scenes was set to a fixed amount that could not be paused once the demo started playing. This feature limited users' engagement and reduces the chance of completing the tasks. This constraint will be fixed for the next user study.

> *The second problem that you were having is processing time. For someone with dementia and developmental disability, the minimal processing time they need to get information is 50 seconds. So if you're going to give instruction, you have to wait at least a minute, for that instruction to be followed through, and because you have the dementia side as well. Their ability to retain that instruction is very poor. So when you're going to give an instruction, you're going to give the instruction using single words, very simple instructions, give wait time up to one minute- I would say minimal one minute, with that picture of the square in front, and then you need to repeat that instruction. So it was simply moving too fast. You basically don't want to move ahead until the next step until she's accomplished the first step... So it was simply moving too fast. You* *basically don't want to move ahead until the next step until she's accomplished the first step.* - P1, Caregiver, Female

In this study, we observed that one of the residents did not have much interaction with the robot. The caregiver explained why:

*How C1 is responding to the robot is how she responds to everything. I think that her response to it is the same as to a less familiar person in the room. I don't think the response would have been any different depending on whatever you put in front of her.* - P1, Caregiver, Female

## 6.7 Discussion

### 6.7.1 Similarities and Differences between HHI, HCI, and HRI

From three individual tests, we noticed that users' personalities, characteristics and skills played critical roles in social interaction, regardless of the type of communication, information, or interaction. This individual difference had a clear impact on the evaluation of our social interaction study. For example, for R1, she tends to get distracted and anxious very easily, even when she is with a caregiver. It is extremely difficult to keep her focus to the robot for a long time. However, we can work on attracting her attention, and make this happen more frequently. From this study, the longest time that we could keep her focus on the robot was 15 seconds, and it was a satisfactory result for people like R1.

We also found that certain types of information were more likely to engage residents. Specifically, amongst all types of information, *questions* seemed to be the most effective in terms of getting users' response. We noted an increasing chance of getting a response when the question was getting easier. For example, a user would respond to the questioner eight out of ten times when the question can be answered simply by a "Yes" or "No." By contrast, when the robot was giving an *instruction*, usually users would ignore this information. However, it is also worth pointing that instructions triggered more responses when they were given by a caregiver, compared to a computer or a robot. The least effective type of information was *introduction*. The only positive responses we got from giving *introduction* was repeat (3*), gestural response (2), gaze (2), and verbal response (1). The frequencies of these responses were much lower than the responses from other types of information.

*It indicates the number of responses as found in Table 6.1

There were also a lot of differences between HCI, HCI, and HRI. This frequency of gaze in HRI (4) was much higher than HHI (1) and HCI (2). Also, residents had slightly more gestural responses (2) compared to HCI (1). Among all residents, only C1 verbally responded to the robot.

However, oftentimes they got distracted by the objects that the robot was referring to. An example is that Resident 3 was occupied by the board game on the table while the robot was narrating the steps. She failed to grasp the visual information displayed on the screen. This occurrence indicates that their acceptance of step-by-step instructions is rather limited. Furthermore, users always processed the verbal information first, before perceiving others. This stands out as a significant pattern for designers to keep in mind. Different kinds of information may not be parallel all the time. Depending on user's perceptive abilities, there could be a layered and ordered relation among them.

Nevertheless, this is by no means suggesting that the introduction of technology makes social interaction less feasible or effective than HHI. From the *Interaction* section in Table 6.1, we noted that the involvement of the computer actually reduces the recurrence of disengagement. Based on our observation, residents tended to get distracted frequently by other residents or guests. They had got used to caregivers' company and thus they often ignored their caregivers. Ordinary objects, such as an iPad or a computer, appeared to be more appealing to them. This being said, a computer that embodies a virtual agent or a robot is certainly novel to residents. The fact that they had less engagement, fewer verbal responses, but more gaze and interest suggest the strengths and weaknesses of deploying the robot agent. Thus, the design of the robot can certainly benefit substantially from HHI. In addition, trust is lacking at this stage and residents had invisible psychological resistance to the virtual agent in some scenarios. This needs to be improved to increase engagement.

Through coding the video data, we had some interesting findings on nodes like "anxiety" and " confusion". For example, we did not observe any anxiety or confusion in the HHI part. We believe that was partly because that "human" was someone that residents were very familiar with. Provided it was a stranger talking to the residents face-to-face closely, we could probably get a different observation. In the HCI part, we did not see much anxiety, either. However, for S1, apparently she was having some confusion when the computer was playing the demo. To her, sitting in front of a screen was basically like operating a computer – there was a lot of information for her to process. For HRI, both S1 and R1 showed a bit of anxiety. It was interesting to see that S1 breathed a sigh of relief after the user study, when she left the room. Therefore, anxiety had different manifestation among HHI, HCI, and HRI.

### 6.7.2 Design Implications

Our results indicate that the overall design of this primitive robot system fulfilled many of the objectives that we proposed. We acknowledge that our design is not polished enough to handle certain scenarios at this stage.

Through testing our prototype, we learned that many commonly accepted design principles could not be applied to our users. Animating transitions is an effective approach to reducing cognitive perturbation when a user is trying to follow a change in an adapting GUI (Dessart et al., 2011). Transitions we implemented in our prototype included resizing, fading, movement, and etc. These transitions were designed to be smooth and gradual for users to cope with new contents, especially new types of contents. It is a widely accepted design principle that contrast colours of subdued tone contribute substantially to the readability of textual and graphic information, and the gradual change of colours can help users shift their focus from one element or scene to another. However, this technique does not apply to individuals with severe cognitive impairments, as they could be struggling a lot with any subtle change of element. Hence, users would need extra time to process the change of colour before noticing the change of a pictorial element.

The GUI design embodied strong and concise visual hierarchies to attract and keep users' gaze most of the time. When users lost their focus on the screen, it might indicate an incompatibility or discontinuity of information from the previous scene. From the user study, we conclude that the GUI on the robot plays an essential role in guiding the users and in reinforcing the engagement.

Lastly, I would like to discuss the importance of speech in HRI. Although we are not designing a speech interface, the robot's speech function needs to be simple and efficient due to the fact that our users are people with cognitive challenges. One of the most practical and efficient approaches for this speech problem is that we can design the robot to ask only general and simple questions without correct answers, and give users timely feedback as acknowledgement or response. In this way, users will not get discouraged.

Learning the language patterns of both the residents and their caregivers at DDA were very helpful for researchers to understand the lives and communication patterns of people with developmental disabilities. We learned that caregivers like offering a few options when they ask residents questions, instead of giving a broad question that forces residents to think and choose. This way, residents only need to say "Yes" or "No," or simply make a choice from the given options. We also found that people with developmental disabilities may change their minds fast. For example, maybe a resident just talked with you peacefully five minutes ago, but suddenly he would go to his room and lock the door, and yell at you, "Get out" or "Leave me alone!" Maybe he will be fine after another ten minutes. Therefore, a social robot needs to be designed to be capable of dealing with this kind of situations.

At the same time, the communication patterns that we learned from the HHI part also benefits the GUI design. For example, the caregiver usually gives no more than 3 options (mostly 1 or 2) to residents, and the robot's GUI should be designed to provide one ("Do you want this or not?") or two ("A or B?") pictures when a resident needs to make a choice. Another example is that caregivers use a lot of questions in the conversation. The GUI can be possibly designed to use a specific colour to indicate the robot is asking a question. This helps reinforce residents' perception of questions, and hence they will know that the robot is asking a question whenever the screen or the background colour becomes red, yellow, or another highly stimulative colour.

## Chapter 7

# Designing Social Interaction for Engagement: User Study 3

Following the previous two user studies emphasizing the visual factors of HRI, we were also interested in other aspects of the social dimension, especially the physical factors. We were curious about how space and the physical representation of the robot impacts users' experience. We learned that trust is a critical factor in HRI. However, the development of acceptance and trust is a long process that is beyond the scope of our study, so we aimed to explore the impetus and deterrents to engagement in HRI in the context of proxemics.

## 7.1 Part 1

Besides the commonly acknowledged use of robots for physical assistance and industrial productivity, in recent years there is also a great societal demand for sociable robots that have empathy. The projection of the user's traits and personality to the robot design process has been validated in the course of adjusting the robot's abilities and patterns (Tapus & Matarić, 2008).

Changes in normal vision due to aging (Table 7.1) have been found by the Dementia Services Development Centre (DSDC) (University of Stirling, 2012). Although these changes are generally affecting most people, they do appear to have a more marked impact on individuals with dementia.

Table 7.1: Changes in normal vision due to aging

| | |
|---|---|
| + | Sensitivity to glare |
| + | Need for additional light |
| - | Peripheral vision |
| - | Sensitivity to contrasts |
| - | Speed of adapting to change in light level |
| - | Visual acuity |
| - | Depth perception |

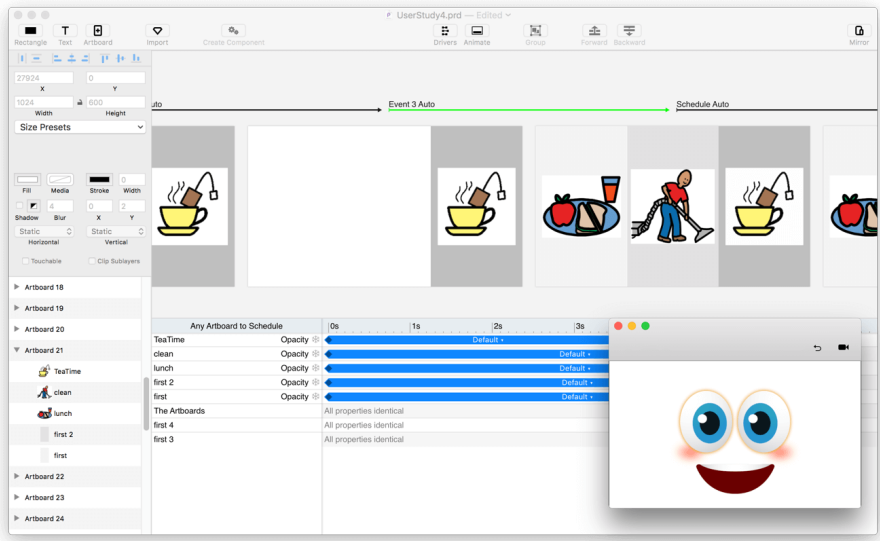

Figure 7.1: Prototyping process of user study 3

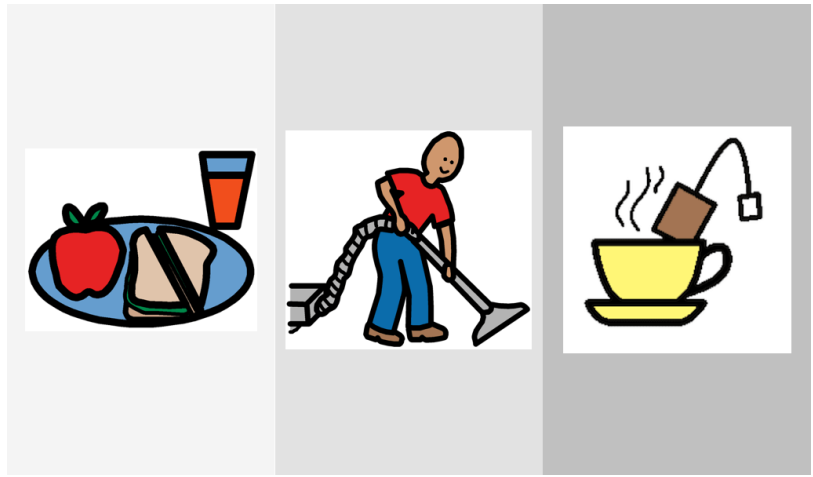

Figure 7.2: Design of scheduling scenario

The resident participants in this user study are the same as the previous study. Two caregivers participated in this study.

In User Study 2, we observed that C1 was amused to see Aether move close to her and bump into her wheelchair. We want to gain more insights into users' experience through modulating space, especially close distance. This is due to the fact that most of the social interactions that we designed for our user group happen in the personal or intimate zone, so these two zones are more critical for us to study at this stage.

### 7.1.1 Procedure

A 220-second-long demo was prepared on *Aether*. At 10.30 a.m., we invited each participant to come to the Snoezelen room, sitting in their wheelchairs or on the couch. The position of each participant remained still during the user study and we used two cameras to record and observed their head movement and shift of focus while the robot was showing the demo on the screen or moving in the room. Two caregivers whom the participants were familiar with stood behind the door of the Snoezelen room but still reachable and visible if the participant turned their heads. There is no other human intervention while the study went on.

The demo consisted of three pictures representing three activities on participants' calendars. Schedules at the local group home were pretty flexible and not always followed, depending on resident needs, staffing levels and other activities happening in and around the home that day. A rough outline of the daily schedule is shown in Table 7.2. We selected the most frequent daily activities and matched them with symbols that had been taught to residents on a daily basis.

Table 7.2: Daily schedule at the group home

| | |
|---|---|
| **8:30am** | Breakfast |
| **9am** | Exercise |
| **9:30am** | Visual calendar |
| **10am** | Outing in the community |
| **noon** | Home for lunch |
| **1pm** | Household duties |
| **1:30pm** | Watering plants |
| **2pm** | Puzzles |
| **2:30pm** | Tea break |

After we initiated the demo, we controlled the robot to slowly approach residents using an Xbox Adaptive Controller. The distance was approximately calculated based on the square pattern of the mat (18 in x 18 in for each small square) on which Aether was moving. We adjusted the rotation of Aether to make sure that it would be facing directly to participants. The demo was designed to start shortly after we finished these adjustments. After 200 seconds, the demo was over and *Aether* started to ask questions with a smiley face on the display: "Could you tell me what three events are on your calendar?" Because some residents had an extremely short-term memory, they might forget this question in a minute. Therefore, we asked one of the caregivers to ask them the same question again. We then recorded the results of this small memory-recall test.

It is worth pointing out that we wanted to do this informal memory test, as opposed to simply observe what happened, because this test could provide us with more firsthand information about users' experience. Contrary to other methods, this test does not require a formal set of metrics or need residents to externalize their experience. By checking whether users manage to complete the simple test we proposed, we could gain design insights and collect their reactions and feedback. In this social interaction study, the amount of information we could obtain directly from the users except for on-site observation, was extremely limited. Thus, any measure that we could take to enrich the primary source of data would benefit this exploratory study.

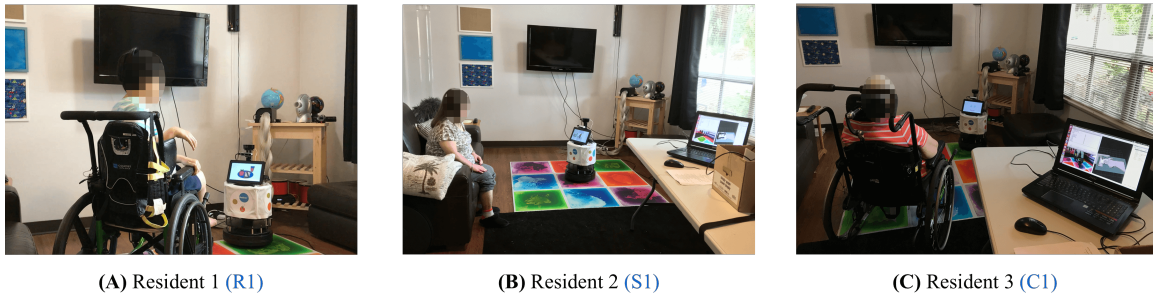

**(A)** Resident 1 *(R1)*        **(B)** Resident 2 *(S1)*        **(C)** Resident 3 *(C1)*

Figure 7.3: Setup for User Study 3 - Part 1

### 7.1.2 Data Collection and Analysis

I video recorded the study and audio recorded the interviews with caregivers. I highlighted the significant frames of the study video (e.g. when the resident seemed to get alerted) in NVivo and noted my observation. I then proceeded to use open coding for these two sources of data.

### 7.1.3 Results

None of the residents could memorize all three events in the correct sequence. We did not ask the first participant, R1, the question as she was not verbal and had to rely on memory aids. The second participant, S1, could remember the last activity but misremembered the detail ("Tea Time" instead of "Coffee Time"). For participant C1, she could recall the last two events out of the three, but not in the correct order.

In terms of users' responses, R1 only said "Yeah" when *Aether* stated the first activity–breakfast. She kept looking at the caregivers standing behind the door. She occasionally came back to Aether for several seconds and then moved her focus to her caregivers.

C1 replied to every question, such as "Fine, thanks" to *Aether*'s "How are you today?" She seemed to be a bit anxious and even impatient while waiting to hear the next question or instruction from *Aether* – she even started to tap her fingers on the lap. The time gap was designed to be long enough so that every resident could have time to process the information. However, while Aether was talking or asking, C1 got very focused. She had a big smile when being asked, "Guess what we are eating today?" C1 was trying to recall somethings when Aether asked "Could you clean your room later, please?" and later replied, "No, I have not cleaned it yet." While waiting to hear the last two questions from Aether, she became more nervous as if she was taking an exam. When Aether was checking if C1 could remember what three tasks that she had, she tried hard to think about them and replied "Coffee" ("tea" in fact). Unfortunately, Aether ended the conversation shortly, not leaving enough time for C1 to recall. When the researcher asked her again in person, she corrected the first answer to "having tea", and she could also remember "something in the room" vaguely.

S1 only responded to one question in the demo. She said, "I don't have a calendar." when Aether asked her whether she could remember the tasks on her calendar. And replied, "No, no tea." to *Aether*'s "Do you want a cup of tea?" She even asked the robot a question "Should I make the bed?" after it listed all three tasks to her. She looked very nervous before the demo, and was looking for her caregiver. Instead of answering the questions from Aether, she was thinking of something all the time. She was looking at the monitor on Aether half of the time. Right after the demo, her caregiver asked her if she could recall the tasks, but she could not remember any of them. While leaving the room, she breathed a sigh of relief.

Through the user studies, we found that the motion of the robot made its presence more noticeable– and sometimes more appealing, and that the change of social distance did trigger different responses. All three residents visually followed *Aether*'s movement, especially when they had sensed that *Aether* was coming towards them. Two of the participants, S1 and C1, were tracking the robot the entire time as long as it was in their visual field, whereas R1 only paid attention at times. As for the social zones, all three residents did not show any sign of resistance when *Aether* was at in the public and personal zone. When it moved to the intimate zone, R1 showed obvious resistance and disfavour, especially when it touched the wheelchair footrest.

When we designed this user study, we had an expectation that none of the residents could remember most of the details in the demo. Even for the general population, having people maintain their focus for minutes is not always an easy thing. It is common that they would miss something if they shift the focus for a few seconds. However, two of three participants could partially remember the information from the interaction.

> *We're dealing with people who have dementia. Their ability to retain and recall information is so impaired that we want them to be able to see the next 3 things on their schedule, and be able to go back and reference it as many times they want. So our goal would never be to for them to memorize the schedule* *but we would be teaching the skill of "If you don't know what's happening, here is how you get that information."* - P2, Caregiver, Female

Maintaining residents' mental functions is a big mission for facilities such as DDA. This required regular and frequency training in cognition and memory. At the group home, caregivers use images of various degrees of abstraction (Figure 7.4) to improve or maintain residents' comprehension and memory of repeated events or objects in their daily lives. The goal of this kind of training is to enable residents to make connections and be able to reference proactively.

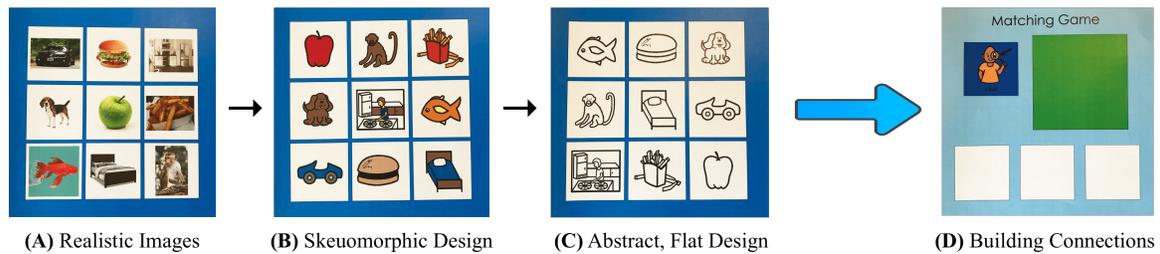

**(A)** Realistic Images    **(B)** Skeuomorphic Design    **(C)** Abstract, Flat Design    **(D)** Building Connections

Figure 7.4: User Study 3 Images

While exploring the social interaction, we found that the initiate zone is an interesting area to investigate. In contrast to our assumption, residents were visually tracking the robot when it moved to the back of them. Thus, we conclude that the presence of the robot can still be effective in the same radius. Although participants stared at the robot both at the personal zone and at the intimate area, there was a psychological difference. At the initiate zone, participants got alert and anxious, and tended to seek help or attention from their caregivers or people they trust. This kind of attention appeared to be a negative occurrence at this time. It will be a problem to tackle through regular training if we want to have the robot as a close companion. Perhaps this issue might also disappear when the robot becomes a part of the group home.

*Our goal is to reduce anxiety by allowing them to know what's currently happening and sometimes it also is coming up. For some of our clients like C1, they can see, but it's always going to be visual reference for her; she's not going to remember that. She can see what's happening that day. She can see 2 or 3 activities ahead but for a number of folks, it's just simply what's happening now and perhaps what's happening next. But we rely very heavily on visuals*

*and never want to remove those – often we're adding more.* - P1, Caregiver, Female

Certainly, there are many reasons resulting in residents' anxiety. In the previous user study, we found that residents did not have enough time to process the information from *Aether*. There should be more wait time after every bit of information. The robot should also be a tool that residents can use when they need to ask for information.

In this user study, we experienced the same problem as previously: the robot was too short to be noticed by some residents. We could not modify the physical dimension as the screen was mounted on the robot.

*One of the things that we do for clients that struggle with referencing or cannot reference is moving things up to their line of sight. Taking away all of those other things that would be barriers for them to engage in that. For R1, having a robot that comes to her at her line of sight is going to be significantly more effective than having when where she's looking down at the floor.* - P1, Caregiver, Female

Given these engineering challenges, we asked the caregivers whether adjusting the screen tilt angle would help. One of them gave recognition, knowing the limitation we had.

*Anything that you can do to limit that would probably be better. The way you guys are thinking like if that (robot) was here every day, we would be using it every day. It wouldn't really matter – there's ideal heights, ideal voices, ideal volumes... but beyond that, again, it's just that allowing that repetition (of interaction) to happen over and over again.* - P2, Caregiver, Female

### 7.1.4 Discussion

Although the majority of the residents we dealt with had at least mild visual impairments, visual communication, surprisingly, still outweighs speech. The reasons are that visual information is more "reliable" and "tangible" than verbal information in real environments. People can refer to visual information as much as they want, in order to reinforce their understanding or memory of this information. Another reason is that caregivers rely on visual tools, gadgets, and programs to give training. Residents were familiar with this type of daily routine, and thus depend on visuals first. This is by no means to say that verbal communication is unimportant; we just need to synthesize these two primary types of communication effectively to achieve more desirable HRI.

In our interaction design, we tried to recreate an HRI experience based on the communication patterns that we learned from HHI. Based on our observation, we noticed that caregivers mostly use imperative sentences when they issue an instruction or request to residents. In the beginning, we found this way of communication is a bit harsh. However, "It seems harsh but that is how we are transmitting information." (P1, Caregiver, Female) It turned out that using directives is a highly effective way of notifying or guiding residents. By contrast, in our design, the verbal prompts were very polite and subtle, like "Could you get the book from your room?" This does not work for our users – for them, this kind of expression causes confusion – is that a question or a command? Hence, we have to compress verbal prompts to the extent that residents can get a clear command directly. In the meantime, we need to keep in mind that we are designing a sociable *service* robot. If we use 100% directives, it would seem that we are going opposite to the aim. For this reason, we need to closely observe residents' reactions and keep an eye on their experience.

## 7.2 Part 2

So far, we had conducted three studies in a controlled environment – the Snoezelen room where users did not have other residents as company. To various extents, these environments were all somewhat simulated and controlled to reproduce the user's environment (Portet et al., 2013). However, these environmental settings may not help us identify some usability problems that could be found in real-life conditions. In Part 1, there were some limitations when we explored close distance due to the room size. The robot could not move to the public zone or come to residents from their back. Also, the orientation of the robot was kept faced to residents. Thus, in this latter part of the study, we would like to observe how close distance differs far distance in terms of users' reaction and experience. We further continued User Study 3 as an *in-situ* (in the real environment of the participants) study, in order to discover the impacts of proximity in HRI. We created and used a new GUI demo and had a more flexible control of Aether's motion and location.

When it comes to proxemics, there are four dimensions that can be taken into the investigation: the distance between a user and a robot, speed, orientation and movement status (moving/static). Given that task completion has not been working as an effective approach to objectively examine the user's engagement with the robot, we have decided to use a questionnaire and interviews with caregivers who observe the study. This study is not a strictly experiment with controlled variables due to the size of the participant group. Instead, we will observe the effects of the following factors concerning social distance in this user study:

- Physical distance
- Speed
- Orientation
- Motion (Continuously moving / Move & Pause / Time to move )

Listing 1 shows the narrative script used to generate audio using IBM TTS service (IBM Corp, 2017). This TTS service was one the best we could find, as it provides a range of parameters for developers to adjust based on their needs. We selected a warm, female voice for these verbal prompt as most of the caregivers at the group home were female.

```
<voice-transformation type="Custom"  rate="-80%" >
Hello, this is Aether! Nice to meet you again! How are you today?
</voice-transformation>

<voice-transformation type="Custom"  glottal_tension="-100%" rate="-70%" >
I would like to show your activities this afternoon. Do you have any plans?
</voice-transformation>

<voice-transformation type="Custom"  glottal_tension="-50%"
pitch="-40%" rate="-100%" >
You have three things coming up, just like last week.
</voice-transformation>

<voice-transformation type="Custom" pitch="100%" pitch_range="100%"
 glottal_tension="-30%" rate="-80%" >
Alright, now let's get started!
</voice-transformation>
```

Listing 1: Sample Narrative Script in SSML (for the introduction scene with a smiley face)

For caregivers, the robot is a new role introduced to the process of their everyday work. As we were looking at design implications, we noticed the significance of the roles and functions of the robot and how they catalyzed the incorporation of the robot into caregivers' process. Compared to those previous research exploring the effects of social distance, as discussed in Chapter 2, we had a rather different situation. Therefore, simply altering the factors would not produce valid experimental results. Through the first part of this user study, we had three representations of interaction that residents were already familiar with, so we kept them for this user study. Due to the demographics of our participant, we did not have a chance to explore all factors, or to validate hypotheses methodologically. Besides, normal experimental manipulations, which usually used summative tests as the evaluation method, did not apply to our user study – it would not be feasible to run any statistical test for a sample size of three. For these reasons, we adopted the formative testing method which allowed us to get a closer and more precise observation of users' behaviours and responses.

We held the following conjectures:

- Participant R1 tends to get intimidated by people – especially strangers, at the intimate distance. Thus, she will glance at the robot once in a while when the robot is not very close (distance $> 0.46$m ). However, she will show noticeably resistance and anxiety when the robot passes that boundary.

- Participant S1 will not show any patent sign of fondness or discomfort. She might answer a few of the many questions in the verbal dialogue.

- Participant C1 would answer every question she hears. She is generally fine with a robot's movement and will stare at the robot the entire time. She might get a bit uncomfortable when the robot is approaching her beyond the personal zone.

### 7.2.1 Procedure

The user study took place at the group home shortly after 2 pm. Unlike the environment setting in Part 1, this study was conducted at the common area, next to the kitchen and residents' bedrooms. There were 5 caregivers on duty, along with 6 residents resting, eating, or watching the TV in this public space. That being said, the environment is quite noisy. Nonetheless, we did not make any change to the room setting because that was really their everyday space.

Prior to the test with the residents, we demonstrated the entire process to 3 caregivers at the Snoezelen room and asked about their thoughts. The pilot study with caregivers would help them get more comfortable with having a robot in their process. These caregivers were considered advisors to us for this study. Besides a quick informal interview, we also designed a questionnaire as following –

On a scale of 1 to 10, with 10 being the most effective or necessary, how would you evaluate:

- the visual readability of the graphical elements on the display;
- the wording of the verbal dialogue, including pitch, tone, speed, etc;
- the motion of *Aether*, including smoothness, speed, and orientation;
- the necessity of transforming graphical elements based on the change of distance;

- residents' trust and acceptance of the robot, now that the distance can be controlled (would there be an obvious increase?)

Per the previous inquiry, we learned that residents usually water plants, play puzzle games, and have a tea break in the afternoon. Due to the time, we changed the **lunch** scene to **game**, the symbol of which is as familiar as that of lunch to residents. Since **game** and **tea time** are two desired scenarios to residents, we kept **chores** deliberately to observe the possible differences of response when this scene is triggered. Due to the limitation of room size, we had adjusted the distance of public zone – the robot came from the back of each participant. We had fixed the speed issue when controlling the robot, so *Aether*'s movement could take place slowly.

We drove *Aether* to the front of the participant. *Aether* started from the social zone with a smiley face. Each symbol, including the smiley face, started from a small size with a height of 200 px out of 720 px on the display. We initiated the zoom transition by using the keyboard on the server machine, making the size of the icon three times larger. All images resources were designed to be fixed in the "absolute centre" throughout the study. Each complete transition took 30 seconds to complete, accompanied by verbal dialogues. Immediately after the transition started, we moved the robot closer to the resident from the social zone to the personal zone. At the boundary of the intimate zone (i.e. 0.46 m), the transition completed, and the symbol faded away. We then changed the symbol back to the "face" at the height of 200px. We started the same process again, after moving *Aether* back to the social zone. The difference is that for the last scene, we started the interaction from the intimate zone. The symbol would start at its maximum size (Height: 620px), and the transition went on when we moved the robot back to the personal zone, and then to the social zone. For the design of this study, the senior caregiver suggested that we implement both the face symbol and activity icons into the demo. The switch between these two types of symbols helped residents identify the robot's anthropomorphic identity and understand the difference between these two states.

As we made modifications after the previous user study, we now had flexible control of Aether's motion and graphic transition. We controlled Aether's movement using an Xbox game controller, and switched scenes using the keyboard. Thus, we could always try one of the scenes again if there was no response from the participant in the first attempt. The user study with residents was followed by an open-ended interview with caregivers.

### 7.2.2 Data Collection and Analysis

We used the same technique for data collection and analysis as Part 1 (Section 7.1.2). As we could not measure the distance during the study, we approximated calculated the distance based on *Aether*'s size and proportion (14 W x 14 L x 34 H in) when analyzing the video data.

### 7.2.3 Results

Figure 7.5 shows the caregivers' responses to the pre-study questions after they observed the demo. Overall, their perspectives roughly align on the visual readability and the necessity of transforming graphical elements based on the change of distance. As their responses have divergent views on the other three aspects, I adopt their feedback critically for these three. We rely on the after-study interviews to gain more insights into these considerations.

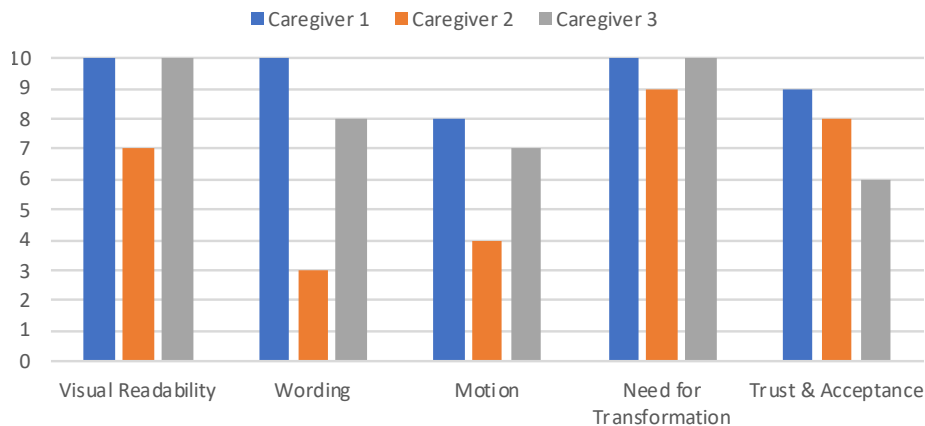

Figure 7.5: Results of caregivers' responses from the pre-study questionnaire

As *Aether* was only half residents' height who were sitting on wheelchairs, Participant R1 did not even notice the existence and slow movement of the robot in the beginning. She was looking at the caregivers preparing food and seemed to be thinking of something. After about half a minute, she finally saw *Aether*, but moved her focus to the room after a 5-second glance. R1 did have another glimpse of *Aether* when it is moving away, at about a meter away from her. However, that glimpse was so short that it was almost unnoticeable. Basically, she stopped looking at the robot for the rest of this demo. For the second round, we had to move *Aether* right in front of her, about 20 cm from her wheelchair. She had another look for a couple of seconds and turned her head away. In sum, she only looked at *Aether* for more than a few seconds when it is right in front of her and very close to her.

By contrast, the other two participants were more engaged in the demo. C1 was trying to find the robot when she heard its motion – she was already engaged before *Aether* showed up. Due to the fact that C1's wheelchair was right in front of a table, we could not move *Aether* to be right in front of her. While *Aether* was moving back and forth between C1's intimate and personal zones, she smiled and even replied "Good morning!" to *Aether*'s greeting. C1 nodded when Aether was giving the instruction: "Get the book from your room." She was watching *Aether* move away to the next participant.

The third participant, S1, paid attention to *Aether* when it was entering her intimate zone far from the social zone. She looked a bit confused, but recognized *Aether*'s presence with a murmur"Good morning", replying to its greeting. Afterwards, she looked at her caregiver for a couple of seconds. She seemed to get a bit more nervous when Aether moved so close that it was almost having physical contact. Throughout the study, S1 would "monitor" the robot as long as it was in her visual field. Later, she became very happy when she heard *Aether* mentioning "lunch", and even tried to talk to Aether proactively (when *Aether* was in her personal zone). There was an unexpected postman coming to the group home, and S1's attention was fully attracted to him, until *Aether* started to give verbal prompts.

We had obtained a baseline for comparison after inquiring about the responses that caregivers get when they told residents, "It's time for lunch." We found the frequency that caregivers need to remind them of scenarios like chores and tea breaks. In our previous study (User Study 3, Part 1) we found that the interactions had imperceptibly associated purpose other than as visual stimulus. Overall, the robot was incompetent to engage participants into anything or to trigger any response.

For the conjectures that we had prior to the user study, only the second and third ones were partially correct. We did not notice any resistance from R1, or any discomfort from C1. S1 showed obvious pleasure at one point, although there were no other facial emotions most of the time. S1 was

A caregiver commented on constructing contexts to increase residents' engagement and experience. The contexts help to make the interaction between users and the robot more personal and intentional.

*If we can make it really relevant to what their day is, I think that can become a part of their day, just like anything. We have visuals, schedules and calendars*

*that become part of their day. They can continue to engage with it. Then you switch it up a little bit, like adding an activity and change the colour. That usually helps.* - P1, Caregiver, Female

During the user study, C1 was very engaging in it, whereas R1 seemed to be distracted as always. This was well expected, based on our understanding of users' characteristics. Still, there might be some engagement for R1 even though she seemed to be distracted.

*R1 always turned sideways. Even if I came up to her and said hi, she was excited and she would turn sideways – she doesn't do a lot of eye contact. So when she gets excited, her hands are like this and she turns, I interpret that as she is excited. Eye contact for her or listening to people is always hard for her. But she still acknowledges that, even when it is a bit lower for her, you know it's not her eye level.* - P1, Caregiver, Female

The caregiver remarked that S1 was "was very curious" (P1) because S1 was visually tracking the robot even when it was moving behind her.

In the end, we asked caregivers about the significance of motion and distance in HRI. They seconded the introduction of proxemics to the interaction, which makes Aether different from many of the service or concierge robots that are mostly static.

*If it is static, individuals tend to not be there or just simply turn away. But it's moving, it kind of catches their attention. "Hey there, what is going on?" So it is engaging, right?* - P1, Caregiver, Female

### 7.2.4 Discussion

In this user study, we aimed to observe differences, rather than modulating the social distance to get emergent common results. The focus of this study is on behaviour. For example, a question we had was: what happens when we introduce the robot into dynamic social interaction? Therefore, we used the simplest morphing transitions to observe participants' behaviours for the social interaction given that they were familiar with the icons. We replicated caregivers' daily procedures to remind the residents of certain activities. Regarding the social distance, we had the following main research questions to examine:

- Does *Aether* need to be closer? If so, when does it?
- What happens when the robot gets closer?

We had the goal to explore the impetus and deterrents to the development of trust. In this part of User Study 3, we found that residents' interest and curiosity boosted their engagement, and thus they started to build a more intimate and friendly relationship with the robot. Certain level of spacial stimulation (e.g. unexpected appearance or movement) can be applied to residents who exhibited higher acceptance to increase the engagement, whereas for residents who tended to show resistance or anxiety, smooth and slow interaction is the key to building trust in the long term. On the other hand, distraction, annoyance, and alarm obstructed residents' engagement and acceptance. The robot should avoid having unexpected proxemic interactions with residents with low levels of intellectual capabilities because they are more likely to struggle with understanding such interactions that they rarely encountered previously.

In the course of building and maintaining a positive relationship with the robot, users get affected by multiple factors causing or reducing the sense of reliability (Hancock et al., 2011). Those factors consist of several different dimensions, including the robot's performance and attribute, user's ability and expectations, and cultural awareness of modern technologies. Not all factors of engagement are controllable by designers, but we have identified several critical ones as the top priority: robot's level of automation, personality, proximity, and behaviour. User's engagement in HRI is a manifestation of psychological safety, which means that the social interaction with the robot does not bring anxiety or resistance in the long term (Lasota et al., 2014). For our research project, a challenge in the close-proximity HRI is user's passive acceptance of a robot's service or company. Unlike many other social robots that have been designed for customers, residents or people who might actively need assistance, our robot is developed for residents who may not express their need proactively. Or they tend to turn to their caregivers for help and inquiries. How to engage our residents in unexpected scenarios, such as the robot approaches user from far and ask if they need any help, is a question that we will explore in the future.

# Chapter 8

# Conclusion

This thesis describes an exploratory effort to examine HRI in the context of service, assistance, and companionship. The long-term goal of this research is to develop a socially assistive robot for the vulnerable population suffering from developmental disabilities. We aimed to come up with meaningful social interaction design that could reinforce robots' role as a social agent between people with DD and their caregivers.

Overall, our results from three user studies have met the main research objectives:

- **Research Objective 1: Design GUI prototypes to identify residents' challenges and needs.** I designed GUI prototypes based on caregivers' advice and feedback, and observed how residents behaved and reacted to our design. I selected several fundamental design principles and tested them in prototypes. I examined the effects and desired design of each aspect I selected. After studying the effect of each visual or interactive factor, I drafted preliminary propositions for designing robust human-robot interaction for users with visual or cognitive impairments. We have gained findings regarding designing visual language for our user group. For instance, people with DD are very sensitive to colours. Through user studies, caregivers suggested that highly contrast visuals and consistency of using colours in the GUI help residents perceive and understand the information. It is advisable that the background colour of the GUI is kept one colour – it could be simply white or a pale colour (e.g. orange) throughout the demo. A smooth, gradual transition of colour is not a problem for the general population, but for our user group, it raises users' confusion.

- **Research Objective 2: Investigate how residents experience the technical interactions we designed.** I designed new GUI prototypes, compared the effects of different types of interaction, and identified lessons that we can learn from residents and caregivers' interaction. We confirmed that the robot plays a unique role in social interaction with residents. Compared to HHI, the introduction of technology creates possibilities of enhancing residents' social lives, although it also brought problems: the disengagement due to the lack of trust. There were many factors contributing to this shortcoming, including the physical embodiment and the "loose" communication. Residents knew that it was simply a computer when it was playing the demo, and the information from the demo was quite general. By contrast, caregivers, with whom residents were very familiar, could have personalized communication with residents. On the other hand, the robot increases the trust because of its human-like features such as motion.

- **Research Objective 3: Explore how motion and social distance affect trust and acceptance in HRI.** Trust and acceptance are of critical significance in social interaction. I observed how residents reacted when their spatial perception got stimulated and assessed their experience when the personal space was being "violated". I also compared their responses (e.g. gaze) when the robot keeps moving versus being stationary. Our exploration of proxemics showed that the motion of social robots enriches robots' social behaviours and enhances users' engagement through modulating users' spatial perception. This finding suggests that motion design of social robots should be taken into consideration as many of existing social robots are static, such as *Valerie the Roboceptionist* (Gockley et al., 2005).

We presented our iterative design process in the user studies, and this approach has helped us achieve a user-centric design. Through three user studies, we gained insights to this very special population, who shared so many similarities with other demographic groups like people with Alzheimer's, MCI (mild cognitive impairments), autism, or simply the elderly. At the same time, we found what makes our user population special: their challenges, needs, expectations, etc. We had preliminary verification of some of our assumptions regarding interaction design in the context of human-robot interaction. For example, we assumed that a robot increases users' engagement compared to a computer screen showing the same content, and this aligns with the results in User Study 2. At the same time, we also found some of our assumptions opposite to reality. For example, prior to User Study 1, we had a conjecture that the character type on GUI would significantly affect users' acceptance of the robot, which was contrary to our observation – the character type did not matter to users.

Our study reveals the significance of visual hierarchy, verbal communication and proxemics in HRI. In the past, those three aspects had rarely been taken into consideration collectively by researchers of social robotics. The synthesized use of them determines the quality of social interaction. Visual hierarchy is a fundamental design principle. We discovered it at the beginning of our project, but only learned of its importance after the user studies. The user studies made us realize that we actually have a rather limited visual vocabulary in terms of designing for people with DD. Their cognitive impairments meant that we had less flexibility to design the GUI. Verbal communication refers to the wording, tone, and type of communication (derivatives, questions, etc). Although our study did not focus on speech at the beginning, we found it indispensable and critical to the user's acceptance.

As designers, we need to keep the overall objective of the workflow in mind while tweaking details of each part. Besides, a single user group may contain many individual disparities. For example, even though we had a rather small user group of 3, we could still notice the evident divergence of cognitive abilities, skills, characteristics, preferences, etc. We should also think out all levels or categories of users in the same group, and create responsive design for them. For example, some people with higher skills are able to handle a richer visual language with animation. In this case, using animation would be a good way to communicate with them. For the rest of users, keeping a simple visual language would be a better choice.

One of the reasons that residents and group homes need SARs is to reduce residents' anxiety and confusion. Many group homes in North America are understaffed, and the frequent change of caregivers' schedules brings instability to residents' lives. (B. Morris, 2017) The SAR was not designed to replace the caregivers; instead, it was developed to help residents better engage in their social lives. Thus, it needs to have the ability to provide feedback whenever a resident wants to inquire about their schedules or emergent upcoming events. Certainly, a prerequisite for reducing residents' anxiety and confusion is the trust and acceptance from them.

By conducting user studies, we have sensed the potential of socially assistive robots for improving the live quality of our users. We also have learned much more about the special needs and behaviour patterns of our users from their caregivers. Caregivers' insights are invaluable to our future designs as we reiterate the prototyping development. Therefore, this study lays a solid foundation for future studies.

## 8.1 Research Contribution

Currently, there are not many robots specifically designed for people with developmental disabilities. Our robot can potentially have a great impact on their daily lives. Our graphical interface design and exploration of robot's behaviour will contribute to the research area of social interaction design in social robotics. Our results from studying the communication patterns of people with developmental disabilities could possibly help other researchers or psychiatrists focusing on the same or similar user population. This study, as research work of using socially assistive robots, can also benefit researchers and developers working on apps, computer programs, or other assistive technologies for people with mild or severe cognitive impairments.

In this study, we made many propositions for HRI design for people with developmental disabilities. For example, the visual language needs to be concise and be based on users' real-life experience. For this user group, mentioning anything that is off the context will be invalid due to their limited abilities to dual and cross reference. Their memory functions are generally low, and so are their cognitive skills due to mental impairments. Therefore, any piece of information from the robot needs to be very clear, simple, and repeated to reinforce users' perception and acceptance. The visual information should be based on simple photos and graphics that residents have already learned from their regular training. The introduction to new graphical elements will be a risky move because of their low cognition.

We also had discoveries regarding speech in HRI. For our user group, using questions can be an auxiliary type of information, but it is less effective than using directives when it comes to asking residents to perform tasks. Sometimes, they will respond to the questions, but the more they interact with the robot, the less likely they will respond, as sooner or later they will notice that the robot is not able to return a personalized response for theirs. Using directives can be a feasible solution to engage residents in social communication, as it suggests in User Study 2 - Part 3. However, the verbal communication, similar to the visual language, has to be extremely simple to the extent that each directive contains no more than five words, such as "take a shower now."

## 8.2 Limitations and Future Work

A limitation of our study was the time delay between scenes. Although we made the process of presenting information fairly slow in our prototype, it was still not slow enough. The caregiver pointed out that on average an elderly with cognitive impairments need at least 50 seconds to process a piece of information.

To enhance the replicability and to reduce the complexity, the WOZ experiments in the user study were semi-automatic, which meant that the wizard did not have maximum control of the prototype while the user study was ongoing. The wizard could only initiate or pause the prototype, and move the robot. There could not be any improvisation during the user study. This had been a bottleneck for the robot to adjust its pace to cater to users' sluggishness.

Besides this above-mentioned challenge, we were also faced with some engineering difficulties. One of the most important challenges was the height of *Aether* – it was unadjustable. The robot was simply not tall enough to be noticed by some residents in a real environment. Although this issue was pointed out by a caregiver in the part of User Study 3, we could not make any change for the next study. A similar issue we found was the speaker of the robot – its volume was a bit low so that the verbal communication became less effective. We will need to work with the engineering team of our research project to fix these issues.

Admittedly, a limitation in our study was the size of our user groups, although the three participants have different cognitive abilities. Nevertheless, as the manager of the group home suggested, individuals with developmental disabilities are a very special user group, to whom regular usability tests may not apply. It is more significant and effective to design and evaluate interaction models with several individuals for a long time than to carry out standard tests with a large group. This being said, this study is not ready for generalization at this stage, although it was not a goal at the beginning anyway given the exploratory nature of this research.

We would consider continuing the future work by fixing existing issues first, such as the height and the audio set-up of the robot. Many of the design implications mentioned earlier are worth more exploration. For example, we are curious what happens when we implement a "binary" verbal communication mechanism into our current HRI design. In this case, the robot would simply rephrases all questions to a simple "Yes or No" question instead of an open-ended question. This could probably trigger more responses from some residents, but the frequency could decline over time. We need to reframe conjectures like this into an integrated study and explore social interaction design in depth.

# Appendix A

# Forms & Questionnaires

# Study Information

**Title:** Graphical User Interface Design for Social Robots Assisting People with Developmental Disabilities
**Principal Investigator:** X. Wu
**Research Supervisor:** L. Bartram
**Department, School or Faculty:** School of Interactive Arts & Technology, SFU
**Contact:** xavi▮▮▮▮▮▮u.ca
**Research Sponsor**: JDQ Systems Inc., Vancouver, BC, Canada

**Purpose of the study:**

The purpose of this study will be to understand the language patterns of both participants with developmental disabilities and their caregivers. Our research team is currently exploring and evaluating different design possibilities for creating a context-based graphical user interface (GUI) on social robots assisting participants. The overall and long-term objective is to achieve a better design of socially capable robots and thus to substantially improve the quality of participants' lives.

**Study Goals:**

In this study, you or your caregiver will be interviewed as a preliminary investigation and analysis on how caregivers and you communicate. Through this process, we will record and evaluate different speech patterns, including pace, tone, and specific phrases. Results will contribute to better understanding of comprehension, reassurance and acceptance by you through a series of in-situ studies with you and your caregivers. The results obtained from this study will help us with the design of user interfaces for assistive care robots for people with cognitive disabilities.

After the interviews, you will be provided with a prototype to interact with verbally and we will begin by interviewing you about your experience of using our prototype to evaluate the communication efficiency using our designed prototype. Through this process, we will evaluate your understanding and experiences interacting with a user interface on a screen installed on a mobile robot base. Results will contribute to better understanding of comprehension and acceptance by you through a series of studies with you and caregivers.

**Method:**

The investigator will start a semi-structured interview with caregivers, or you with your caregivers being present. Interviews are estimated to take 30 minutes or less for each interviewee. Questions in this interview may include, but not restricted to the following topics: your daily activities/jobs, social life patterns, types of assistance that you need from the caregivers, preferences for media entrainment (YouTube, Netflix, TV shows, etc), and how you make decisions for trivial things like choosing a shirt. At all times, you will be accompanied by a Developmental Disabilities Association (DDA) caregiver. Data including your responses and comprehension, and caregivers' assessment will be recorded.

This study will comprise a series of prototype-based experiments with you with the companion and supervision of your caregivers. Studies are estimated to take 60 minutes or less. You will be provided with a prototype device with a GUI implemented for use and tests. You will interact with an audio or visual simulation of a GUI which may be human or may be generated by a computer. A post-session interview will be conducted to collect your impressions of the prototype's utility and design suitability, which may be related to health, fitness, or other kinds of personal data of you. You may be asked to try to do simple tasks according to normal day-to-day interactions as identified by the caregivers. Examples may include you choosing a shirt based on recommendations from our designed social robot, or asking the robot questions in natural language like "Hey check the weather for me." The task completion time, cognitive and emotional responses, and user acceptance will be tracked during or after the studies. Users' responses will be observed and recorded for qualitative analysis after the study using video and audio. Data including your responses and comprehension, and your caregivers' assessment will be recorded. Through these qualitative and observational studies and interviews, we will execute several studies that explore visual and gaze-based interaction with you with the aim of determining more targeted design requirements.

**Data Collection**

The data collection of this study will take place mainly in the form of semi-structured interviews. There is a list of questions and topics that may be covered during the conversation, usually in a particular order. The interviewer will follow the interview guide, but is able to follow topical trajectories in the conversation that may stray from the guide when he finds this is appropriate. You may pause or terminate the interview at any time, and you can ask questions or add additional thoughts and feedback during the interview. Responses will be noted and short notes will be taken. Interviews will be audio recorded and transcribed.

The data collection of this study will take place mainly in the form of interviews. There is a list of questions and topics that may be covered during the conversation, usually in a particular order. The interviewer will follow the interview guide, but is able to follow topical trajectories in the conversation that may stray from the guide when he finds this is appropriate. You may pause or terminate the interview at any time, and you can ask questions or add additional thoughts and feedback during the interview. Responses will be noted and short notes will be taken. Interviews will be audio recorded and transcribed.

**Potential Harms, Risks or Discomforts:**

It is not likely that there will be any harms or discomforts from our study. The questions in our interview guide are insensitive in nature. You do not need to answer questions that they do not want to answer or that make them feel uncomfortable.

**Confidentiality:**

Research records will be kept confidential to the extent allowed by law. A research archive will be created containing data collected during these research procedures, and will be separated from all identifying information (such as your name and contact information).

**Benefits:**

You will not receive any payment for your participation in this study. However, you will have the access to our prototypes and study progress at any time during this study. After the final prototype or product gets completed, you will have the priority to use it.

**Dissemination of results:**

Your identity will not be found directly or indirectly when we provide the results of the research, but the small number of you in this study may compromise strict confidentiality, as we will need to describe certain aspects of participant's daily lives, such as the types of activities you do regularly and the community events that they are interested in. We will not disclose your address or other personal details, but in some cases, the description of the living environment may potentially provide some small clues to your identities. If you are interested to know the results, it will be made available by publishing our study in SFU server and made available online. You can contact the principal investigator listed above if you are interested to study the final results of the research.

**Participation and Withdrawal:**

This study is conducted with the permission from Developmental Disabilities Association (DDA). You need to be over 20 years old in order to participate in this study. Your participation in this study is voluntary. It is your choice to be part of the study or not. If you decide to be part of the study, you can stop/withdraw, from the interview for whatever reason, even after signing the consent form or part-way through the study or up. There will be no consequences to you if you decide to withdraw. In cases of withdrawal, any data you have provided will be destroyed unless you indicate otherwise. If you do not want to answer some of the questions you do not have to, but you can still be in the study. Should researchers need to re-contact you as part of the study or after the study, you will be given an option to approve that re-contact request.

By completing this form, you:

1) understand to your satisfaction the information provided to you about your participation in this research project, and

2) agree to participate as a research participant

In no way does this waive your legal rights nor release the investigators or involved institutions from their legal and professional responsibilities. To accept this form, please print your name below, sign, and date the form.

Participant's Name: _______________________________

Participant's Signature: _____________________________

Date: _____________________________ (YYYY/MM/DD)

**Questions/Concerns**:
Should a participant have any concerns or complaints about the way he/she has been treated as a participant or questions about the research project, please contact:

*Dr. Jeff Toward, Director, Office of Research Ethics, 778-@sfu.ca*

For other questions and inquiries about this study, please contact:

*Dr. Lyn Bartram, Associate Professor & Graduate Program Chair, SIAT, SFU, 778-7sfu.ca*

**Links:**
1. JDQ Systems Inc: www.jdq.com

2. Developmental Disabilities Association: http://develop.bc.ca

**Research Interview Guide (For Caregivers)**

Study Number:

Study Title: Graphical User Interface Design for Social Robots Assisting People with Developmental Disabilities

Principal Investigator: Wu, X.          Supervisor: Bartram, L.

SFU Position: Graduate Student Faculty/Department: Interactive Arts and Technology

1. Introduction (specifies the topic, the setting, the number of participants)
   1. **Name your Research Interview Guide**

      A Preliminary Research Interview Guide for Caregivers and Volunteers for People with Developmental Disabilities

   2. **Type of Interview Guide (Semi-structured, Focus-Group, Open-Ended)**

      The interview guide will be semi-restrictive open-ended. A general outline of interview questions will be used, but other questions generated spontaneously can also be considered and used according to the responses of the participant.

   3. **Describe the Setting (location and reason for choosing this location).**

      The location will be the following DDA (Developmental Disabilities Association) home:

      ***Location:***
      Camsell Group Home,
      Richmond, BC V7C 2M9  CANADA

   4. **Describe the Participants (how many, cultural characteristics, gender, age-range, etc, if applicable)**

      There will be roughly 5-10 participants who are all over 25 years old. They are all from Greater Vancouver.

   5. **Documentation (e.g., video, audio, digital photography)**

      Video and audio recording will bring us high degree of reliability, validity, and legal defensibility. They are helpful for identifying who is speaking and for improving the accuracy of the research by replaying sessions during analysis.

   6. **Use of Additional props (digital props, low- or high-fidelity prototypes, cards, images, bodymaps, etc)**

      No. For this project, pops are not necessary to improve the efficiency of this interview.

2. **Research Questions (outline the research questions that you are interested in answering through this interview)**

Before this I will do a brief self-introduction:

"Hello, my name is ▓▓▓▓▓. I am a master's student studying human-robot interaction at Simon Fraser University. I am currently doing a research that uses social robots to help people with developmental disabilities achieve better lives. We are interested in studying the language patterns of people around these patients, just like you. Do you mind me taking you a few minutes to answer several questions? Before we go ahead, do you have any questions?"

(Imagine the questions are asked by interviewees)

- What is a social robot?
- How can social robots improve the life quality of our patients?
- Will they replace caregivers completely?         (No…)
- Which area of social robots are you researching?
- Is there any similar research or previous demonstrations of social robots? (Yes.)
- How can my responses help your research?
- What are speech patterns and are they really impacting our work as caregivers?

3. Research Topics (what research topics are defined by your research question)

   Social robots and speech patterns are defined by the questions above.

4. Interview Questions (from the Research Topics list the Interview Questions that you plan to use in the interview)

   - Could you tell me what your everyday job includes?
   - What are the most common kinds of assistance for people with developmental disabilities?
   - How do you talk to your patients/clients? Do you need to change tone or words?
   - Could you talk about your training? How did they train/teach caregivers?
   - Do you have any problem understanding them while communicating with them? (followed by the next question)
   - If so, what do you do to solve this problem?
   - What's your primary concern or difficulty for your job?
   - Do you think robots can help you to do a better job?
   - What do your patients do every day? (i.e. common daily activities)
   - Do you use any kind of technology to assist you with your work?

5. Activities (include activities that you are planning to use during the interview or focus group). Activities can help to direct attention to a design problem, or to the experience of the interviewee. For example: for a semi-structured interview of performers, this could be having the performer review video footage, for a wearable design focus group, this could be having the participants view images, video or an actual prototype such as a wearable interactive garment itself.

   I will play two video clips about social robots for the participants after they ask me questions（between III and IV） so that they can better understand our research project.

   - Sweden: 'Social Robot' designed to recognise very early stages of Alzheimers
     https://www.youtube.com/watch?v=w2Y7yPYPFbE
   - Jibo: The World's First Social Robot for the Home
     https://www.youtube.com/watch?v=3N1Q8oFpX1Y

# DDA Group Home Resident Information Form

## Resident's Information

Resident Name: ______________________ Age: ____________ Gender: M / F

- On a scale of 1 to 5, how do you evaluate resident's interest in the following motivators?

| Type of Motivators | 1 (Ineffective) | 2 | 3 | 4 | 5 (Most Effetive) |
|---|---|---|---|---|---|
| Food | | | | | |
| Outdoor Activities / Outings | | | | | |
| Board Games | | | | | |
| Other Simple Games, e.g. Puzzle | | | | | |
| iPad Games | | | | | |
| Indoor Socials | | | | | |
| Music | | | | | |
| Movies / Videos /TV | | | | | |
| Personal Attention & Company | | | | | |

- Could you summarize this resident's characters? For example, does he/she talk a lot or prefer to stay alone? Is this client outgoing/social?

- Could you list things appealing to this client, besides those listed above?

- How does this client use iPad? What kind of prompts does he/she use or need?

- What type and level of assistance does this client need?

[                                                                          ]

## General Questions

The following questions are not specific for the resident individuals.

- Do you use wall charts to record progress at this DDA home? "Many adults with developmental disabilities will enjoy seeing their names on the board and following their progress as they increase their physical activity levels. A wall chart can show a photo of the person exercising with blank boxes next to his or her name. Upon completion of the exercise session, the person places a checkmark in the box. Some adults like to keep track of how many boxes are filled with checkmarks. (NCHPAD)"
- How can we improve the reward system? Should we also have a competition system by which residents can compare their progress with others at the same group home?
- There are situations where the robot is unable to assist the client, or is only able to provide partial support, including: guiding the robot to take a shower or to go out. Even for simple tasks like doing laundry, there might still be safety concerns because of water and electricity. Do you have suggestions that we can tackle this challenge?
- Are there prompt tools besides the pictures and the iPad app that were shown before?
- If we need to define two or three main tasks to assist, what are the best ones to begin with?

Reference:
http://www.nchpad.org/104/795/Developmental~Disability~and~Fitness

Version: June 26, 2017# Interview Questions for DDA Caregivers

- Have you ever come across a situation where a client tested/challenged your patience? (E.g. refusing to take the medication.)

- What are the typical skills necessary to perform the caregiver job at DDA?

- As part of this job, you'll have to do routine tasks. What are some of the irregular tasks and how comfortable are you with that?

- When you are working with clients with developmental disabilities, what steps would you take to ensure their well-being?

- From your experience and observation, what are the differences between developmental disabilities, Alzheimer's, and dementia? Does any of your clients have dyslexia?

- What types of clients with developmental disabilities have you worked with? (Assuming there are multiple kinds/categories of developmental disabilities.)

- What is the most difficult part about working with people with developmental disabilities?

- Does your organization (i.e. DDA) use any kind of technology for personal care/hygiene tasks that are needed for an elderly client? (E.g. Roomba Robot Vacuum) If not, what do you think are the reasons that it is not being used here, besides financial affordability.

- Could tell me about a mistake you made while caring for a client? (Is it caused by any misunderstanding of communication?)

- Do you have experience with community inclusion activities with the developmentally disabled?

- In terms of company, to what extend do clients need people to stay with them (not for immediate help, but only for company). Do you think a social robot can help or worsen this need?

- What are your concerns about social robots being used in institutions like this? Do you have doubts of their functionalities or efficiency?

Version: July 2017# Interview Questions for DDA Caregivers

### Verbal

1. How would you assess clients' capabilities to read and comprehend textual information? Do they read newspapers or books regularly?

2. What do you think is lacking in your communication with clients? Are they able to grasp most of the important information you said or you have to repeat occasionally?

3. What kind of entertainment technology do clients use? For example, watching YouTube videos on a tablet.

4. Does any of your clients have visual limitations or impairment? Do they prefer to read texts with large/extra-large fonts?

5. What kinds of interactions do you usually have with the clients where communicating information is critical? For example, do you write things down and show them?

### Other Cognitive Skills and Abilities

6. Are clients able to discriminate between different colours promptly?

7. Do clients have severe memorizing difficulties? (This way we can decide if we need to use similar functions for doing different jobs or to add descriptive texts and guidelines for tools.)

8. From your experience and observation, what are the differences between developmental disabilities, Alzheimer's, and dementia? Does any of your clients have dyslexia?

9. What kinds of visual imagery do the clients respond to? Again, are there types of images or scenes that certain clients prefer? (fuzzy kittens, beach scenes, …)

10. What are the non-verbal cues and elements in responding to/communicating with the client that you think are important?

    a. Do these differ by client, and if so how? For example, do you need to gesture with a client, or make sure your facial expression is a certain way?

    b. For the non-verbal clients, what are the most important aspects of nonverbal communication from you? From them? What do each respond to best? E.g. tone of voice, slow gesture, facial expression, all.

### Personality and Individual Differences

11. Is there any individual difference among your clients in terms of reading and other cognition abilities, or are they about the same?  (To make it more specific, how many of your clients can read. And if they are verbal, what kind of verbal interaction/phrases work well?) What are the different constraints each client has?

12. How do clients ask for things? When the client initiates conversation, what are the different situations that determine how you respond? What are the verbal and nonverbal approaches you use?
    a. Ask for help
    b. Indicate upset, etc.
13. What kinds of things in a conversation trigger clients' responses, and how do you change your way of interacting in these situations? E.g.
    a. Speak more slowly
    b. Face client
    c. Introduce new aspects to the conversation (laugh, talk about or show something the client likes, etc.)
14. Please list the important different "atmospheres" or situational emotional contexts you need to be aware of in your interactions with the client. For example, when do you need to be firm? To soothe? To encourage or motivate? To instruct and guide??
    a. What are the different ways you adapt your communication to these situations?
    b. Are there individual client attributes that need to be considered in these emotional contexts, and what are they?
    c. How do you bring them in to your decision of how to proceed with the interactions? In other words, do you respond/initiate completely differently for each client?
15. How do you interact with a group of clients (or do you) differently than with the individuals?
16. In which situations is it critical to give feedback to the client, e.g.
    a. To encourage
    b. To prompt or redirect
17. Are there situations in which the robot can help with "show me" feedback (exercises, activities, etc)?
18. What kinds of non-verbal communication are important for the client as his/her main types of communication?

### Others

19. What is the role of the robot in the Snoezelen room? What kinds of visual and auditory elements are associated with it, and how might we bring it in to the more general set of communication aspects the robot has with each client?
20. In terms of companionship, to what extend do clients need people to stay with them (not for immediate help, but only for companionship). Do you think a social robot can help or worsen this need?
21. What are your concerns about social robots being used in institutions like this? Do you have doubts of their functionalities or efficiency?
22. What kinds of communication do you think the robot will initiate? (In what different kinds of circumstances?) E.g.:
    a. To change what a client is doing?
    b. To prompt or prepare for a change (reminders about appointments)?

# User Study 3: Pre-study Review

On a scale of 1 to 10, with 10 being the most effective or optimum, how would you evaluate:

1. **The visual readability of the graphical elements on the display**
   *Mark only one oval.*

   |   | 1 | 2 | 3 | 4 | 5 | 6 | 7 | 8 | 9 | 10 |   |
   |---|---|---|---|---|---|---|---|---|---|----|---|
   | Poor | ◯ | ◯ | ◯ | ◯ | ◯ | ◯ | ◯ | ◯ | ◯ | ◯ | Great |

2. **The wording of the verbal dialogue, including pitch, tone, speed, etc**
   *Mark only one oval.*

   |   | 1 | 2 | 3 | 4 | 5 | 6 | 7 | 8 | 9 | 10 |   |
   |---|---|---|---|---|---|---|---|---|---|----|---|
   | Poor | ◯ | ◯ | ◯ | ◯ | ◯ | ◯ | ◯ | ◯ | ◯ | ◯ | Great |

3. **The motion of Aether, including smoothness, speed, and orientation**
   *Mark only one oval.*

   |   | 1 | 2 | 3 | 4 | 5 | 6 | 7 | 8 | 9 | 10 |   |
   |---|---|---|---|---|---|---|---|---|---|----|---|
   | Poor | ◯ | ◯ | ◯ | ◯ | ◯ | ◯ | ◯ | ◯ | ◯ | ◯ | Great |

4. **The necessity of transforming graphical elements based on the change of distance**
   *Mark only one oval.*

   |   | 1 | 2 | 3 | 4 | 5 | 6 | 7 | 8 | 9 | 10 |   |
   |---|---|---|---|---|---|---|---|---|---|----|---|
   | Unimportant | ◯ | ◯ | ◯ | ◯ | ◯ | ◯ | ◯ | ◯ | ◯ | ◯ | Very Important |

5. **Residents' trust and acceptance of the robot, now that the distance can be controlled (would there be an obvious increase?)**
   *Mark only one oval.*

   |   | 1 | 2 | 3 | 4 | 5 | 6 | 7 | 8 | 9 | 10 |   |
   |---|---|---|---|---|---|---|---|---|---|----|---|
   | Decreasing | ◯ | ◯ | ◯ | ◯ | ◯ | ◯ | ◯ | ◯ | ◯ | ◯ | Inreasing |